\documentclass[twocolumn,aps,pra,amsmath,amssymb,showpacs,superscriptaddress]{revtex4}

\usepackage{graphicx}
\usepackage[utf8]{inputenc}
\usepackage{latexsym}
\usepackage{amsmath}
\usepackage{amsfonts}
\usepackage{amssymb}

\usepackage{ucs}
\usepackage[utf8]{inputenc} 
\usepackage{hyperref}  

\begin{document}

\title{
Orbital stability and the quantum atomic spectrum from Stochastic Electrodynamics 
}

\author{David Rodr{\'\i}guez}
\affiliation{
Departamentos de F{\'\i}sica Aplicada II y III, Universidad de Sevilla, E-41012 Sevilla, Spain}
\email{drodriguez@us.es}

\date{\today}


\begin{abstract}

\begin{center}
\textit{(slight revision, soon some numerical estimations)}
\end{center}

\textbf{High order terms in the electromagnetic multi-pole development expose a stabilizing mechanism
for the atomic orbitals in the presence of a random background of electromagnetic fluctuations.}
Boyer and Puthoff set forward the idea that for the Bohr orbits in the hydrogen atom, radiation
losses could be compensated by absorption from the QED-predicted Zero-Point Field (ZPF) background.
This balance is, on average over the orbit, a necessary condition for stationarity of the orbits,
imposing a relation on the pair $R_{0}$ (orbital radius), $\omega_{0}$ (orbital angular velocity);
such relation is simply what we have for long known as angular momentum (AM) quantization ($l=1$).
Nothing has been said yet, however, on how could this balance be attained on a quasi instantaneous
basis, in other words, how could the orbit accommodate the instantaneous excess or defect of energy
so as to keep constant the (at least average) values of its parameters ($R_{0}$, $\omega_{0}$).
Using classical electromagnetism, we explore some high order interactions between realistic particles,
exposing a mechanism (a feedback loop between variables) that makes that stability possible.
\emph{
Puthoff's work led necessarily to the quantization of AM: ``if stable orbits exist...
then their AM must be quantized''; now we are able to do a much stronger statement:
``the equations of the system, in the presence of ZPF background, lead necessarily
to a discrete set of stable orbits''.}

The job is done in two steps.
First, we work with an ideal electromechanical model composed of two (charged) particles:
one is a point-like magnetic source and the other one consists of a spatially ``extended''
charge distribution, orbiting around the point like one.
In the second place, a relocation of the inertial reference system and other minor changes
are done to reinterpret the former picture in terms of a realistic hydrogen atom, the two
situations turning out to be formally identical, but for some factor of proportionality.
\emph{
As a novelty in the subject, here it is the inner structure of the
nucleus (and not of the electron) that plays the crucial role in our model.}
The following step regards the existence (for each $l$) of an infinite and discrete
spectrum: 
\emph{the presence of a secondary feedback loop in the equations is crucial
for the feasibility of excited states}.
Up to the order of our equations and under some further approximations, our model
admits a continuum of stationary trajectories described by four parameters:
$\{R_0,\omega_0, R_1,\omega_1\}$;
the first pair $(R_0,\omega_0)$ corresponding to the orbital movement, the latter
$(R_1,\omega_1)$ to what we have called a ``secondary oscillation''
(may there from here a connection to Cavalleri and others' model of spin as a residual
helical oscillation of a point-like electron \cite{Cavalleri10}?).
Stationarity determines, for a given $R_0$, both $\omega_0$ and $\omega_1$ (imaginary
poles in the linearized frequency description of the system), while $R_1$ remains free,
allowing, once radiation/absorption (rad/abs) processes are considered, the power
balance that is necessary for stability,
while discreteness of the spectrum is now retained via an additional condition,
a phase relation $\omega_1 = n/\omega_0$ ($n$ integer) that we regard also necessary
for stability.
Such phase relation also allows us to arrive to an E-spectrum that could be made
to resemble, under some approximations, the well known $1/n^2$ one.
Of course stationarity is so far only a necessary condition for stability: more as an
intermediate step than a rigorous proof,
\emph{
we introduce a stochastic description of the rad/abs processes, in a way that makes
(at least) possible the negativity of the real part of the eigenvalues (poles in the
frequency domain) of the linearized description of the system.}

Obviously, circular trajectories can only give rise to $p$-orbitals (or at least
$l > 1$ ones):
\emph{an additional (infinite and discrete) set of trajectories with no net AM
can also be found, that would account for the $s$-spectrum.}
Both in the $s$ or $p$ cases,
\emph{an atomic transition implies an emission/absorption of a wave-packet
with energy $\hbar(\omega_i - \omega_j)$, $\omega_i,\omega_j$ being the (main)
frequencies of the oscillation (primary osc.) of the orbitals involved.}
Besides, for the feasibility of our model, we are led to think in \textbf{\textit{fully
relativistic orbitals}}, ``nodes'' absent at least for $l>0$ \cite{Szabo69}; anyway, the
unrealistic nature of pure states would also solve this problem, even for $l=0$ (Fritsche).
Finally, the action of the stochastic background on the oversimplified, entirely deterministic
orbits that we provide here would produce a probability distribution extended to the whole space,
an estimation which is left for elsewhere.
We also barely touch other issues like the extension of the results to 3D, as well as to $l>1$
orbits; a preliminary exploration is, in any case, the ultimate goal of this paper.
\end{abstract}



\maketitle






\tableofcontents

\section{\label{Prologue}Prologue}


The aim of this paper is to show that an elementary mechanism exists, in the
context of SED (classical electromagnetism plus a zero-point background of
radiation), 
that can be a possible explanation for the discrete and stable character of
the hydrogen orbital spectrum, and by extension the atomic one in general.
Such a stabilizing mechanism arises in a higher order description of the system,
where an elementary model of the nucleus with some inner structure is included.
When complexity is present in the structure of a systems, it is not a rare
thing that its dynamics presents some stability and/or attraction phenomena,
and very often far beyond the obvious.
A basic analysis of orders of magnitude will be added soon; so far it does
not seem to invalidate my ideas here.

\subsection{Some foundations}

We will be dealing here, exclusively, with classical Maxwellian electromagnetism.
We have charges (or distributions of them) and fields, amongst them a random background:
fluctuations of the value of the fields in the vacuum.
The following ideas are both a point of departure and arrival:

(I)
To be able to relate the concepts of quantum and classical angular momentum (AM),
we will assume that each projection of quantum (orbital) AM corresponds to an
average on (the projection of) the standard classical AM along a closed, periodic
trajectory.
This is justified in Sec. \ref{Rel_AM}: it seems almost obvious given the
corresponding addition rules in QM.

(II)
I assume (I) also applies to the inner AM or spin of a particle.
I will often talk of a ``classical spin'' when referring to a rotational
movement of a particle (or a distribution of particles) around an axis
of symmetry.
Whether if quantum spin appears in a representation where an isomorphism can be
established (or not) to the group $SO(3)$ of rotations in ordinary space is a
question that we have analyzed somewhere else \cite{DR_Spin};
here  we are only interested in the fact that, whenever a particle has
spin, at least some sort of classical AM can be associated to it (and
in particular in regard to its magnitude).
Besides, we are also aware there are some relativistic difficulties with the
picture of a classically spinning sphere of charge and include some comments
on that.

(III)
Accordingly, I interpret QM as a ``semi-static''
theory that masks a richer dynamics underneath, involving perhaps a higher number
of (hidden) classical degrees of freedom (that ``complexity''):
for instance, the quark dynamics (as point-like entities carrying charge and spin,
but configuring a distribution, with an associated dynamics, in space) inside a
nucleon.

(IV)
In coherence with (III), while average values of the projections of those
classical AM will stay attached to a discrete spectrum as in QM, their instantaneous
values (and in a micro-dynamics that is transparent to atomic transitions) can, at
least, oscillate around the quantum mechanical one.
Interaction with a random background could cause much of those fluctuations, and,
incidentally, spontaneous state transitions that are nothing new in orthodox QM.
For instance the orbital AM of the quarks inside a proton (therefore adding to
the particle's inner AM or spin) can oscillate around its quantum value, in
response to perturbations coming from the interaction with the background.
Because those oscillations are more or less minimal, we propose the term ``residual
spin'' (RS).
The reference to the quark model is necessary (is solves some difficulties of
the model with special relativity), but nevertheless nothing more than tangential.

(V)
The concept of a ``photon'' ($E=\hbar \omega$) could be perhaos be understood
as the natural constraint on the spectrum of these possible ``discrete'' exchanges
of energy-momentum between metastable states of a system of charges.

\subsection{An overview of results and some general context}

We provide a very brief summary of the paper.
Basically, what we do is: we begin this effort by analyzing an elementary electromechanical
model. Later, we make the necessary changes to apply our results to an elementary model of
a hydrogen atom.

Under the action of an ideal magnetic dipole, a moving charge with some spatial extension
(non-vanishing second order moment) experiences a force and a torque, mediating between
its ``orbital'' and ``self rotation'' degrees of freedom, this last what we will call a
(classical) ``residual spin'' (RS).
That pair (force, torque) introduces a bidirectional coupling between the (instantaneous)
values of two classical angular momenta corresponding to those two degrees of freedom,
a coupling that acts on their magnitude, as a difference, for instance, with the spin-orbit
(LS) coupling in atomic physics, which acts only on their relative orientations:
this is no surprise but only a consequence of the fact that our mechanism corresponds to terms
of a higher order in the multipole development.
For instance, too, the LS coupling only involves a punctual value of the field, therefore being
unable to feel its gradient, which is another fundamental difference with the interaction we
expose here.
As we said before, we have named it, provisionally and for the lack of a better choice,
\textit{Inverse Magnitude Spin Orbit coupling} (IMSO).

An interpretation of our model in terms of the hydrogen atom is more elaborate, but we
conclude that similar results could be applicable to the orbitals of the electron around
the nucleus, when a spatially ``extended'' charge distribution is associated to this last,
and always in the presence of a background.
On the other hand, later, to be able to interpret our result in a more realistic
situation (an atomic model), the role played here by that classical spin will be taken
by what we have called \emph{residual spin} (RS) of the nucleus:
in the simplest case of a proton, a residual (inner) AM coming from orbital movement of
quarks inside its structure.
In the presence of a certain background compensating on average the radiative losses,
we are already able to prove that the identified  ``feedback loop'' (FL) between orbit and
the RS-N provides a necessary  and sufficient condition for the existence of stable
``orbitals'' in the classical configuration space of the system.
These orbitals are produced by oscillations around a set of privileged trajectories
or ``attractors'', that (we have proven) can either bear or not net (average) AM.
For instance, this last would be the case of the $s$-orbitals, with vanishing average AM.
Both the stability of atomic orbitals and the discrete character of the spectrum can be,
then, at least potentially explained.

Curiously enough, our preliminary calculations show that the dependence of the
associated energy correction in the (classical) orbit radius agrees with that $L\cdot S$
term in atomic physics.
Nevertheless, those energy contributions are not net ones: they are only translation of
energy from one degree of freedom to another, and moreover, their average (seems to)
vanish on a whole cyclic trajectory (an orbit).
This is related to the fact that all new terms arise from the Lorentz law, whose
associated (non-conservative) work on a particle is always zero.
As a result of this, no new observable contribution to the spectra appears.


\subsection{Relation with the Bohr model}

We depart from the assumption that quantum angular momentum (AM) is related to an average
classical AM, along a closed trajectory.
This is justified in Sec.~\ref{Rel_AM}. Aside from other differences, our
approach veers from the Bohr approach in two important points:

(i) The Bohr approach works with the purely Coulomb potential.

(ii) The Bohr approach only seeks for circular or elliptical orbits.
Clearly, this cannot account for the actual ground state of the real hydrogen (an
s-orbital in QM), because Bohr's trajectories show net AM. 
In our framework, we do find a set of stable trajectories with vanishing (average)
AM.
Those trajectories, including both the ground state, introduced in Sec.~\ref{s_stability},
and its excited states, see Sec.~\ref{Ex_states}, are not possible in a problem with
a purely electrostatic potential.
We have to resort here to terms related to the Lorentz force acting on an ``extended''
model of a particle.
Details on all this will be given in the following pages.

\subsection{Other additional questions}

To complete this momentarily vague picture, at least two other questions should
also be mentioned:

(I)
\emph{Relativistic considerations: the need to consider a (classical) ``residual''
spin (RS).}
We are aware of the difficulties that arise from assuming, for the proton, the model
of a uniform sphere of charge, rotating with angular velocity proportional to its
phenomenological magnetic moment.
These difficulties, of relativistic origin, can be safely ignored if a more realistic
model of quarks (carrying both charge and magnetic moment) is assumed.

If the quark model is assumed, the contribution of the angular velocity of the proton
as a whole, around its axis of symmetry, to its overall inner AM is
either negligible or at least vanishes on average.
However, classically, the magnitude of this magnetic moment can still oscillate.
We have renamed this oscillation dynamics as ``residual spin'' (RS), and it can
play the role that, in our first idealized situation, conventional (classical)
spinning movement played itself.
This RS would eventually correspond to oscillations of the orbital AM of the quarks
inside the proton (oscillations with no quantum counterpart, but with complete sense
in our classical reformulation of the system).

Besides, we make several references to the ``rigidity'' of this distribution, but
this is simply a device to make our argument clearer.
Indeed, the mechanism we expose is dominant up to a certain order, whenever the two
objects remain sufficiently apart.
Higher orders may deform the shape of the distribution, or have other effects that,
whenever the range of distance is the appropriate one, do not have to bother us.

(II)
\emph{The 3D problem.}     
We will provide some preliminary ideas about the extension of the argument to three
dimensions (3D).
In particular, that extension would need to include precisely the (also well known)
classical counterpart of that quantum LS interaction, acting as a modulator of the
strength of our ``stabilizing mechanism'' (IMSO).
On a leading order, their combined effect would allow for a ``modulated precession''
of the axis of the classical orbit, necessary to account for $l>0$ orbitals (therefore
associated to a certain privileged direction in space).

\subsection{Other considerations}

(a) A model theory built to account for the mechanism we expose (for instance based on
an effective potential) would not suffer from renormalization problems:
the kind of mechanism that we describe here constitutes a ``leading order'' behavior.
In fact, it disappears amongst higher order terms when the point particle and the
distribution of charges get too close.
When too far apart, naturally, the interaction also becomes negligible.
This kind of behavior is very suggestive, as imposing a lower limit on the distances
is equivalent to introducing the conventional cut-off in momentum transfer which is
a common rule (trick) in Quantum Electrodynamics (QED).

(b) Moreover, and though we do not plan to go anywhere near for the moment, we are
led to think:
could we get rid of all renormalization problems in QED (and classical electromagnetism)
if we regarded it as an effective description of a simplest interaction law, but
involving more degrees of freedom?
For instance, could we generate electromagnetism (with its vectorial potential)
from a bare scalar potential acting on particles with some extra inner degrees of
freedom (for instance, instead of being point-like entities, allowing them to
deform elastically)?

(c) Implications of this work about spin are not yet sufficiently analyzed.
Anyway, \emph{could our ``secondary oscillation'' be related to Cavalleri and others'
model \cite{Cavalleri10} of the electronic spin as a helical oscillation of
a structureless electron around its main trajectory?}




\section{\label{Introduction}Atomic stability and SED}


In this section, we provide some academical background where to set our work,
as well as describe the content of the rest of this paper.

\subsection{SED: Stochastic Electrodynamics}

\begin{center}
\textit{This account of SED is not very accurate in regard to chronology 
and contains important omissions: an exhaustive review can be found in \cite{QDice}.}
\end{center}

The electromagnetic radiation of an accelerated charge predicted by Maxwell's laws is
often cited as the (obvious) reason why the hydrogen atom cannot be classically stable,
and one should then invoke the quantum mechanical Pauli principle to account for the
existence of ordinary matter, at its common atomic phase.
In spite of this fact, the reader will surely agree that at least the question of how
unstable the atom is, or, how quasi-stable it may be, from strictly classical grounds,
still retains interest.

Indeed, much in the spirit of the, already old, theory of Stochastic Electrodynamics
(SED), started by Braffort, Marshall and Boyer (amongst others), 
several attempts have been made to explore this question in detail.
SED's main addition to the classical picture of electromagnetism is the presence of a
random, homogeneous, isotropic and Lorentz invariant electromagnetic radiation that
permeates all space even at a zero temperature (refs), the so-called zero-point field
(ZPF).
This ZPF background coincides with QED predictions for the vacuum.

Marshall \cite{Marshall65} and Boyer \cite{Boyer69a}, based solely on that Lorentz
invariance, established the spectrum (with a cubic dependence on frequency) of this ZPF,
and, as early as 1969, T.Boyer was able to derive the famous black body radiation law
of Planck, without the need to assume any quantization \cite{Boyer69a}.
This very recommendable paper has been followed, over the years, by some other very
suggestive ones, mainly by Marshall and Boyer themselves, and D.C. Cole.
For instance, we are aware of \cite{Boyer69b, Boyer75a, Boyer75b, Boyer84, Marshall81,
Rueda80} and also of \cite{Cole90a, Cole90b, Cole92, Cole00}.

While, on theoretical grounds, the very origin of the ZPF was also investigated by
Puthoff in \cite{Puthoff89}, there is also some evidence on the experimental ones.
Specifically, the observed Casimir force between two neutral metal plates in the vacuum
is itself evidence for the existence of some kind of background.
In any case, as we said, the existence of ZPF constitutes the main hypothesis of what
is come to be known as Stochastic Electrodynamics (SED). We will also refer to
\cite{Puthoff87} for a brief account of SED ``successes''.

\subsection{SED (alone) is not enough}

However, up to now, these ideas have not been enough to completely build a convincing
bridge from the classical to the quantum theory, as serious difficulties appear with
a broad range of quantum phenomena.
The stability of atomic orbitals is one of them: in this context, recent attempts have
been made to examine in detail the interaction of a simple system, such as the hydrogen
atom, with certain kinds of radiation.
Some very interesting attempts use a numerical approach 
\cite{Cole_et_al03a, Cole_et_al03b, Cole_et_al03c, Cole_et_al04a, Cole_et_al04b}.
Although an impressing piece of work, they cannot be considered anything further than
preliminary explorations.
That is also the stage, a preliminary one, where we are content to place this work.

From our point of view, a particularly relevant effort is the one by Puthoff \cite{Puthoff87},
who explains the ground state energy of the hydrogen atom from a dynamic equilibrium between
the radiation emitted by the accelerated electron and the radiation absorbed from the
zero-point field (ZPF) fluctuations of this electromagnetic background.
It is particularly remarkable that discreteness of the AM spectrum arises here
as a natural and almost obvious consequence. However, in spite of that and the
fact that it avoids, on average over time, the radiative collapse of the Bohr atom,
it does not explain how that equilibrium may exist on a quasi-instantaneous
basis.
It does not explain, either, why there should exist a discrete set of energies
for the orbitals with a given AM. To this matter we aim most of our efforts here.

Before we continue, we feel the need to insist once more on the fact that we will
be using here just classical Maxwellian (hence fully relativistic) electromagnetism,
and, really, nothing more.
Indeed, the picture of SED described above comes here simply as an inspiration to us.
The kind of mechanism we propose retains its interest (and may work, too) regardless
of what particular spectrum we choose for our radiation field, the only element
needed is the capacity of a system to exchange power with a certain background,
both for dissipation and absorption.

Yes, an accelerated electron radiates, but also absorbs radiation (as already pointed
out by Puthoff).

According to SED, this could be the ZPF, but it is not at all a necessary assumption
in the rest of this work: any kind of radiation, for instance electromagnetic noise
coming from the rest of the charges in the universe, would work.
Therefore, in principle the loss of energy could be balanced by the absorption of power
by contact with a bath of radiation.
As we said, the idea is not new, and others have tried to study and simulate the behavior
of a system in limited ranges of this kind of circumstances.
We find, indeed, a very graphical description of results in \cite{Cole_et_al04b}:
\textit{
``a detailed simulation of the effects of classical electromagnetic radiation acting on a
classical electron in a classical hydrogen potential, results in a stochastic-like motion
that yields a probability distribution over time that appears extremely close to the ground
state probability distribution for hydrogen''}.

But a bath of radiation, whether or not balancing the loss of energy from the
radiating process, is not enough, and cannot be enough, regardless of any other
feature (amplitude, frequency, polarization, etc.), to give rise to a strict stability
of the orbits around the nucleus.
Certainly, as it already had somehow been made apparent after the work of Puthoff
\cite{Puthoff87}, the demand of strict compensation of radiated and absorbed power
leads naturally to quantization (of AM, but not, at least simultaneously, on the
values of the orbital radius and angular velocity) as a necessary condition for
stability.
In no way, though, it constitutes a sufficient one, and the presence of a mechanism
such as the one we will show here will turn out to be a necessary condition for that
``sufficient'' implication.
Insisting on this issue a bit more, Puthoff's impressive derivation of the Bohr's
ground state energy assumes equilibrium with the ZPF random background as a main
hypothesis.
He does not provide a real mechanism through which to establish this equilibrium.

\subsection{What may be missing}

By radiative loss, a wider orbit will gradually converge to the ``stable'' one, and it
is also true that, on average, a narrower orbit will gradually grow, by absorption from
the background, converging back again to that privileged one.
But the key word here is ``on average''. This average is equivalent to considering that
the system is in an ``statistical equilibrium''. Nothing, nevertheless, says how this
equilibrium is achieved, or why instantaneous mismatches in radiated/absorbed power
would yield ``the right reaction'' in the system.
For instance, the excess of absorbed power in a ``suboptimal`` orbit could well make
the particle return to the optimal one, or could well, recalling that the ZPF is
completely stochastic, make it collapse even faster towards the nucleus.

From the point of view of a system theoretician, the extra element still needed in
this picture is quite clear: we need some mechanism that allows the system to accept
the energy of a perturbation, store it, and then give it back again in the form of
emitted radiation.
This perturbation would be associated with an instantaneous mismatch between the
emitted (radiative loss) and absorbed (background) radiation.
This kind of mechanism could now potentially allow an electron to orbit for an
indefinite period of time around the nucleus, on a particular circular, elliptical
or more complicated trajectory that does not degrade over time.
Naturally, at least in an average over a certain scale of time, radiative losses and
absorption from the background must compensate, if the orbit is to be stable.
We believe that this kind of possibility is not at all generally acknowledged within
the physics community.

Aside from all that, the allowance for the system to dissipate is a strictly necessary
condition for the appearance of what we call ''attractors``.
This appears very clearly from energy conservation considerations.
Indeed, the relation between radiated and absorbed power constitutes itself another
feedback loop (FL), this time on average values.
This is why the new dynamical FL that we will present here is so relevant:
it applies to instantaneous values of the dynamical variables.

Very graphically, Cole and others' previous work illustrates a background for
these ideas.
For an electron moving in a near circular orbit, under the effect of a circularly
polarized plane wave, normal to the plane of the orbit, we find \cite{Cole_et_al04b}: 
\textit{``The result is a constant spiraling in and out motion of the electron,
with the spirals growing larger and larger in amplitude, until finally a critical
point is reached and the decay of the orbit occurs''}.
For elliptical orbits, \textit{``this behavior also occurs for more general, but more
complicated, elliptical orbits, where now an infinite set of plane waves is required
to achieve the same effect, where the plane waves are harmonics of the periods of
the orbit''}.
We need some kind of mechanism to force the electron to hold to its initial orbit.

How can we do such a thing? We must find an internal degree of freedom, which
is the one that will store the energy.
This degree of freedom will need to be able to ``communicate'', to exchange energy
with the ``other'' degree of freedom, the orbit, the one that is subject to external
perturbation.
The particular way in which this would result in a stabilization mechanism of the
value of both is more a matter of the theory of dynamical systems.

We will finish this little dissertation saying that spin, a ``classical'' spin,
the rotational AM of the particle, seems, in a first guess, a perfect
candidate for that ``internal'' degree of freedom we seek.
Nevertheless, the interpretation of either the electron or the nucleus as a rotating
uniform charge distribution leads to some difficulties with special relativity that
were first encountered some eighty years ago, and will force us to be more concrete
in our choice:
we propose an interaction between the orbital degree of freedom and
what we have called ``residual spin'' (RS, the amount of AM coming from
the orbits of quarks inside the proton - if we had more than one nucleon, we could
associate this RS with orbital AM of the nucleons inside the nucleus).

As we had already hinted before, the picture is not complete, nevertheless, until
we include some kind of background, because the system must be given a source to
be able to compensate the radiative losses (dissipation).
Otherwise there is no way of making the system stable.
But, still, how do we connect, in the classical formulation, those two degrees of freedom?

\subsection{\label{At_Hamiltonian}
Electromagnetic interactions within the hydrogen atom}

The quantum atomic Hamiltonian for the hydrogen atom consists of several terms.
If we establish a relation with its classical counterpart, it is clear that the first
two terms represent the coupling of the charge of the electron to the scalar (electrostatic
term, electric field) and vectorial (Lorentz term) part of the potential.
These two terms configure the ``coarse grain'' atomic structure.
The electron and proton lack on dipolar electric moment, the following (in order) non
vanishing moment is their dipolar magnetic moment, but straight interaction (dipoling)
of these does not appear as a primary term in the quantum development (the term of the
kind $\hat{\sigma} \cdot \hat{\sigma}$ is not very relevant).


Another term finally appears in the quantum formulation, giving rise to the ``fine''
structure of the atomic levels, namely the one known as LS coupling.
At least two routes to this term are well known.
The first of them departs from ordinary non-relativistic QM, and reasons semi-classically
in the following way:
the electron ``sees'' the proton orbiting around, and therefore ``feels'' a certain
magnetic field created by this current.
Different orientations of the spin of the electron in this field yield the LS contribution,
an additional factor 2 being introduced to account for the so-called Thomas precession.
The second route to the LS term results from the combination of the Dirac equation with
the principle of minimal coupling substitution.
It is, therefore, fully relativistic, as so is the Dirac equation.

All those terms are related either with coupling of fields to a point-like charge, or
coupling of a field (a dipole field) with a dipole. All these interactions are somehow
more ``primordial'' than the one we want to uncover.
They leave untouched the magnitude of (magnetic) dipoles, its action therefore reduced
to a change of their orientation is space.

\subsection{Interaction between two systems of charges}

Classically, two magnetic dipoles of equal sign tend to anti-align. The magnetic
field created by one of them causes a torque on the other.
Moreover, the electric charge associated with the dipole also gives rise to a force,
through the Lorentz law, $\mathbf{F}=q\mathbf{E} + \mathbf{v} \wedge \mathbf{B}$,
with its purely electrostatic and magnetic terms, respectively.
These two effects completely determine the interaction between two charge distributions,
at least whenever moments of higher order are all vanishing.
Specifically, the first one of those effects, the classical dipoling interaction, would
introduce an spin-spin term in the corresponding quantum Hamiltonian, where the second
would stand for an ``orbital'' term, independent of the ``spin'' (in this case the classical
self-rotation of the particle experiencing the field created by the source).

In any case, our picture is more general to that of two charges with magnetic dipolar
moment interacting electro-statically, through the Lorentz force, and also through the
dipoling interaction.
An ``extended'' distribution of charges has an associated magnetic dipolar moment,
but it is also capable of feeling the ``gradient'' of the field.
This gives rise to some other effects that, had we considered an ideal dipole, would
have remained unapparent.
These new effects, as we will later show, do arise exclusively from the assignation
of a spatial extension to the charge distribution, that should for simplicity be
regarded as a solid rigid (its movement being thus reducible, for any instant of
time, to a pure translation with the velocity of the center of mass and a rotation,
in this case around an axis of symmetry).
In this last situation, the calculation of forces and torques no longer depend on
the value of the fields at one point (the center of mass) but they involve the
evaluation of an integral, where the action over each differential ``element of
charge'' must be computed and added.

\subsection{A previous stage}

In Sec.~\ref{Qual_game}, initially, we focus on a somewhat simplified situation,
and through a graphical argument, we expect to convince the reader that, under
the action of a dipolar magnetic source, a (spherical) distribution of charge
with non-vanishing second order moment (a spatial ``extension'') experiences the
following:

(i)
a net torque applied on its center, as a result of its ``orbital'' (center of mass,
collective) velocity, in the plane normal to the dipoling source,

(ii)
a net resultant force, as a result of a non-vanishing AM of self-rotation,
in the direction that joins the center of mass with the the dipoling source.



In Sec.~\ref{2D} and subsequent ones, we do an already quantitative analysis of the
$2$-dimensional problem. 
This situation corresponds to what we have called ``planar'' initial conditions: both
the source magnetic dipole and the AM of the distribution are initially
aligned, and remain like that because the initial velocity is strictly normal to them.
Our intention is to be completely systematic, and to give quantitative expressions
(although symbolic) for the two effects already introduced in Sec.~\ref{Qual_game}.

Later, in Sec.~\ref{General_ic}, we will complete our analysis allowing for general
initial conditions. We will find out that the new contributions are (as expected)
perpendicular to the ones in Sec.~\ref{2D}.
This is a highly satisfactory result because it means that feedback loop (FL) that
we already had foreseen in \ref{2D} is not destroyed by the new terms arising
from the new freedom in the initial conditions.
Sec.~\ref{p_stability} stands for an interpretation of the former results, from
the point of view of the theory of systems dynamics (it is perhaps here convenient
to say that the author has some background in this field).
The theory of systems dynamics provides us with tools to determine necessary and
sufficient conditions for the existence of stable ``orbitals'' in the classical
configuration space of the system, without the need to address more specific
calculations that in principle would lay beyond the scope of this paper.
Something that may be clarifying is that we would be using here the term ``orbital''
in a much broader sense than just a kind of stationary/stable orbit: we mean just
a portion of the configuration space of the system, where, once it is placed inside,
there is a very little probability of leaving it.


\subsection{Model for a realistic hydrogen atom}

In Sec.~\ref{Mass_reloc}, we face now a slightly modified scenario: now the
inertial system is anchored to the center of the charge distribution, and the point-like
source is moving.
This choice is, as we say, necessary to make the analogy with a real atom possible,
where an almost massless point particle orbits around an almost static (because of
its almost infinite relative mass) distribution of charge, the nucleus.

\subsection{A feedback loop (FL)}

Two variables ``A'' and ``B'' are under a FL if, upon a sudden change on ``A'',
this influences ``B'' changing its value, and finally the change in ``B''
modifies again ``A''.
For instance, a time-varying electric field and its associated magnetic field
evolve under the action of a loop of influence.
This kind of thing becomes apparent when we decouple the Maxwell equations
for a perturbative calculation.
Another example is the relation between the current of a coil and the charge
of a capacitor in an oscillating LC circuit.
In our case, the ``loop of influence'' in which we are interested relates
the following variables:
\begin{equation}
\mathbf{v}^{(o)} \Rightarrow \boldsymbol{\tau} \Rightarrow \boldsymbol{\omega} \Rightarrow
\mathbf{F} \Rightarrow \mathbf{v}^{(o)}, \label{FL}
\end{equation}
where $\mathbf{v}^{(o)}$ is the ``orbital'' velocity of the center of mass of
the charge distribution, $\boldsymbol{\tau}$ is a torque applied at its center
of mass, $\boldsymbol{\omega}$ is an angular velocity vector, expressing a
rotation around an axis of symmetry.
By $\mathbf{F}$ we mean a resultant force, applied on that same center of mass.
There are some subtleties regarding the definition of $\boldsymbol{\omega}$,
that we will treat later; for the moment we are content just to provide the
reader with some ``flavor'' of what we are dealing with.

Perhaps it is already convenient, even at this initial point, to enter a bit
more into detail: actually, our equations for the (2-dimensional) comprise
\textit{two simultaneous FLs}. 
The first of them will be responsible for the stationarity of the main orbital
movement, and, once both the radiation losses and the interaction with a
background are included, for its stability.
Meanwhile, a second FL will introduce a secondary oscillation (2nd-Osc) that
will allow us to provide a feasible explanation for excited states, hence for
a full discrete and infinite, spectrum of (completely classical) stable states
for the system.
Finally, when doing the extension of the analysis to 3D, a third FL appears,
this last corresponding to the classical counterpart of the well known
quantum LS.

On the other hand, the association of these loops with particular ``natural''
frequencies of oscillation (which is crucial here) is proven (approximately,
as the system is non-linear), via a frequency analysis of the linearized
equations of the system.

\subsection{Linearization}

Given the dynamical equations of the system, what we do first if to look for
stationary trajectories: if a suitable choice of dynamical variables is made,
we can then associate them with (classical) eigenstates of the system.
We know, from system theory, that any harmonic function is indeed an eigenstate
of the linearized system, and this will be of great use later (though we use,
however, more physical arguments to derive our stationary solutions, this idea
will be of great use).
A second (and last, for now) step is to identify which of those stationary
trajectories are stable. 
To prove stability, we linearize the system around a ``stationary'' point
$\mathbf{x}_{st}$ in the form $\dot{\mathbf{x}} = A \cdot ( \mathbf{x} - \mathbf{x}_{st} )$, where
$A \in {\cal M}_{nxn}$ and $\mathbf{x} \in {\cal R}^n$ is a vector of dynamical
variables.
Now, at least for that linearized description of the system, a necessary and
sufficient condition for stability is simply that the real parts of the
eigenvalues of the matrix $A$ must be negative.


\section{\label{Qual_game}An electromechanical game}


We work on a purely electromechanical model. This model has little relation to
a real hydrogen atom: we are going to calculate the action of the field created
by a point-like magnetic dipole on a distribution of charge with zero (monopolar)
and second (quadrupole?) order non vanishing moments.
The source is attached to an inertial system. The charge distribution can move
as a rigid body. A reinterpretation the results that arise from this previous
treatment to a more physical situation will be the main task of subsequent sections.


We present now a very simple, qualitative argument, that will surely shed necessary
light over all that we will calculate afterwards.
Let us consider a point-like magnetic source with non-vanishing magnetic dipolar
momentum, hence an ideal magnetic dipole.
This source is attached (hence it does not move) to the origin of an inertial
reference frame that we will call $RF_{0} \equiv \{\hat{x}_{0},\hat{y}_{0},\hat{z}_{0}\}$.

This magnetic dipole is aligned with the z-axis of  $RF_{0}$, and hence,
far away from it, creates a magnetic field $B$ in the direction $-\mathbf{z}_{0}$
that will act on any charge that may be moving with a certain velocity with respect
to $RF_{0}$, but it is not affected by them (this is equivalent to consider, for
instance, that its mass is infinite).

Now, we also consider a spherical distribution of (positive) charge whose center
of mass is instantaneously moving with velocity $\mathbf{v}$ in the plane OXY with
respect to $RF_{0}$, and rotating around its symmetry axis in the direction $\mathbf{z}_{0}$.
We invite the reader now to a qualitative examination of two different situations.
We have singled out points $A$ and $B$ to make the argument clear.\\

\noindent
1) \textit{Figure 1.}
The distribution moves as a whole with a linear ``orbital'' velocity $\mathbf{v}^{(o)}$
with respect to the source. There is no rotation ($\omega=0$).
As a result of the radial component of this velocity $v^{(o)}_{t}$ (the component
in $\mathbf{x}_{1}$), forces on $A$ and $B$ (and also at the rest of the points of
the distribution) are induced.
Clearly, because $|\mathbf{B}_{A}| > |\mathbf{B}_{B}|$, $|\mathbf{F}_{A}| > |\mathbf{F}_{B}|$
and there is a net torque $\boldsymbol{\tau}$ applied at the center of the charge distribution.
The tangential component of this orbital velocity, $v^{(o)}_{t}$, also gives rise to a net force,
but we do not represent it for clarity.
\\

\begin{figure}[ht!] \includegraphics[width=0.90 \columnwidth,clip]{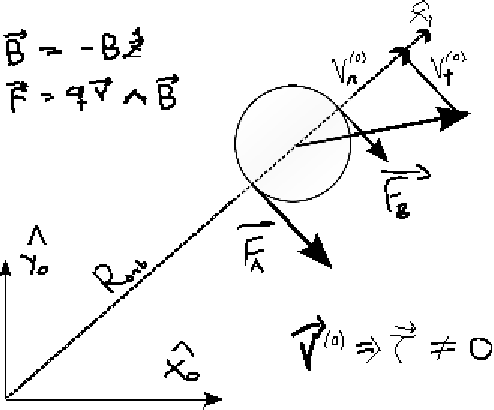} 
\caption{
Under the action of a magnetic source in the origin, an orbital motion (in the radial
direction, $v^{(o)}_{r} \neq 0$) generates a net torque $\mathbf{\tau} \neq 0$, due
to the fact that $|\mathbf{F}_{A}| > |\mathbf{F}_{B}|$.
} \label{Fig1} \end{figure}

\noindent
2) \textit{Figure 2.}
The distribution has no orbital motion now, but rotates around its center of symmetry.
As a result of an angular velocity $\omega \neq 0$ (measured with respect to the inertial
system $RF_0$), the Lorentz law induces forces in points $A$ and $B$, and all the points
of the distribution.
Again, because $|\mathbf{B}_{A}| > |\mathbf{B}_{B}|$, $|\mathbf{F}_{A}| > |\mathbf{F}_{B}|$,
and (if we integrate for all points), there is a net resultant force acting on the
mass center.

\begin{figure}[ht!] \includegraphics[width=0.90 \columnwidth,clip]{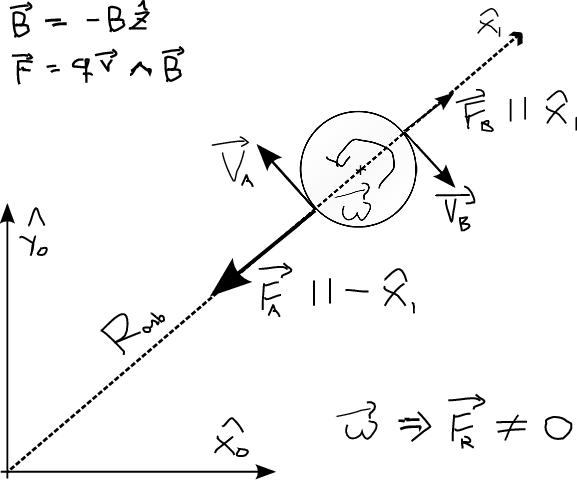} 
\caption{
Under the action of a magnetic source in the origin, a  self-rotational momentum
$\mathbf{\omega} \neq 0$ originates a net resultant force $\mathbf{F}_{R} \neq 0$,
modifying the orbit, due again to the fact that $|\mathbf{F}_{A}| > |\mathbf{F}_{B}|$.
} \label{Fig2} \end{figure}

\textit{Attention to sign consistency:} from the graphical argument, we see that a positive
(counterclockwise) angular velocity $\omega > 0$ yields a negative force on the radial
direction ($\mathbf{F}\cdot\hat{x}_{1} < 0$, and a positive (outwards) radial velocity
$v^{(o)}_{r} > 0$ yields a positive (favors counter-clockwise rotation) torque
($\mathbf{\tau} \cdot \hat{z}_{0} > 0$).

\section{\label{2D}
Quantitative analysis in 2D}



\subsection{Reference frames}

As before, we consider an inertial reference frame $RF_{0}$. The magnetic source stays attached to it, and
the distribution moves.
For convenience, we also consider a second system of reference $RF_{1}$, with its origin attached to the
center of the distribution. The axes of $RF_{1}$ are not rigidly attached to the distribution, so there
can be a net angular velocity $\boldsymbol{\omega}$ for $RF_{1}$ in respect to $RF_{0}$. Indeed, we will let
it rotate around an axis of symmetry of the distribution.
Moreover, $\mathbf{x}_{1}$, the x-axis of this system, $RF_{1}$, will be always aligned along the positive
direction of the vector that joins the point-like source and the center of mass of the distribution. We
therefore define, formally
\begin{equation}
\vec{R}_{orb} = - R_{orb}\ \mathbf{x}_{1}, \label{R_orb}
\end{equation}
The auxiliary system $RF_{1}$ will rotate instantaneously with angular velocity $\omega$ with respect to $RF_{0}$.
Obviously, the fact that $RF_{1}$ can rotate in relation to $RF_{0}$ clearly excludes its inertiality. Nevertheless,
we will express almost any vector in this frame $RF_{1}$. This is done for reasons of convenience.

\emph{Very important:} this vector system $RF_{1}$ is only used as a reference in the mathematical sense,
it is only a mathematical device. This means it is never used in the physical way: no force law (for instance,
the Lorentz law), no mechanics (for instance, Newton's second law) is ever evaluated in this frame.
What we can do, however, is express any vector (any position, velocity, AM, force or torque) as
referred to its basis vectors $\mathbf{x}_{1},\mathbf{y}_{1},\mathbf{z}_{1}$, which in turn are themselves
functions of $\mathbf{x}_{0},\mathbf{y}_{0}, \mathbf{z}_{0}$.
This use of $RF_1$ will be very convenient to make the necessary changes that will allow to apply our picture
to the more realistic situation of the hydrogen atom.

\subsection{Charge distribution}

For the present work, we will consider a charge density $\rho(\mathbf{r}_{1})$ such that $\rho(\mathbf{r}_{1})>0$
only for $|\mathbf{r}_{1}|\leq R_c$, being the radius of the (spherical) distribution. We define a differential
element of charge $dq = \rho(\mathbf{r}_{1})\ dv$, with $dv$ the differential element of volume.

\subsection{\label{kin_con}
Purely kinematic considerations}

For each differential element of the charge distribution, we have a velocity
that we will divide in two components:
\begin{equation}
\mathbf{v} \ =\ \mathbf{v}^{(o)} \ +\ \mathbf{v}^{(R)}(\mathbf{r}_{1}), \label{v_dec}
\end{equation}
where the superscript 'o' will stand for ``orbital'' and 'R' will stand for
``rotational''.
For convenience, we will write all expressions in terms of the versors of
$RF_{1}$.
This poses no problem as long as we regard $\mathbf{x}_{1},\mathbf{y}_{1},
\mathbf{z}_{1}$ as functions of $\mathbf{x}_{0},\mathbf{y}_{0},\mathbf{z}_{0}$,
and keep in mind $RF_{0}$ is our (only) inertial system.
We will also include the dependence in $\mathbf{r}_{1}$ when this is indeed
present, as we do in $\mathbf{v}^{(R)}(\mathbf{r}_{1})$.
For the ``orbital'' component we will barely write $\mathbf{v}^{(o)}$, as
we will see it is not dependent in $\mathbf{r}_{1}$.
We define, for each differential element of charge in the distribution:
\begin{equation}
\mathbf{v}^{(o)} \ 
=\ v_{r}^{(o)}\ \mathbf{x}_{1} \ +\  v_{t}^{(o)}\ \mathbf{y}_{1}, \label{v_orb}
\end{equation}
with the components
\begin{eqnarray}
\mathbf{v}^{(o)}_{t} 
= v^{(o)}_{t} \mathbf{x}_{1} &=& \boldsymbol{\omega}_{(o)} \ \wedge\ \vec{R}_{orb},
\label{v_o_t}\\
\mathbf{v}^{(o)}_{r} = v^{(o)}_{r} \mathbf{y}_{1} &=& \frac{d \vec{R}_{orb} }{ dt },
\label{v_o_r}
\end{eqnarray}
where $\omega_{(o)}$ is defined by (\ref{v_o_t}) as the angular velocity
of $RF_{1}$ with respect to $RF_{0}$ and
\begin{eqnarray}
\mathbf{v}^{(R)}
&=& \boldsymbol{\omega}_{(R)} \ \wedge\  \mathbf{r}_{1},  \ \ |\mathbf{r}_{1}| \le  R_{c},
\label{v_R}
\end{eqnarray}
with $\boldsymbol{\omega}_{(R)} = \omega_{(R)} \ \mathbf{z}_{1}$ (remember
that in this section we restrict the initial conditions to the plane $OXY$),
defined as
\begin{equation}
\omega_{(R)} = \omega - \omega_{(o)}, \label{omega_R}
\end{equation}
and $\omega$ is the angular velocity of the distribution with respect
to $RF_{0}$ (see \cite{defs_omega}).
Vectorially, $\boldsymbol{\omega}_{(R)} = \boldsymbol{\omega} - \boldsymbol{\omega}_{(o)}$,
with the three vectors directed in the $\mathbf{z}_{1} \equiv \mathbf{z}_{0}$ axis.
It is important to remark here that $\omega_{(R)}$, as defined above, is not
an angular velocity as measured in $RF_{0}$, in contrast with $\omega$ and
$\omega_{(o)}$.
This definition of $\omega_{(R)}$ is justified for convenience, as it makes
all our reasonings far more apparent.
%
For clarification, and also because it will be useful, we add the following
expression using spherical coordinates in $RF_{1}$:
\begin{eqnarray}
\mathbf{v}^{(R)} 
&=& \omega_{(R)} r_{1}\sin\theta_{1} \ \left[\ \sin\psi_{1} \mathbf{x}_{1} - \cos\psi_{1} \mathbf{y}_{1} \ \right],
\label{v_rot_curv1} 
\end{eqnarray}
where $r_{1} = |\mathbf{r}_{1}|$ and $\omega_{(R)}$ is positive for a
counter-clockwise rotation.

\subsection{Fields}

We start by recalling that the electric field created by a point-like
particle with charge $-q$ (electric monopole):
\begin{eqnarray}
\mathbf{E}(\mathbf{r_0}) &=& - \frac{q}{4\pi\epsilon_0} \cdot \frac{ \mathbf{r_0} } {|\mathbf{r_0}|^3 },
\label{E_r0}
\end{eqnarray}
but as a rule, we said we will express all in $RF_{1}$:
\begin{eqnarray}
\mathbf{E}(\mathbf{r}_{1}) &=& - \frac{q}{4\pi\epsilon_0} \cdot
\frac{ (\mathbf{r}_{1} - R_{orb}\ \mathbf{x}_{1}) } {|\mathbf{r}_{1} + R_{orb}\ \mathbf{x}_{1}|^3 },
\label{E}
\end{eqnarray}
with $\mathbf{r_0} = \mathbf{r}_{1} + R_{orb}\ \mathbf{x}_{1}$, with $R_{orb}$
the orbital radius.
Acting on each differential element of charge there is a purely electrostatic force:
\begin{eqnarray}
d\mathbf{F}_{e} &=& dq(\mathbf{r}_{1}) \  \mathbf{E}(\mathbf{r}_{1}), \label{dF_e}
\end{eqnarray}
and obviously
\begin{eqnarray}
\mathbf{F}_{e} = \int{ d \mathbf{F}_{e} }. \label{F_e}
\end{eqnarray}
In the following we will face an analogous calculation with the Lorentz force.
Here we have not developed (\ref{dF_e})-(\ref{F_e}) in a multipole expansion,
as this would yield no special distinction among terms of a different nature.
This will be, however, our main tool in the following calculations.

Now, the magnetic field created by a point-like magnetic dipole
$\boldsymbol{\mu}=\mu\mathbf{z_1}$, if the source is sufficiently
far to consider the observed field only has a $\mathbf{z}$ component:
\begin{eqnarray}
\mathbf{B}(\mathbf{r}_{1}) &=& - B(\mathbf{r}_{1}) \  \mathbf{z}_{1}, \\ \label{B_vec}
B(\mathbf{r}_{1})
&=& \hat{\mu} \ \frac{1}{|\mathbf{r}_{0}|^3} 
= \hat{\mu} \ \frac{1}{|\mathbf{r}_{1} + R_{orb}\ \mathbf{x}_{1}|^3}, \label{B}
\end{eqnarray}
defining the quantity $\hat{\mu} = \frac{\mu_0}{4\pi} \mu$, with $\mu > 0$
the value of the magnetic moment of the point-like source.

\subsection{Multipole expansion}

It is time to recall the first terms of a multipole expansion (around
$\mathbf{r}_{0}=\vec{R}_{orb}$, or equivalently, $\mathbf{r}_{1}=O_{1}$):
\begin{eqnarray}
\frac{1}{|\mathbf{r}_{0}|}
&=& \frac{1}{|\mathbf{r}_{1} \ -\ (\ -\  R_{orb}\ \mathbf{x}_{1})|}
\nonumber\\ 
&=& \frac{1}{|\mathbf{r}_{1} + R_{orb}\ \mathbf{x}_{1}|}
\nonumber\\ 
&=& \frac{1}{R_{orb}} + \frac{ (- R_{orb}\ \mathbf{x}_{1})\cdot\mathbf{r}_{1}} {R_{orb}^3} + \ldots
\nonumber\\ 
&=& \frac{1}{R_{orb}} - \frac{\mathbf{x}_{1} \cdot \mathbf{r}_{1}} {R_{orb}^2} + \ldots,
\label{multipole1}
\end{eqnarray}
\begin{eqnarray}
\frac{1}{|\mathbf{r}_{0}|^2}
&=& \frac{1}{|\mathbf{r}_{1} \ -\ (\ -\  R_{orb}\ \mathbf{x}_{1})|^2}
\nonumber\\ 
&=& \frac{1}{|\mathbf{r}_{1} + R_{orb}\ \mathbf{x}_{1}|^2}
\nonumber\\ 
&=& \frac{1}{R_{orb}^2} + \frac{2 (- R_{orb}\ \mathbf{x}_{1})\cdot\mathbf{r}_{1}} {R_{orb}^4} + \ldots
\nonumber\\ 
&=& \frac{1}{R_{orb}^2} - \frac{2 \mathbf{x}_{1} \cdot \mathbf{r}_{1}}
{R_{orb}^3} + \ldots,
\label{multipole2}
\end{eqnarray}
\begin{eqnarray}
\frac{1}{|\mathbf{r}_{0}|^3}
&=& \frac{1}{|\mathbf{r}_{1} \ -\ (\ -\  R_{orb}\ \mathbf{x}_{1})|^3}
\nonumber\\ 
&=& \frac{1}{|\mathbf{r}_{1} + R_{orb}\ \mathbf{x}_{1}|^3}
\nonumber\\ 
&=& \frac{1}{R_{orb}^3} + \frac{3 ( - R_{orb}\ \mathbf{x}_{1})\cdot\mathbf{r}_{1}} {R_{orb}^5} + \ldots
\nonumber\\ 
&=& \frac{1}{R_{orb}^3} - \frac{3 \mathbf{x}_{1}\cdot\mathbf{r}_{1}}{R_{orb}^4} + \ldots,
\label{multipole3}
\end{eqnarray}
that we will now apply to the scalar $B(\mathbf{r}_{1})$, where
$\mathbf{B}=B(\mathbf{r}_{1})\mathbf{z}_1$ is the magnetic field created
by the point-like source:
\begin{eqnarray}
B(\mathbf{r}_{1}) &=&
\hat{\mu} \ \left[ \frac{1}{R_{orb}^3} - \frac{3\ \mathbf{x}_{1}\cdot\mathbf{r}_{1}}{R_{orb}^4} + \ldots \right].
\nonumber\\ 
\label{B_esc_multipolar} 
\end{eqnarray}

\subsection{On the electrostatic force}

We already settled the expression for the purely electrostatic force
in (\ref{dF_e}) and (\ref{F_e}). 
That formula already accounts for the inclusion of higher order moments for
the charge distribution. None of these terms, aside from introducing
corrections to the overall magnitude of the resultant force $F_e$, adds
any other effect from our point of view.
In fact, there is no component of $F_e$ in the tangential ($\mathbf{y}_{1}$)
direction, neither there is any kind of resultant torque over the center
of mass of the distribution.
For these reasons, we will not hereafter refer to this electrostatic term
$F_e$, if it is not strictly necessary.

\subsection{Magnetic forces}

From here on, we concentrate on the forces arising from the ``second term''
within the Lorentz force, that of the kind $q\mathbf{v} \ \wedge\  \mathbf{B}$.
Acting on each differential element of charge, we have a force:
%
\begin{eqnarray}
d\mathbf{F}_{m}
&=& dq(\mathbf{r}_{1}) \ \mathbf{v}(\mathbf{r}_{1}) \ \wedge\ - B(\mathbf{r}_{1})\ \mathbf{z}_{1}
\nonumber\\
&=& dq(\mathbf{r}_{1}) \ \left[\mathbf{v}^{(o)} + \mathbf{v}^{(R)}(\mathbf{r}_{1}) \right]
\wedge - B(\mathbf{r}_{1})\ \mathbf{z}_{1}.
\nonumber\\ 
\label{dF_m_no_decomposition}
\end{eqnarray}
where we have applied our decomposition in (\ref{v_dec}),
and now we can write:
\begin{eqnarray}
d\mathbf{F}_{m} &=& d\mathbf{F}_{m}^{(o)} \ +\  d\mathbf{F}_{m}^{(R)}.
\label{dF_m}
\end{eqnarray}
defining:
\begin{eqnarray}
d\mathbf{F}_{m}^{(o)}
&=& dq(\mathbf{r}_{1}) \ \mathbf{v}^{(o)} \wedge - B(\mathbf{r}_{1})\ \mathbf{z}_{1},
\label{dF_m_o} \\ 
d\mathbf{F}_{m}^{(R)}
&=& dq(\mathbf{r}_{1}) \ \mathbf{v}^{(R)}(\mathbf{r}_{1}) \ \wedge\  - B(\mathbf{r}_{1})\ \mathbf{z}_{1}.
\label{dF_m_R} 
\end{eqnarray}

\subsubsection{Orbital component} 

We start by analyzing the first of those two terms (a force from an
``orbital'' origin):
\begin{eqnarray}
d\mathbf{F}_{m}^{(o)}
&=& dq(\mathbf{r}_{1}) \ \mathbf{v}^{(o)} \wedge  - B(\mathbf{r}_{1})\ \mathbf{z}_{1} \nonumber\\
&=& dq(\mathbf{r}_{1}) \ \mathbf{v}^{(o)} \wedge \nonumber\\
&&\quad
- \hat{\mu} \ \left[ \frac{1}{R_{orb}^3} - \frac{3 \mathbf{x}_{1}\cdot\mathbf{r}_{1}}{R_{orb}^4}
+ \ldots \right]
\ \mathbf{z}_{1},
\label{dF_m_o_1}
\end{eqnarray}
and now, integrating for the whole distribution:
\begin{eqnarray}
\mathbf{F}_{m}^{(o)} &=& \int{ d\mathbf{F}_{m}^{(o)} },
\label{F_m_o}
\end{eqnarray}
and taking into account (\ref{B_esc_multipolar}), we clearly see the first term
(electric monopole) already gives the leading contribution to the integral, and
therefore:
\begin{eqnarray}
\mathbf{F}_{m}^{(o)} &=& \int{
dq(\mathbf{r}_{1}) \ \mathbf{v}^{(o)} \ \wedge\ \hat{\mu}  \cdot
\left[ \frac{1}{R_{orb}^3} + \ldots \right]
\ \mathbf{z}_{1}
} 
\nonumber\\ 
&=& (\mathbf{v}^{(o)} \ \wedge\ \mathbf{z}_{1}) \cdot \frac{\hat{\mu}}{ R_{orb}^3 }
\ \int{ dq(\mathbf{r}_{1}) }.
\label{F_m_o_truncated} 
\end{eqnarray}
If we now isolate the leading order, defining:
\begin{eqnarray}
\mathbf{F}_{m}^{(o)} &\approx&
\frac{ \hat{\mu}\ Q_{0} }{ R_{orb}^3 } \cdot (\mathbf{v}^{(o)} \ \wedge\ \mathbf{z}_{1}),
\label{F_m_o_multipolar_lead_order} \\
Q_{0} &=& \int{ dq(\mathbf{r}_{1}) },
\label{Q_0}
\end{eqnarray}
where $Q_{0}$, a zero order moment of charge around $O_{1}$, i.e., $Q_{0}=q$, with $q>0$
the total charge of the distribution.
Let us now use (\ref{v_orb}) and give a more detailed expression:
\begin{eqnarray}
\mathbf{F}_{m}^{(o)} \approx
\frac{ \hat{\mu}  \  q } { R_{orb}^3 } \cdot \left[ v^{(o)}_{t} \mathbf{x}_{1} - v^{(o)}_{r} \mathbf{y}_{1} \right],
\label{F_m_o_multipolar_lead_order_detailed}
\end{eqnarray}
though, in this case we could simply use the full order,
\begin{eqnarray}
\mathbf{F}_{m}^{(o)}
&=& |B(\mathbf{r}_{0} = \vec{R}_{orb})|
\left[ v^{(o)}_{t} \mathbf{x}_{1} - v^{(o)}_{r} \mathbf{y}_{1} \right] \nonumber\\
&=& |B(\mathbf{r}_{1} = \mathbf{0})|
\left[ v^{(o)}_{t} \mathbf{x}_{1} - v^{(o)}_{r} \mathbf{y}_{1} \right].
\label{F_m_o_full}
\end{eqnarray}

\subsubsection{Rotational (``spinning'') component} 

Recalling (\ref{v_dec}), we set now the focus on $\mathbf{v}^{(R)}$,
\begin{eqnarray}
d\mathbf{F}_{m}^{(R)}
&=& dq(\mathbf{r}_{1}) \ \mathbf{v}^{(R)} \ \wedge\  - B(\mathbf{r}_{1})\ \mathbf{z}_{1} \nonumber\\
&=& dq(\mathbf{r}_{1}) \ (\mathbf{\omega_{(R)}} \wedge \mathbf{r}_{1})\ \wedge\  - B(\mathbf{r}_{1})\ \mathbf{z}_{1}.
\nonumber\\ 
\label{dF_m_R_1}
\end{eqnarray}
We can calculate a bit more:
\begin{eqnarray}
d\mathbf{F}_{m}^{(R)} 
&=& dq(\mathbf{r}_{1}) \ \omega_{(R)}\ \mathbf{z}_{1} \ \wedge\  \mathbf{r}_{1}\ \wedge\
- B(\mathbf{r}_{1})\ \mathbf{z}_{1} \nonumber\\
&=& dq(\mathbf{r}_{1}) \ \omega_{(R)} B(\mathbf{r}_{1})\ \mathbf{z}_{1} \ \wedge\ \mathbf{z}_{1}
\ \wedge\  \mathbf{r}_{1}^{(h)} \nonumber\\
&=& dq(\mathbf{r}_{1}) \ \omega_{(R)} B(\mathbf{r}_{1})\ ( - \mathbf{r}_{1}^{(h)}) \nonumber\\
&=& -\ dq(\mathbf{r}_{1}) \ \omega_{(R)} B(\mathbf{r}_{1})\ \mathbf{r}_{1}^{(h)},  \nonumber\\
\label{dF_m_R_2} 
\end{eqnarray}
where each time we do $\mathbf{z}_{1} \ \wedge\ $ on the left we ``rotate''
a 2D vector by an angle $\pi/2$ around the $\mathbf{z}$.
We have also introduced, for a horizontal projection of a vector, the
following notation, for any vector $\mathbf{A}$:
\begin{eqnarray}
\mathbf{A}^{(h)} &=&
(\mathbf{A}\cdot\mathbf{x}_{1}) \ \mathbf{x}_{1} \ +\  (\mathbf{A}\cdot\mathbf{y}_{1}) \ \mathbf{y}_{1},
\label{h_projection}\\
\mathbf{A}^{(z)} &=& (\mathbf{A}\cdot\mathbf{z}_{1}) \ \mathbf{z}_{1}.
\label{v_projection}
\end{eqnarray}
We return to the integration now
\begin{eqnarray}
\mathbf{F}_{m}^{(R)} &=& \int{ d\mathbf{F}_{m}^{(R)} },
\label{F_m_R}
\end{eqnarray}
we can see the first term (\ref{B_esc_multipolar}) gives rise to a
contribution of first order to $\mathbf{F}_{m}^{(R)}$ (be aware of
the last factor $\mathbf{r}_{1}^{(h)}$!), thus vanishing (as the
dipolar electric moment of the distribution does).
The second term in (\ref{B_esc_multipolar}) gives rise, however,
to a non vanishing contribution (as the second order moment does not
vanish), due again to that last factor $\mathbf{r}_{1}^{(h)}$.
We can then write:
\begin{eqnarray}
\mathbf{F}_{m}^{(R)}
&=& \int{
dq(\mathbf{r}_{1}) \ \omega_{(R)}
\hat{\mu} \ \left[\ -\ \frac{3 \mathbf{x}_{1}\cdot\mathbf{r}_{1}}{R_{orb}^4} + \ldots  \right]
\ ( - \mathbf{r}_{1}^{(h)})
} \nonumber\\
&=& \int{
dq(\mathbf{r}_{1}) \ \omega_{(R)}
\hat{\mu} \ \left[ \frac{3 \mathbf{x}_{1}\cdot\mathbf{r}_{1}}{R_{orb}^4} + \ldots \right]
\ \mathbf{r}_{1}^{(h)}
} \nonumber\\
&=& \int{
dq(\mathbf{r}_{1}) \ \omega_{(R)}
\hat{\mu} \ \left[ \frac{3 |\mathbf{r}_{1}^{(h)}|^{2} } {R_{orb}^4} + \ldots \right] \ \mathbf{x}_{1}
} \nonumber\\
&=& \omega_{(R)} \hat{\mu} \ \int{
dq(\mathbf{r}_{1}) \left[ \frac{3 |\mathbf{r}_{1}^{(h)}|^{2} } {R_{orb}^4} + \ldots \right] \ \mathbf{x}_{1}
} \nonumber\\
&=& \omega_{(R)} \hat{\mu} \ \int{
dq(\mathbf{r}_{1})
\left[ \frac{3 |\mathbf{r}_{1}^{(h)}|^{2} } {R_{orb}^4} + \ldots \right] \ \mathbf{x}_{1}
},\nonumber\\ 
\label{F_m_R_multipolar} 
\end{eqnarray}
where we have used $\vec{R}_{orb} = R_{orb}\ \mathbf{x}_{1}$.
We now truncate to leading order, and solve the integral:
\begin{eqnarray}
\mathbf{F}_{m}^{(R)} &\approx&
\frac{3 \hat{\mu} Q_{2}} {R_{orb}^4} \ \omega_{(R)} \ \mathbf{x}_{1},
\label{F_m_R_truncated}
\end{eqnarray}
where $Q_{2}$ is a second-order axial momentum around the $\mathbf{z}_{1}$-axis:
\begin{eqnarray}
Q_{2} &=& \int{ dq(\mathbf{r}_{1}) \  |\mathbf{r}_{1}^{(h)}|^{2} }.
\label{Q2}
\end{eqnarray}

\subsubsection{Remarks}

We remark that both $dF_{m}^{(o)},dF_{m}^{(R)}$, under these 2D initial
conditions, only have components in the plane OXY:
\begin{eqnarray}
\mathbf{F}_{m} =
\int{ (\ d\mathbf{F}_{m}^{(o)} + d\mathbf{F}_{m}^{(R)}\ ) } = \int{ d\mathbf{F}_{m}^{(h)} },
\label{F_m_planar}
\end{eqnarray}
and we have also seen the first non vanishing term in both $dF_{m}^{(o)}$,
in (\ref{F_m_o}), and $dF_{m}^{(R)}$, (\ref{F_m_R}), belongs, respectively,
to zero and second order (desarrollar esto!!!).

\subsection{Magnetic torque}

We analyze now the torque, that we previously introduced, from a very simple
but intuitive graphical argument (reference to graphics).
As a summary of this section, let us say we will show that this torque does
indeed comes exclusively from the orbital movement, as we had already foreseen
through the graphical argument. 
We define:
\begin{eqnarray}
\boldsymbol{\tau}_{m}^{(o)} &=& \int{ d\boldsymbol{\tau}_{m}^{(o)} }, \quad
d\boldsymbol{\tau}_{m}^{(o)} = \mathbf{r}_{1} \ \wedge\  d\mathbf{F}_{m}^{(o)},
\label{eqn12} \\
\boldsymbol{\tau}_{m}^{(R)} &=& \int{ d\boldsymbol{\tau}_{m}^{(R)} }, \quad
d\boldsymbol{\tau}_{m}^{(R)} = \mathbf{r}_{1} \ \wedge\  d\mathbf{F}_{m}^{(R)},
\label{eqn13}
\end{eqnarray}
and obviously $\boldsymbol{\tau}_{m}=\boldsymbol{\tau}_{m}^{(o)}\ +\ \boldsymbol{\tau}_{m}^{(R)}$.

\subsubsection{Orbital component} 

Using (\ref{dF_m_o}) we have
\begin{eqnarray}
d\boldsymbol{\tau}_{m}^{(o)} &=& \mathbf{r}_{1} \ \wedge\  d\mathbf{F}_{m}^{(o)} \nonumber \\
&=& \mathbf{r}_{1} \ \wedge\  dq(\mathbf{r}_{1}) \ \left[\ \mathbf{v}^{(o)} \wedge - B(\mathbf{r}_{1})\ \mathbf{z}_{1} \ \right].
\label{dT_m_o_1}
\end{eqnarray}
Now, with $\mathbf{v}^{(o)}=v^{(o)}_{r}\mathbf{x}_{1} + v^{(o)}_{t}\mathbf{y}_{1}$
because our planar initial condition, we do
\begin{eqnarray}
d\boldsymbol{\tau}_{m}^{(o)} 
&=& \mathbf{r}_{1} \ \wedge\  dq(\mathbf{r}_{1}) \
\left[\ v_{r}^{(o)}\ \mathbf{x}_{1} + v_{t}^{(o)}\ \mathbf{y}_{1} \ \right]
\wedge - B(\mathbf{r}_{1})\ \mathbf{z}_{1} \nonumber\\
&=& dq(\mathbf{r}_{1})\ B(\mathbf{r}_{1}) \ \mathbf{r}_{1} \ \wedge\
( - v_{r}^{(o)}\ \mathbf{y}_{1} \ +\  v_{t}^{(o)}\ \mathbf{x}_{1} ),
\nonumber\\   
 \label{dT_m_o_2}
\end{eqnarray}
Also, due to reflection symmetry around the OXY plane, all contributions to
$\boldsymbol{\tau}_{m}^{(o)}$ in that plane vanish when integrated:
\begin{eqnarray}
\boldsymbol{\tau}_{m}^{(o)} = \int{ d\boldsymbol{\tau}_{m}^{(o)} } =
\int{ (d\boldsymbol{\tau}_{m}^{(o)})^{(z)} },
\label{T_m_o}
\end{eqnarray}
and using (\ref{dT_m_o_1}) we do
\begin{eqnarray}
d\boldsymbol{\tau}_{m}^{(o)}
&=& \mathbf{r}_{1} \ \wedge\  dq(\mathbf{r}_{1}) \
\left[\ - \mathbf{v}^{(o)} \ \wedge\  - B (\mathbf{r}_{1})\ \mathbf{z}_{1} \ \right] \nonumber\\
&=& \mathbf{v}^{(o)} \ \wedge\ dq(\mathbf{r}_{1}) \
\mathbf{r}_{1} \ \wedge\  - B (\mathbf{r}_{1})\ \mathbf{z}_{1} \nonumber\\
&=& - \mathbf{v}^{(o)} \ \wedge\ dq(\mathbf{r}_{1}) \
\mathbf{r}_{1}^{(h)} \ \wedge\  B (\mathbf{r}_{1}) \ \mathbf{z}_{1},
\end{eqnarray}
and applying the multipole expansion for the magnetic field (\ref{B_esc_multipolar}),
\begin{eqnarray}
&& d\boldsymbol{\tau}_{m}^{(o)} =
- \mathbf{v}^{(o)} \ \wedge\  dq(\mathbf{r}_{1}) \ \times\ \nonumber\\
&& \quad\quad\quad (\ \mathbf{r}_{1}^{(h)} \ \wedge\
\hat{\mu} \ \left[ \frac{1}{R_{orb}^3} - \frac{3 \mathbf{x}_{1}\cdot\mathbf{r}_{1}}{R_{orb}^4} + \ldots \right]
\ \mathbf{z}_{1} \ ),
\nonumber\\ 
\end{eqnarray}
With the ``additional'' factor $\mathbf{r}_{1}^{(h)}$, the first term
of the expansion goes from zero to second order.
Clearly it gives no net contribution to the integral. The second term goes
from first to second order, and it is the leading contribution.
Keeping this leading term and truncating:
\begin{eqnarray}
d\boldsymbol{\tau}_{m}^{(o)} &\approx&
- \mathbf{v}^{(o)} \wedge\ dq(\mathbf{r}_{1}) \  \mathbf{r}_{1}^{(h)} \wedge\
\hat{\mu} \ \frac{3 \mathbf{x}_{1}\cdot\mathbf{r}_{1}}{R_{orb}^4}\ \mathbf{z}_{1} \nonumber\\
&=& -\ \mathbf{v}^{(o)} \wedge\ dq(\mathbf{r}_{1}) \ \mathbf{r}_{1}^{(h)} \wedge\
\hat{\mu}\ \frac{3 \mathbf{x}_{1}\cdot\mathbf{r}_{1}}{R_{orb}^4}\ \mathbf{z}_{1} \nonumber\\
&=& \-\mathbf{v}^{(o)} \wedge\ \hat{\mu}\ dq(\mathbf{r}_{1}) \
\frac{3 |\mathbf{r}_{1}^{(h)}|^{2} \  \cos\psi_{1} } {R_{orb}^4} \nonumber\\
&& \ \ \ \ \ \ \ \ \ \ \ \ \ \ \ \ \ \ \ \ \ \ \ \ \times\
(\ \sin\psi_{1}\mathbf{x}_{1} - \cos\psi_{1}\mathbf{y}_{1} \ ) \nonumber\\
&=& \-\mathbf{v}^{(o)} \ \wedge\
\hat{\mu}\ dq(\mathbf{r}_{1}) \ \frac{3 |\mathbf{r}_{1}^{(h)}|^{2} } {R_{orb}^4} \nonumber\\
&&\ \ \ \ \ \ \ \ \ \ \ \ \ \ \ \times\
(\ \sin\psi_{1}\cos\psi_{1}\mathbf{x}_{1} - \cos^{2}\psi_{1}\mathbf{y}_{1} \ ),
\nonumber\\  
\end{eqnarray}
We could not avoid having to invoke explicitly some coordinates here,
$\phi_1$ being the azimuthal angle in system 1.
The term in $\sin\phi_{1}\cos\phi_{1}$ clearly vanishes if we integrate
for $\psi \in [0,2\pi]$, but no the one in $\cos^{2}\psi_{1}$, and finally:
\begin{eqnarray}
\boldsymbol{\tau}_{m}^{(o)} &\approx& \int{
\mathbf{v}^{(o)} \ \wedge\ \hat{\mu}\ dq(\mathbf{r}_{1})  \
\frac{3 |\mathbf{r}_{1}^{(h)}|^{2} \ \cos^{2}\psi_{1} } {R_{orb}^4} \ \mathbf{y}_{1}
} \nonumber\\
&=& \mathbf{v}^{(o)} \ \wedge\ \frac{3 \hat{\mu}}{R_{orb}^4} \cdot \int{
dq(\mathbf{r}_{1}) \ |\mathbf{r}_{1}^{(h)}|^{2} \cos^{2}\psi_{1} } \ \mathbf{y}_{1} \nonumber\\
&=& \mathbf{v}^{(o)} \ \wedge\  \frac{3 \hat{\mu} Q_{2}^{(*)}}{R_{orb}^4} \ \mathbf{y}_{1}
\nonumber\\
&=& \frac{3 \hat{\mu} Q_{2}^{(*)}}{R_{orb}^4} \ \mathbf{y}_{1} \ \wedge\ \mathbf{v}^{(o)}
= \frac{3 \hat{\mu} Q_{2}^{(*)}}{R_{orb}^4} \cdot v_{r}^{(o)} \ \mathbf{z}_{1},
\nonumber\\ 
\label{T_m_o_truncated}
\end{eqnarray}
in the $z$-direction, as expected, and where
\begin{eqnarray}
Q_{2}^{(*)} = \int{ dq(\mathbf{r}_{1}) \cdot |\mathbf{r}_{1}^{(h)}|^{2} \cos^{2}\psi_{1} },
\label{Q2b}
\end{eqnarray}
is a second order axial moment of the distribution (we
have to check this calculations).

\subsubsection{Rotational component} 

Now we analyze $d \boldsymbol{\tau}_{m}^{(R)}$, and we prove that
$\boldsymbol{\tau}_{m}^{(R)}=0$, i.e., the only contribution to
this torque of magnetic origin comes from the orbital term.
We see:
\begin{eqnarray}
d\boldsymbol{\tau}_{m}^{(R)} &=& \mathbf{r}_{1} \ \wedge\  d\mathbf{F}_{m}^{(R)} \nonumber\\
&=& \mathbf{r}_{1} \ \wedge\  dq(\mathbf{r}_{1}) \ (\ \mathbf{v}^{(R)} \ \wedge\  - B(\mathbf{r}_{1})\ \mathbf{z}_{1}\ ) \nonumber\\
&=& dq(\mathbf{r}_{1}) \ \mathbf{r}_{1} \ \wedge\ (\mathbf{\omega_{(R)} } \ \wedge\  \mathbf{r}_{1}) \ \wedge\  - B(\mathbf{r}_{1})\ \mathbf{z}_{1} \nonumber\\
&=& dq(\mathbf{r}_{1}) \ \mathbf{r}_{1} \ \wedge\ \omega_{(R)}\mathbf{z}_{1} \ \wedge\  B(\mathbf{r}_{1})\ \mathbf{z}_{1} \ \wedge\  \mathbf{r}_{1} \nonumber\\
&=& dq(\mathbf{r}_{1}) \ \mathbf{r}_{1} \ \wedge\  \omega_{(R)}  B(\mathbf{r}_{1}) \ \wedge\  ( - \mathbf{r}_{1}^{(h)}) \nonumber\\
&=& dq(\mathbf{r}_{1}) \ \omega_{(R)} B(\mathbf{r}_{1})\ \mathbf{r}_{1}^{(h)} \ \wedge\  - \mathbf{r}_{1}^{(h)} \nonumber\\
&=& - dq(\mathbf{r}_{1}) \ \omega_{(R)} B(\mathbf{r}_{1})\ \mathbf{r}_{1}^{(h)} \ \wedge\  \mathbf{r}_{1}^{(h)} \nonumber\\
&=& 0. \label{eqn22}
\end{eqnarray}
Therefore,
\begin{eqnarray}
\boldsymbol{\tau}_{m}^{(R)} = \int{ d\boldsymbol{\tau}_{m}^{(R)}} = 0. \label{T_m_R}
\end{eqnarray}
It is worth to remark this last result: the torque on the z-direction
(to be applied on the center of mass of the distribution of charge),
comes exclusively from the ``orbital'' contribution, that contribution
with origin in the ``orbital'' AM of the distribution as a point particle
(therefore, represented by a center of mass).


\subsection{Summary of results for the 2-dimensional problem}

The following shows the first contribution in the electrical multipole expansion for each of the forces or
torques. To second order (Q), no other forces or rotational momenta arise.
\begin{table}[ht] \centering \begin{tabular}{ | c || c c c c | }
\hline Order & $\mathbf{F}_{m}^{(o)}$\ \ \ \ & $\mathbf{F}_{m}^{(R)}$\ \ \  & $\boldsymbol{\tau}_{m}^{(o)}$\ \ \  & $\boldsymbol{\tau}_{m}^{(R)}$\ \ \  \\
\hline Zero (M)\ \ \ \ & $Yes$ & $No$ & $No$ & $No$ \\
       First (D)\ \ \ \ & $Yes$ & $No$ & $No^{\ast}$ & $No$ \\
       Second (Q)\ \ \ \ & $Yes$ & $Yes$ & $Yes$ & $No$ \\
\hline \end{tabular} \end{table}

A table of dependencies may be of use, too,
\begin{table}[ht] \centering \begin{tabular}{| c || c c c c | }
\hline Variable \ \ \ \ \ &$\mathbf{F}_{m}^{(o)}$\ \ \ \ \ & $\mathbf{F}_{m}^{(R)}$\ \ \ \ \ & $\boldsymbol{\tau}_{m}^{(o)}$\ \ \ \ \ & $\boldsymbol{\tau}_{m}^{(R)}$\ \ \ \\
\hline $|R_{orb}|^{-1}$\ \ \ \ \ & $Yes$ & $Yes$ & $Yes$   & $ - $ \\
       $\omega_{(R)}$  \ \ \ \ \ & $ - $ & $Yes$ & $ - $   & $ - $ \\
       $ v^{(o)}_{r}$  \ \ \ \ \ & $Yes$ & $ - $ & $Yes$   & $ - $ \\
       $ v^{(o)}_{t}$  \ \ \ \ \ & $Yes$ & $ - $ & $ - $   & $ - $ \\
\hline \end{tabular} \end{table}

\subsection{\label{General_ic}3D initial conditions}

We have a 3D extension of the problem. Here, $\mathbf{x}_{1}$ still
goes in the radial direction of the orbit, $\mathbf{y}_{1}$ in the tangential
direction and $\mathbf{z}_{1}$ is always the instantaneous axis for $\omega_{(o)}$.
We have now:
\begin{eqnarray}
\mathbf{v}^{(o)} &=& v_{r}^{(o)} \mathbf{x}_{1} \ +\  v_{t}^{(o)} \mathbf{y}_{1} \ +\  v_{z}^{(o)} \mathbf{z}_{1},
\label{v_orb_general}
\end{eqnarray}
and
\begin{eqnarray}
\boldsymbol{\omega}_{(R)} 
&=& (\omega_{(R)})_{r} \ \mathbf{x}_{1} + (\omega_{(R)})_{t} \ \mathbf{y}_{1} + (\omega_{(R)})_{z} \ \mathbf{z}_{1}, \nonumber\\ 
\label{omega_general}
\end{eqnarray}
with, now, vectorially (recall Sec. \ref{kin_con}),
\begin{equation}
\boldsymbol{\omega}_{(R)} = \boldsymbol{\omega} - \boldsymbol{\omega}_{(o)}, \label{omega_R_vec}
\end{equation}
Needless to say, we can always decompose the dynamics of the system in
this way. 
Now we complete the systematic analysis of the dominant terms for the forces,
$\mathbf{F}_{m}^{(o)}, \mathbf{F}_{m}^{(R)}$, and torques,
$\boldsymbol{\tau}_{m}^{(o)},\boldsymbol{\tau}_{m}^{(R)}$.
    %
%
First we calculate the new contributions to the forces:
\begin{eqnarray}
\Delta ( d\mathbf{F}_{m}^{(o)} )
&=& dq(\mathbf{r}_{1}) \ v_{z}^{(o)} \mathbf{z}_{1} \ \wedge\ - B(\mathbf{r}_{1})\ \mathbf{z}_{1}
= 0, \nonumber\\
\label{Delta_dF_m_o} 
\end{eqnarray}
so there is no new contribution to $\mathbf{F}_{m}^{(o)}$, and
\begin{eqnarray}
\Delta ( d\mathbf{F}_{m}^{(R)} )
&=& dq(\mathbf{r}_{1}) \ (\boldsymbol{\omega}_{(R)})^{(h)} \ \wedge\  \mathbf{r}_{1} \ \wedge\ - B(\mathbf{r}_{1})\ \mathbf{z}_{1} \nonumber\\
&=& -\ dq(\mathbf{r}_{1}) \  B (\mathbf{r}_{1})\ (\boldsymbol{\omega}_{(R)})^{(h)} \ \wedge\  \mathbf{r}_{1} \ \wedge\ \mathbf{z}_{1} \nonumber\\
&=& \Delta ( d\mathbf{F}_{m}^{(R)} )^{(z)}.  \nonumber\\ 
\label{Delta_dF_m_R} 
\end{eqnarray}
This is rewarding because the new contribution is normal to the one we already have.
We can also write
\begin{eqnarray}
\Delta ( d\mathbf{F}_{m}^{(R)} ) =
dq(\mathbf{r}_{1})\ B (\mathbf{r}_{1})\ (\boldsymbol{\omega}_{(R)})^{(h)}\ \wedge\ \mathbf{r}_{1}^{(h, \pi/2)},
\end{eqnarray}
where $\mathbf{r}_{1}^{(h, \pi/2)}$ stands for a rotation of $\mathbf{r}_{1}^{(h)}$
by an angle $\pi/2$ around the z-axis. We would have the integral
\begin{eqnarray}
\Delta(\mathbf{F}_{m})
&=& \int{ \Delta(d\mathbf{F}_{m}) } \nonumber\\
&=& \int{ \Delta(d\mathbf{F}_{m}^{(o)}) } + \int{ \Delta(d\mathbf{F}_{m}^{(R)}) } \nonumber\\
&=& \int{ \Delta(d\mathbf{F}_{m}^{(R)})^{(z)} } \ =\ \Delta( \mathbf{F}_{m}^{(R)})^{(z)},
\label{Delta_F_m} 
\end{eqnarray}
where we have just identified that $\mathbf{F}_{m}^{(R)}$ only has a component in
the $z$-direction.
%
Now we see the new components for the torques. Clearly from (\ref{Delta_dF_m_o})
we see there is neither any new contribution to the part of the torque that
originates from the ``orbital'' movement:
\begin{eqnarray}
\Delta (d\boldsymbol{\tau}_{m}^{(o)})
&=&  \mathbf{r}_{1}\ \wedge\ \Delta(d \mathbf{F}_{m}^{(o)})
= 0, 
\label{Delta_dT_m_o_prev} 
\end{eqnarray}
and for the rotational part, using that $\Delta(d\mathbf{F}_{m}^{(R)})=\Delta(d\mathbf{F}_{m}^{(R)})^{(z)}$,
we can write
\begin{eqnarray}
\Delta ( d \boldsymbol{\tau}_{m}^{(R)} )
&=& \mathbf{r}_{1} \ \wedge\ \Delta(d\mathbf{F}_{m}^{(R)}) \nonumber\\
&=& \mathbf{r}_{1} \ \wedge\ \Delta(d\mathbf{F}_{m}^{(R)})^{(z)}  \nonumber\\
&=& \mathbf{r}_{1}^{(h)} \wedge\ \Delta(d\mathbf{F}_{m}^{(R)})^{(z)} \ =\ (\Delta d\boldsymbol{\tau}_{m}^{(R)})^{(h)}, \nonumber\\
\label{Delta_dT_m_o} 
\end{eqnarray}
again with the rewarding result that this new contribution $\Delta(d\boldsymbol{\tau}_{m}^{(R)})$
is a ``horizontal'' vector (vector in the plane OXY) and hence normal to the contribution
we already had for the ``planar'' initial conditions.

Again we attempt to summarize everything in a table.
These results are generalized to any -planar or not- initial conditions.
First line says whether a net contribution exists. Second line establishes a dependence.
Last line says which is the leading order: zero (M), first (D) or second (Q).


\begin{table}[ht] \centering \begin{tabular}{| c c | c c | c c | c c |}
$\mathbf{F}_{m}^{(o,h)}$\  & $\mathbf{F}_{m}^{(o,z)}$\  & $\mathbf{F}_{m}^{(R,h)}$\  & $\mathbf{F}_{m}^{(R,z)}$\ &
$\boldsymbol{\tau}_{m}^{(o,h)}$\  & $\boldsymbol{\tau}_{m}^{(o,z)}$\  & $\boldsymbol{\tau}_{m}^{(R,h)}$\  & $\boldsymbol{\tau}_{m}^{(R,z)}$\ \\
\hline
$Yes$ & $-$ & $Yes$ & $Yes$ & $-$ & $Yes$ & $Yes$ & $-$\\
$v_{r}^{(o)}$ & $-$ & $(\omega_{(R)})^{(z)}$ & $(\boldsymbol{\omega}_{(R)})^{(h)}$ & $-$ & $v_{r}^{(o)}$ & $(\boldsymbol{\omega}_{(R)})^{(h)}$ & $-$\\
\hline
$M$ & $-$ & $Q$ & $Q$ & $-$ & $Q$ & $Q$ & $-$ \\
\end{tabular} \end{table}
%
On the other hand, the new terms should correspond to what is known as the classical
counterpart of the quantum LS or  spin-orbit coupling.
This interaction is also a stabilizing one, as it tends to keep parallel the spinning
axis of particle and the axis of the orbit (the direction of the dipoling source).
Nevertheless, for the moment we will only pay limited attention to questions regarding
the problem in 3D.


\section{\label{Mass_reloc}
Towards a realistic model of the H-atom: relocation of mass}


We had begun presenting an idealized electromechanical game. Now we establish a
bridge from that situation to a realistic model for the classical H-atom.
This implies a relocation of the inertial system, and a proof that the equations of
the system remain the same in the new situation.
Later, the value of the parameters in the model must be adjusted to resemble the
actual charges, masses, etc of the hydrogen atom, but this will not be, for
convenience, done yet.

\subsection{\label{Relocation}Relocation of mass}

In this section we do a relocation of the inertial frame of reference, $RF_{0}$.
To indicate this we introduce the notation $RF_{0}|^{(n)}$, in contrast to $RF_{0}|^{(old)}$.
If our situation is to resemble reasonably the real hydrogen atom, a natural choice
is to attach $RF_{0}|^{(n)}$ to the center of the proton (the charge distribution),
taking advantage from its much higher mass in respect to the electron.
Still, the proton can ``rotate'' around $\mathbf{z}_{0}$ (on the planar problem),
with angular velocity
\begin{eqnarray}
\omega|^{(n)} = \omega_{(R)}|^{(n)} + \omega_{(o)}|^{(n)},
\end{eqnarray}
with respect to $RF_0|^{(n)}$.
This definition is consistent with the one already given in (\ref{omega_R}), so,
analogously to what happened there, $\omega|^{(n)}$ and $\omega_{(o)}|^{(n)}$ both
have an ``inertial'' meaning (they are both angular frequencies measured with
respect to an inertial reference frame $RF_0$), while $\omega_{(R)}|^{(n)}$ does
not (but is nevertheless a very convenient dynamical variable from the point of
view of the equations and our whole argument).
Of course, we are only dealing here with the 2-dimensional situation: for the
3D problem, the same would hold but vectorially this time (it is always
possible to decompose the (instantaneous) movement this way):
$\boldsymbol{\omega}|^{(n)} = \boldsymbol{\omega}_{(R)}|^{(n)} + \boldsymbol{\omega}_{(o)}|^{(n)}$.

Again, an auxiliary system $RF_{1}|^{(n)}$ is of good use.
We choose to attach it, as we did before, to the center of the distribution, and
also choose the same orientation, so that $RF_{1}|^{(n)} \equiv RF_{1}|^{(old)}$
(hence we simply write $RF_{1}$), but this time, also, the origin $RF_{1}|^{(n)}$
and $RF_{0}|^{(n)}$ is common.
%
\begin{figure}[ht!]
\includegraphics[width=0.90 \columnwidth,clip]{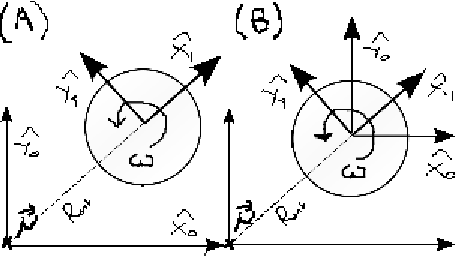} \caption{
Location of systems $RF_{0}$ and $RF_{1}$ in situation (A) and (B).
In (A), origin of $RF_{0}$ coincides with the position of the magnetic point-like
dipole $\boldsymbol{\mu}$, while in (B) both origins coincide.
} \label{Fig3} \end{figure}
On the other hand, the point-like magnetic source (an electron, here) will orbit
with velocity 
$\mathbf{v}^{(o)}|^{(n)} = \mathbf{v}^{(o)}_{t}|^{(n)} + \mathbf{v}^{(o)}_{r}|^{(n)}$
around the proton, and, again in this situation, the components can be expressed
using the versors of $RF_1$, in the way:
\begin{eqnarray}
\mathbf{v}^{(o)}_{r}|_{(n)} &=& v^{(o)}_{r}|_{(n)}\ \mathbf{x}_{1}, \\
\mathbf{v}^{(o)}_{t}|_{(n)} &=& v^{(o)}_{t}|_{(n)}\ \mathbf{y}_{1}.
\end{eqnarray}

\subsection{\label{Equivalence}Formal equivalence of the equations}

The problem of a moving magnetic dipole is a classical one in electromagnetism.
Due to Lorentz covariance, a magnetic dipole $\boldsymbol{\mu}$ that moves with velocity
$\mathbf{v}$ in respect to an inertial frame is seen, by an observer in that inertial
frame, as an electrical dipole of value $\mathbf{p}$ proportional to
$\boldsymbol{\mu}\wedge\mathbf{v}$.
For more details in this result, we can cite , for instance, \cite{Panofsky_Philips}
(18-5, page 334).
This electric ``effective'' dipole creates, at the point
$\mathbf{r}_{0} = R_{orb}\ \mathbf{x}_{1} + \mathbf{r}_{1}$, an electric field
\begin{eqnarray}
\mathbf{E}|_{(n)}\
&\propto& \frac{ \mathbf{p} }{ |R_{orb}\ \mathbf{x}_{1} + \mathbf{r}_{1}|^3 }  
= \frac{ \boldsymbol{\mu} \wedge \mathbf{v}|_{(n)} }{ |R_{orb}\ \mathbf{x}_{1} + \mathbf{r}_{1}|^3 },
\end{eqnarray}
therefore exerting a force
\begin{eqnarray}
d\mathbf{F}|_{(n)}\ =\
dq \ \mathbf{E}|_{(n)} \ \propto\
dq \frac{ \boldsymbol{\mu}\wedge\mathbf{v}|_{(n)} }{ |R_{orb}\ \mathbf{x}_{1} + \mathbf{r}_{1}|^3 }, 
\label{dF_reinterpreted}
\end{eqnarray}
on every element of charge $dq$ of the distribution.
Now, we recall that in our previous situation we had, sufficiently away from the source,
\begin{eqnarray}
\mathbf{B}|_{(old)}\ 
\approx \frac{\mu_0}{4\pi} \times \frac{ \boldsymbol{\mu} }{ |R_{orb}\ \mathbf{x}_{1} + \mathbf{r}_{1}|^3 },
\end{eqnarray}
and therefore
\begin{eqnarray}
d\mathbf{F}|_{(old)}
&=& dq \frac{ \mathbf{v}|_{(old)} \wedge \mathbf{B}|_{(old)} }{ R_{orb}^3 } 
= dq \frac{ \mathbf{v}|_{(old)} \wedge \boldsymbol{\mu}}{ |R_{orb}\ \mathbf{x}_{1} + \mathbf{r}_{1}|^3 }.
\nonumber\\ \label{B_recall}
\end{eqnarray}
But now is the source that moves in respect to the element of charge $dq$.
Clearly, $\mathbf{v}|_{(n)} = - \mathbf{v}|_{(old)}$.
Moreover, for clarity we also state:
\begin{eqnarray}
\mathbf{v}^{(o)}|_{(n)} &=& - \mathbf{v}^{(o)}|_{(old)}, \label{change_v_o} \\
\mathbf{v}^{(R)}|_{(n)} &=& - \mathbf{v}^{(R)}|_{(old)}, \label{change_v_R}
\end{eqnarray}
and also clearly,
\begin{eqnarray}
\omega_{(R)}|^{(n)} &=& \omega_{(R)}|^{(n)}. \label{change_w}
\end{eqnarray}
But it does suffice to perform the substitution $\mathbf{v}|_{(n)} = - \mathbf{v}|_{(old)}$ in
(\ref{dF_reinterpreted}), and so we have, again,
\begin{eqnarray}
d\mathbf{F}|_{(n)}\ \ \propto \
dq \frac{ \mathbf{v}|_{(old)} \wedge \boldsymbol{\mu}}{ |R_{orb}\ \mathbf{x}_{1} + \mathbf{r}_{1}|^3 },
\label{dF_reinterpreted_II}
\end{eqnarray}
This is the same as we had in (\ref{B_recall}) and so it was exactly what we were
looking for.


Therefore, at least for the term that depends on a wedge product on the velocity
vector, the expression of the force $\mathbf{F}$ on each element of charge is
completely equivalent (modulus a certain constant) to the one in our previous
situation, the first electromechanical model we presented here.
Moreover, all previous expressions are applicable (modulus a possible constant),
and no further changes needed, as we take advantage here of the fact that all of
them were referred to $RF_{1}$, that has not changed in the new picture (using
this trick has saved us a lot of calculations).

\begin{figure}[ht!] \includegraphics[width=0.95 \columnwidth,clip]{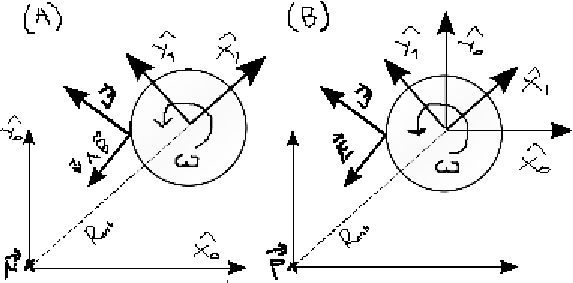}
\caption{
Direction of the force in situation (A): $\mathbf{dF}=dq\mathbf{v} \wedge \mathbf{B}$,
and (B): $dF = dq\mathbf{E}$.
From $RF_{0}$, the magnetic dipole  $\boldsymbol{\mu}$ is seen as an electric dipole
$\mathbf{p}$, source of the corresponding electric field.
} \label{Fig4} \end{figure}



\section{\label{Dynamical_eq_g}
Dynamical equations (2D)}


We have seen that the equations of the system are invariant about whether it
is the point-like negative charge or the extended distribution (of positive charge)
that is moving, if we do the convenient redefinition of velocities.
Using that fact, we present a first set of equations, that will be later enhanced
by the inclusion of the radiative correction as well as the absorption from the
background.
This second step is done, anyway, only in an ``approximate'' way that is nevertheless
enough for our purposes.

\subsection{\label{Dynamical_eq}
Dynamical equations in absence of rad/abs}

In the following equations, all quantities ($q$, $E_1$, $B_1$, etc.) are defined
positive.
In absence of dissipation and absorption and for the planar problem, the
dynamical equations of the system are, to leading order,
\begin{center} 
\textit{(correction from previous version in first equation - originally it was
correct in rest of the paper though)}
\end{center}
\begin{eqnarray}
&&m \ \dot{v}^{(o)}_{r} + m \frac{ (v^{(o)}_{t})^2 }{ R_{orb} } \approx 
\left[\ \mathbf{F}_{e} \ +\ \mathbf{F}_{m}^{(o)} \ +\ \mathbf{F}_{m}^{(R)}\ \right]
\cdot \mathbf{x}_{1} \nonumber\\
&&\quad\quad\quad\quad
= \ -\ q E_1 \ -\ q B_1 \ v^{(o)}_{t} \ -\ \frac{3 \hat{\mu} Q_{2}}{R_{orb}^4}\ \omega_{(R)},
\nonumber\\ \label{vr_dynamic}
\end{eqnarray}
\begin{eqnarray}
m \ \dot{v}^{(o)}_{t} &\approx&
\left[\  \mathbf{F}_{m}^{(o)} \ +\  \mathbf{F}_{m}^{(R)} \ \right] \cdot \mathbf{y}_{1} =
\mathbf{F}_{m}^{(o)} \cdot \mathbf{y}_{1}\nonumber\\
&=& \ +\ q B_1 \ v^{(o)}_{r}, \label{vt_dynamic}\\
\nonumber\\ 
M_{2} \ \dot{\omega_{(R)}} &\approx&
\boldsymbol{\tau}_{m}^{(o)} \cdot \mathbf{z}_{1} \ +\ \boldsymbol{\tau}_{m}^{(R)} \cdot \mathbf{z}_{1} =
\boldsymbol{\tau}_{m}^{(o)} \cdot \mathbf{z}_{1} \nonumber\\
&=& \ +\ \frac{ 3 \hat{\mu}\ Q_{2}^{(*)} }{R_{orb}^4} \ v^{(o)}_{r}, \label{w_dynamic}
\end{eqnarray}
with $M_2$ a mass second order moment or ``inertia'' moment.
In the first of the former equations (\ref{vr_dynamic}), we take into account
the fact that $RF_{1}$ is rotating with respect to $RF_{0}$, and that we have
defined this radial velocity as $v^{(o)}_{r}=v^{(o)}_{r}\ \mathbf{x}_{1}$.

As a clarification, we recall $\mathbf{F}_{m}^{(R)}$ has no component in
$\mathbf{y}_{1}$.
Also, we have to notice $E_1$, $B_1$ do depend on $R_{orb}$ (see below).
To the former three equations, we must add the obvious
\begin{eqnarray}
\dot{R}_{orb} &=& {v}^{(o)}_{r}, \label{R_dynamic}
\end{eqnarray}
and they must also be supplemented with the values of two fields evaluated
in $\mathbf{r}_{1} = 0$. 
With
$\mathbf{E}(\mathbf{r}_{1}) = - E (\mathbf{r}_{1})\ \mathbf{x}_{1}$ and
$\mathbf{B}(\mathbf{r}_{1}) = - B (\mathbf{r}_{1})\ \mathbf{z}_{1}$, we have
\begin{eqnarray}
E_1 &=& |\mathbf{E}(\mathbf{r}_{1})| = + \frac{q}{4\pi\epsilon_0 \ R_{orb}^2},
\label{E_1}\\
B_1 &=& |\mathbf{B}(\mathbf{r}_{1})| \propto + \ \frac{\mu_0}{4\pi}\frac{\mu_e}{R_{orb}^3},
\label{B_1}
\end{eqnarray}
with $\mu_e$ the magnetic dipole moment of the electron.
\textit{Note: we can just state proportionality in the last equation as there
are additional factors due to our ``mass relocation''; 
anyway the contribution of $B_1$ is marginal and without implications for stability,
we will even ignore it a future revision, just as some other terms - for instance
the classical counterpart of the LS quantum term - are also ignored in these basic
sets of equations)}.

See (\ref{E})-(\ref{B}). It is important to note that these fields enter
the former equations with the values in the point $\mathbf{r}_{0} = \vec{R}_{orb}$
(i.e., $\mathbf{r}_{1} = 0$).
To leading order, then, any other information about the distribution is
already contained in $q$, $Q_{2}$ and $Q_{2}^{*}$.
The first moment $Q_{1}=0$ vanishes, as all charges in the distribution
are of equal sign, and moments of higher order are not considered.


\subsection{\label{rad_incl}Inclusion of rad/abs}

How do we include the radiative correction?
From the point of view of the equations, it would suffice to include a stochastic term
in (\ref{vr_dynamic}), (\ref{vt_dynamic}), and possibly in (\ref{w_dynamic}).
This stochastic term represents the difference of loss and absorbed radiation in each
instant of time, and its mean value is zero for a stationary orbit (an ``attractor'').
For non stationary orbits the mean value of this stochastic term would not be zero.

The inclusion in (\ref{vr_dynamic}), (\ref{vt_dynamic}) is quite obvious: we must allow
for an energy loss through a fall of the orbital radius and a decrement of the modulus
of the tangential velocity. 
The term in (\ref{w_dynamic}) is not so obvious, and we will extend on this later.
On the other hand, a more rigorous treatment is not necessary for our purposes,
at least for now.


\subsubsection{Rad/abs for a circular orbit}


Puthoff calculated $\langle P_{ab} \rangle_{(orbit)}$ assuming statistical equilibrium
with the orbital degrees of freedom (these are the two spatial coordinates in the plane,
oscillating with frequency $\omega_{(o)}$, so therefore they can be seen as two
one-dimensional harmonic oscillators in quadrature).
Following \cite{Puthoff87}, we had
\begin{eqnarray}
\langle P_{rad} \rangle_{(circ)} = \frac{q^2 R_{orb}^2 \omega_{(o)}^4 }{ 6\pi\epsilon_{0} c^3}, \label{P_rad}
\end{eqnarray}
which is directly obtainable from the Larmor formula with acceleration
\begin{eqnarray}
\dot{v}^{(o)}_{r} = \frac{(v^{(o)}_{t})^2}{R_{orb}} = \omega_{(o)}^2 R_{orb},
\end{eqnarray}
and
\begin{eqnarray}
\langle P_{ab} \rangle_{(circ)} = \frac{q^2 \hbar \omega_{(o)}^3 }{ 6\pi\epsilon_{0} m c^3}, \label{P_ab}
\end{eqnarray}


But here we deal with an instantaneous basis: the particle will suffer the action
of a field with an stochastic instantaneous value (the ZPF), and will also loose
energy whose (expectation?) (instantaneous) value is given by the radiation term
(dependent on the instantaneous velocity).

\subsubsection{Rad/abs for the radial component}

To first approximation, we can introduce a correction just in (\ref{vr_dynamic}),
through the inclusion of an stochastic term $\chi_{r}$, with an expectation value
which depends on $R_{orb}$, and must change sign around the ``stable'' value $R_{orb}^{st}$.
This $R_{orb}^{st}$ is the value of $R_{orb}$ for which the balance, on average over
a cyclic orbit, of loss and absorption takes place, i.e., an equality between eqs.
(\ref{P_ab}) and (\ref{P_rad}) holds. So,
\begin{eqnarray}
\langle\chi_{r}(R_{orb})\rangle < 0,  \quad R_{orb} > R_{orb}^{st}, \\
\langle\chi_{r}(R_{orb})\rangle > 0,  \quad R_{orb} < R_{orb}^{st},
\end{eqnarray}
for example,
\begin{eqnarray}
\langle\chi_{r}(R_{orb})\rangle \approx k_{r} (R_{orb}^{st}-R_{orb}), \quad k_{r} > 0.	\label{linear_chi_r}
\end{eqnarray}
For clarification, we have to say that the instantaneous values of $P_{ab}$ and
$P_{rad}$ are stochastic variables whose distributions depend on the instantaneous
values of $R_{orb}$, $v^{(o)}_{r}$, etc., and therefore, strictly speaking
$\chi(R_{orb}) = f \left( P_{ab} - P_{rad} \right)$, and (\ref{linear_chi_r})
is only justified for mean values (which is what we have done).

\subsubsection{Rad/abs for the tangential component}

Moreover, the gain/loss of energy affects both the radial and tangential components
of the velocity.
For this reason, we consider a second stochastic component $\chi_{t}(R_{orb})$,
satisfying, this time
\begin{eqnarray}
\langle\chi_{t}(v^{(o)}_{t})\rangle < 0,  \quad v^{(o)}_{t} > v^{(o)}_{t}|_{st}, \\
\langle\chi_{t}(v^{(o)}_{t})\rangle > 0,  \quad v^{(o)}_{t} < v^{(o)}_{t}|_{st},
\end{eqnarray}
and therefore, close to the point of equilibrium $v^{(o)}_{t}|_{st}$, we can write
\begin{eqnarray}
\langle\chi_{t}(v^{(o)}_{t})\rangle 
\approx k_{t} (v^{(o)}_{t}|_{st} - v^{(o)}_{t}), \quad k_{t} > 0.	 \label{linear_chi_t}
\end{eqnarray}

\subsubsection{Rad/abs due to ``spinning''}

Our study of the dynamical equations lead us to conclude that we need another
dissipation/gain loss for the degree of freedom represented by $\omega$. 
The rotational movement of the distribution around an axis is also subjected to
loss and absorption from the background. In the way we have done before, we write
\begin{eqnarray}
\langle\chi_{\omega}(\omega)\rangle 
\approx k_{\omega}(\omega^{st} - \omega), \quad k_{\omega} > 0.	 \label{linear_chi_w}
\end{eqnarray}
Naturally, for a realistic hydrogen atom we would have $\omega|^{st}=0$ (the proton
is quasi-attached to an inertial system: it can oscillate with respect to it, but
the mean value of this oscillation is zero). We will comment on this later.

\subsubsection{\label{Complete_eq}Complete dynamical equations}

We now include all the former corrections in the dynamical equations of the system.
First, for the radial component:
\begin{eqnarray}
m \dot{v}^{(o)}_{r} &+& m\ \frac{ (v^{(o)}_{t})^2 }{ R_{orb} } \approx \nonumber\\
&& - q E_1 - q B_1 \ v^{(o)}_{t} - \frac{3 \hat{\mu} Q_{2}}{R_{orb}^4}\ \omega_{(R)} + \chi_{r},
\nonumber\\ 
\label{vr_dynamic_c}
\end{eqnarray}
and reordering (that term on the left...),
\begin{eqnarray}
m \dot{v}^{(o)}_{r} &\approx&
- q E_1 - q B_1 \ v^{(o)}_{t} - m\ \frac{ (v^{(o)}_{t})^2 }{ R_{orb} } \nonumber\\
&&\quad\quad\quad\quad\quad\quad\quad
- \frac{3 \hat{\mu} Q_{2}}{R_{orb}^4}\ \omega_{(R)} + \chi_{r},
\label{vr_dynamic_c_II}
\end{eqnarray}
Secondly, for the tangential component,
\begin{eqnarray}
m \ \dot{v}^{(o)}_{t} &\approx&
+ q B_1 \ v^{(o)}_{r} + \chi_{t},
\label{vt_dynamic_c}
\end{eqnarray}
and to conclude for now, for the ``spinning'' of the distribution,
\begin{eqnarray}
M_{2} \ \dot{\omega_{(R)}} &\approx&
\ +\ \frac{ 3 \hat{\mu}\ Q_{2}^{(*)} }{R_{orb}^4} \ v^{(o)}_{r} \ +\ \chi_{\omega}. \label{w_dynamic_c}
\end{eqnarray}


\section{\label{Feddback}Feedback loops}


\subsection{\label{Feedback_1}A primary FL (1st-FL)}

As a preliminary approach, we are interested in a ``primary'' feedback loop''
(1st-FL),
\begin{equation}
v^{(o)}_{r} \Rightarrow \boldsymbol{\tau}
\Rightarrow \boldsymbol{\omega} \Rightarrow \mathbf{F} \Rightarrow v^{(o)}_{r}, 
\label{FL_2}
\end{equation}
where $\boldsymbol{\tau}$ and $\mathbf{F}$ are a torque and force, as we
already said, mediating between the (orbital) radial component of the instant
velocity and the self rotation of the charge distribution. 
We can also interpret this in terms of the orbital radius,
\begin{equation}
R_{orb} \Rightarrow \boldsymbol{\tau}
\Rightarrow \boldsymbol{\omega} \Rightarrow \mathbf{F} \Rightarrow R_{orb}.
\label{FL_3}
\end{equation}
Now, if we look at the equations (\ref{vr_dynamic_c_II}) and (\ref{w_dynamic_c}),
we can identify a FL with an odd number of ``minus'' signs.
We now explain what this means. Equation (\ref{vt_dynamic_c}) will be left aside
for the moment, for the sake of clarity.
Whenever there exist these kinds of FLs, the negative sign of one of them is a
necessary and sufficient condition for the existence of some kind of stability (this
is a well known result from the theory of dynamical systems).

\subsection{\label{Feedback_2}A secondary FL (2nd-FL)}

There is a ``secondary'' feedback loop (2nd-FL) in the equations,
\begin{equation}
v^{(o)}_{t} \Rightarrow v^{(o)}_{r} \Rightarrow v^{(o)}_{t}, \label{FL_b}
\end{equation}
that would prevent the existence of stationary orbits where $v^{(o)}_{r}(t)=0, \forall t$.
However,we will in principle disregard this effect for simplicity (simply ignoring
the corresponding term in the equations).
Later, we will see that it is precisely this secondary loop the reason why
stationary/stable orbits corresponding to a quantum labeling $n=0$ are not present
in the real spectrum, as so it happens for $n=1$ for orbitals with net AM (there
are no real orbitals for $n=0$, only $s$-orbitals for $n=1$, only $s,p$ orbitals
for $n=2$, etc.).

\textit{
The presence of a natural frequency of resonance is directly linked, and proven,
by the fact that all poles of the system (under linearization) are purely imaginary.
Indeed, the (approximate) linearization of the system around a stationary point
(either a circular or pendulum-like orbit) yields a set of purely imaginary poles
(in the frequency domain). This corresponds to a harmonic oscillatory behavior.}


\section{\label{Particularization}A hydrogen model in 2D}



Up to here our equations were completely general, now, to particularize the equations
to the hydrogen system, we must make the substitutions (check the correspondence in
each equation):
\begin{eqnarray}
m   &\rightarrow& m_e, m_p, \\
\mu &\rightarrow& \mu_e, \\
\hat{\mu} &\propto& \hat{\mu}_{e}=\frac{\mu_0}{4\pi}\mu_e,
\end{eqnarray}
with $m_e$, $m_p$ the masses of the electron and proton, respectively,
and $\mu_e$ the magnetic moment of the electron.
Proportionality in the third equation comes from the analisys in
Sec.\ref{Mass_reloc} (INCLUDE EXACT FACTOR AND REFORMULATE FROM HERE).

On the other hand, for a realistic hydrogen atom, $\langle \omega \rangle=0$, i.e.,
the proton, with an inertial frame attached to its origin, will not move ``on average'',
but can rotate on small oscillations around the axis of the orbit (at least in the
circular case).
This point is for key importance and we will return to it several more times.


\section{\label{p_stability}
An approach to stability with net AM}


\subsection{An overview}


In the first of the following subsections, \ref{Qualitative}, we begin by presenting
a qualitative reasoning, that may be enough to convince the reader but that it cannot
be considered yet a strict proof. 
Then, we do some calculations to show that a stationary point indeed exists for the system
of equations that describe the system. 
This stationary point, however, can be stable, unstable or critically stable.
In the first case, the system will answer any perturbation that drags it (not very)
far from the point with a reaction that tries to attract it again to the stationary point.
In the second, any perturbation (no matter how small) will launch the system on a
trajectory that will gradually distance it from the initial state. The third case,
critical stability, accounts for a non-limited storage of energy, in the form or
perpetual oscillations.

It is the aim of subsection \ref{Strict_proof} to discriminate between those three
possibilities, determining which one of them occurs. 
This is done through a very clear and well defined mathematical condition: the real
part of all eigenvalues of a certain matrix, involved in the dynamical description
of the system, must be strictly negative.
Nevertheless, through our qualitative approach we already have the certainty that,
at least on a certain point, the system is indeed strictly stable (the first of
the three possibilities).

All this, for the moment, applies strictly only to the trivial case of circular
orbits in the 2-dimensional situation, although the extension to the 3D
space, as well as to more complicated cases, like elliptical orbits, can be already,
and somehow, be foreseen from the base that we are settling here. 
Besides, we will not do any estimates of energies yet here, that will be a matter
of future sections.

\subsection{\label{Qualitative}Qualitative approach}

\subsubsection{The 1st-FL makes stability possible}

Now, if we look at the equations (\ref{vr_dynamic_c_II}) and (\ref{w_dynamic_c}),
we can identify a feedback loop, that we already treated in Sec.~\ref{Feedback_1}.
Whenever there exist these (negative) feedback loops, the negative sign of one of them
is a necessary and sufficient condition for the existence of some kind of stability
(this is a well known result from the theory of dynamical systems).

Following (\ref{w_dynamic_c}), we see that an increase in $v^{(o)}_{r}$, the radial
component of the ``orbital velocity'', causes an increase in $\omega_{(R)}$ (and
therefore also in $\omega = \omega_{(R)} + \omega_{(o)}$).
But, through  (\ref{vr_dynamic_c_II}), and increase in $\omega_{(R)}$ gives as a
consequence a decrease in $v^{(o)}_{r}$.
\emph{
We remind the reader here that $\omega$ and $\omega_{(o)}$ are ``inertial'' angular
frequencies, the first of them referring to a self-rotation of the proton (the residual
part of the inner AM coming from the fact that it is a charge distribution,
most of that (inner)AM -therefore spin - being carried by quarks
as point-like entities), and the second referring to the orbital movement of the
electron around.
Meanwhile, $\omega_{(R)}$ has no inertial meaning and is just a convenient dynamical
variable to work with.}

This stability could be in principle strict, in the sense of a strictly stable point,
or could also be a saddle point, or/and, given the nonlinear nature (see below) of the
dynamical equations, it could give rise to more complicated stable ``orbits'' in the
configuration space of the system.
We use ``orbit'' in a somewhat broader sense than in previous sections, meaning this
time any collection of points in the configuration space of the system where the
probability to find it does not change with time.
Stability implies that the system, confronted with a small perturbation, always
responds in a way that tries to compensate that perturbation. The ``smallness''
of the perturbation sets it clear that stability is, in principle, a local concept,
valid in a certain ``environment'' within the configuration space of the system.
As we already said in previous sections, this perturbation would be, for example,
a sudden ``push'', due to a mismatch on the emitted/absorbed energy.

\emph{A simple argument could be this:}
let us suppose the system is in a circular orbit with $v^{(o)}_{r}=0$.
If, by action of this perturbation, the system is ``pushed'' afar to a higher orbit
(there is a positive $\dot{v}^{(o)}_{r}$, and the orbital radius $R_{orb}$ increases),
the system reacts increasing $\omega_{(R)}$, thus making $\dot{v}^{(o)}_{r}$ negative.
Oscillations may happen, and, if the system is allowed to dissipate (in this case, by
radiation prevailing over absorption), the initial configuration will be recovered
after some time.
This behavior can be summarized by saying the excess of energy is temporarily
stored in that inner degree of freedom that we have named ``residual spin''
(residual oscillations of the AM of the proton, coming from the
``orbital'' movement of quarks inside, in our relativistic interpretation).
From there, it will be released as radiation in subsequent instants of time.

If the effect of the perturbation is to make the orbit loose energy, the process is
the opposite one: as the initial $\dot{v}^{(o)}_{r}$ is negative, to reach a lower
energy orbit (lower orbital radius), the system reacts making $\omega_{(R)}$ decrease,
and this makes $\dot{v}^{(o)}_{r}$ increase again and turn positive.
This is equivalent to say that the system compensates the sudden loss of energy drawing
it from the ``residual'' spin degree of freedom, to which it will later be returned,
after the systems absorbs it from the background.
Of course, a sufficient quantity of energy should be disposable in the immediately
following instants of time or this scheme could not work. As we have said, our bath
of stochastic radiation will be the provider of that energy in subsequent times.

\subsubsection{Consequences of the 2nd-FL}

The secondary feedback loop
$v^{(o)}_{t} \Rightarrow v^{(o)}_{r} \Rightarrow v^{(o)}_{t}$ will prevent stationary
orbits where $v^{(o)}_{r}(t)=0, \forall t$.
However,we will in principle disregard this effect for simplicity (simply ignoring the
corresponding term in the equations).
Later, we will see that it is precisely this 2nd-FL the reason why stationary/stable
orbits corresponding to a quantum labeling $n=0$ are not present in the real spectrum,
as so it happens for $n=1$ for orbitals with net angular momentum (there is no $s$ 
neither $p$-orbital for $n=0$, and no $p$ for $n=1$ either).


\subsubsection{Absence of rad/abs: a continuum of stationary orbits}

By a stationary trajectory we mean a closed, cyclic one, such that if the system is
in a particular point of the it, it will describe on and on that same one, coming
back to exactly the same initial condition after each cycle. We will also use the term
``stationary orbits''.
Later we will introduce the concept of ``stationary set'' as a set of points or initial
conditions given that, if the system is initially in one of them, it will evolve without
leaving that set. This generalizes the concept of stationary trajectory.
Stationary trajectories or orbits can be stable or unstable, depending how they respond
to small perturbations, once the stochastic terms are included in the equations of
the system. In absence of radiation, a perturbation causes a transition between two
of our continuum of stationary trajectories.

\subsubsection{Forcing rad/abs balance}

The following step in our logical development is the following: when imposing the
rad/abs balance, the result is that a particular one of the former
family of strictly circular trajectories is now singled out.
It behaves as an attractor in the configuration space of the system.
If we simulate the effect of the random background on a particle traveling in that
privileged orbit, we will get a probability distribution extended to the whole
space, in other words, an orbital.

\subsection{\label{Stationary_II}A set of stationary trajectories} 

\textit{As we have said already, in this section we will disregard the presence of that
2nd-FL, i.e., we simply ignore the term connecting $v^{(o)}_{r}$ and
$v^{(o)}_{t}$ in eqs. (\ref{vr_dynamic_c}) or (\ref{vr_dynamic_c_II}).}
We will show that there is a continuum family of stationary trajectories in respect to
the dynamical equations of the system.
The stationary points in this collection are not stable in the strict sense, though: a
(small) perturbation will cause a transition, mediated by certain oscillations, between
two of these trajectories.
Later, the inclusion of the rad/abs balance will single out one of these
trajectories, also providing it with the feature of stability.


With no need to get into mathematical analysis, we already now there is stability
in the case when $v_{r}^{(o)}=0$ (hence, $\dot{v}_{r}^{(o)}=0$).
If $v_{r}^{(o)}=0$, from (\ref{vt_dynamic}), $\dot{v}^{(o)}_{t}=0$ and
$\dot{\omega_{(R)}}=0$ too.
This means that stationary circular orbits are given, in this circular case,
by the equation that guarantees $\dot{v}^{(o)}_{r}=0$ (again ignoring the term
connecting $v^{(o)}_{r}$ and $v^{(o)}_{t}$ in eqs. (\ref{vr_dynamic_c}) or
(\ref{vr_dynamic_c_II}).
We have, from, (\ref{vr_dynamic}):
\begin{eqnarray}
q E_1 +  q B_1 \ v^{(o)}_{t} + \frac{3 \hat{\mu} Q_{2}}{R_{orb}^4} \ \omega_{(R)}
= m_e \frac{ (v^{(o)}_{t})^2 }{ R_{orb} }, \label{eqn1a}
\end{eqnarray}
where we have used $m_e$, the mass of the electron, at the right hand side
(it is the electron that revolves around the proton), where $\hat{\mu} =
\frac{\mu_0 \mu_e}{4\pi}$ and where $Q_{2}$ is, as usual, a second order charge
momentum (expressing the fact that the proton is seen as an extended distribution
of charges).
We can recognize, on the left hand side, the three leading contributions on our
multipole expansion of the problem:
(i) electrostatic attraction,
(ii) Lorentz force on a point-like charge,
and the third one,
(iii) a residual Lorentz effect on an ``extended'' particle.


Now, with $v^{(o)}_{t} = \omega_{(o)} \ R_{orb}$ for a circular orbit, we can
rewrite the former equation as
\begin{eqnarray}
q E_1 +  q B_1 \ R_{orb}\ \omega_{(o)}
+ \frac{3 \hat{\mu} Q_{2}}{R_{orb}^4} \ \omega_{(R)} = m_e \ R_{orb} \ \omega_{(o)}^2 ,
\nonumber\\ \label{eqn1b}
\end{eqnarray}
with $E_1,B_1$ as defined in (\ref{E_1}\\(\ref{B_1}).
The former equation defines a (still continuum) family of circular trajectories that
are stationary if we disregard (as we have done) the terms causing our ``secondary
oscillation'' (2nd-Osc).

\subsection{\label{Balance}
Rad/abs balance (Puthoff's condition) singles out one particular stationary trajectory}

In the former equation, there are three free variables: $\omega_{(o)}$, $\omega_{(R)}$
and $R_{orb}$.
We have an extra condition on $\omega_{(o)}$ that we have not used. This condition is
related to that balance of radiated/absorbed power.
For a stationary orbit, then, we have the condition that the emitted and absorbed
radiating power must agree, when summed over the whole set of points of the orbit.
A similar approach, as we already pointed out, was adopted previously by Puthoff
\cite{Puthoff87}:
``It is hypothesized that (at the level of Bohr theory) the ground-state orbit is a
ZPF determined state, determined by a balance between radiation emitted due to
acceleration of the electron and radiation absorbed from the zero-point background''.
Therefore,
\begin{eqnarray}
\langle P_{rad} \rangle_{(orbit)} \ = \ \langle P_{ab} \rangle_{(orbit)},  \label{eqn2}
\end{eqnarray}
from where, using the expressions in (\ref{P_rad})--(\ref{P_ab}), finally, Puthoff
arrived to the condition:
\begin{eqnarray}
m_e \omega_{(o)} R_{orb}^2  = \hbar, \label{Puthoff_cond}
\end{eqnarray}
with $m_e$ the mass of the electron, $\omega_{(o)}$ the orbital angular velocity and
obviously $R_{orb}$ the orbital radius.
This equation obviously quantizes the (ground state) value of (each of the projections
of) AM.

Puthoff's condition can be substituted in equation (\ref{eqn1a}) or (\ref{eqn1c}), 
so a to reduce to two the number of independent variables:
$\omega_{(R)}$ and $R_{orb}$ (we already got rid of $\omega_{(o)}$).

\subsection{\label{stat_omega}
Particle structure and $\omega_{(R)}$}

We are concerned with the stationarity of trajectories with a certain stationary value
$\omega_{(R)} = \omega_{(R)}|^{st}$. Once again, it is time to recall our definition of
$\omega_{R}$ as $\omega_{(R)} = \omega - \omega_{(o)}$, as well as note \cite{defs_omega}.
Now, for this stationary behavior, it is natural to assume $\omega|^{st}=0$, because
the mass of the proton (the charge distribution) is almost infinite in relation to the
mass of the electron.
With this choice,
\begin{equation}
\omega_{(R)}|^{st} = - \omega_{(o)}|^{st}.
\end{equation}
Some comments regarding the model we choose for the structure of the particle are due here.
At least for the simplest model of a uniform charge distribution that rotates, $\omega=0$
would imply that there is no net magnetic moment.
Nevertheless, that was only a model simple enough to serve our purposes.
From now on, we will adhere to the quark model:  quarks are the carriers of the overwhelming
proportion of the magnetic moment of the particle, the contribution of our ``residual'' spin
being sufficiently marginal.
In other words, the quarks that compose the proton add no net ``inner orbital movement''
to the overall spin (inner AM) of the particle.
On our classical model, this must only happen ``on average'': that residual classical
AM of the proton may oscillate back and forth, but its mean value will
stay zero.

\subsection{\label{Strict_proof}
Stability criteria for the simplest orbit}

\subsubsection{
Stability of dynamical systems}

We have already shown that there is a stationary trajectory for our dynamical equations,
but we still do not know if this accounts for an ``stable'' one.
Stability implies stationarity, but the opposite is not true. A stable trajectory must
show, aside from being stationary, some ``resistance'' to be driven apart at least against
small perturbations.
Of course this last concept of stability only make sense when the dynamical equations
of the system (up to now, strictly deterministic) are supplemented with some stochastic
terms, representing the net difference between radiated and absorbed power.
These stochastic terms introduce both the possibility of energy dissipation or absorption.
They were already included in Sec.~\ref{Complete_eq}.
Of course, they only intend to represent roughly this processes of rad/abs from
the background.
Much more detailed calculations could be done (for example, we can express instantaneous
radiated power as a function of acceleration), but they are not necessary for our
main purpose here: we want to show that stability is present in this simplified (and
more general) picture.

To study stability, it is convenient here to define a reduced configuration space
(we are not interested in variables such as position or angle or rotation, but only
on the minimum set of then that characterize an orbit).
On that (reduced) configuration space, a vector $\mathbf{x}$ gives the state of the
dynamical system for a given time $t$.
We thus define the following vector of dynamical variables, expressing the (instantaneous)
state of the system:
\begin{eqnarray}
\mathbf{x} =
\left[ \begin{array}{c} R_{orb}\ \\ v_{r}^{(o)}\\ v_{t}^{(o)}\\ \omega_{(R)} \end{array} \right],
\label{possible_conf_vector}
\end{eqnarray}
where, depending on our purpose, we could omit or add dynamical variables.
Now, given the equations of movement remain invariant with time, we can write the`
`autonomous'' system of equations:
\begin{equation}
\dot{\mathbf{x}} =
f(\mathbf{x}), \ \ \ \ \ \mathbf{x}\ \in \ \mathbb{R}^{p}, \ \ \ f:\ \mathbb{R}^{p}\rightarrow R^{p}.
\end{equation}
Given that the function $f$ may be (indeed it is) non linear, we can linearize it
around a point $\mathbf{x}_{0}$,
\begin{equation}
\dot{\mathbf{x}} = M_{l} \cdot (\mathbf{x}-\mathbf{x}_{0}),
\end{equation}
where $M_{l}$ is a matrix of constant coefficients resulting from the liberalization.

\emph{
A necessary and sufficient condition for a stable ``point'', i.e., a circular orbit
in our formulation, is simply that the $p$ eigenvalues of $M_{l}$ must be negative.}

\subsubsection{An ideal circular orbit}

Around a certain point $R_{orb}=R_{orb}^{st}$ we will define the following state
vector:
\begin{equation}
\mathbf{x}_{st} =
\left[ \begin{array}{c} R_{orb}^{st}\\ 0\\ v_{t}^{(o)}|^{st}\\ \omega_{(R)}|^{st} \end{array} \right],
\label{pto_lineariz_p}
\end{equation}
with $v_{t}^{(o)}|^{st}>0$ and, for the case already treated in Sec.~\ref{stat_omega},
$\omega_{(R)}|^{st} = - \omega_{(o)}|^{st}$.

We can now linearize the dynamical equations of the system given in
(\ref{vt_dynamic_c})--(\ref{vr_dynamic_c}) and (\ref{w_dynamic_c}).
Following for example (\ref{linear_chi_r}), close to $R_{orb}=R_{orb}^{st}$ we have
$\langle\chi_{r}(R_{orb})\rangle \approx - k_{r} (R_{orb}-R_{orb}^{st}), \quad k>0$,
where the mean value stands for an average on an interval, but small enough to comprise
just a reasonable small segment of the (cyclic) orbit.
Therefore, our equations, following now, are valid ``on an average'' along a short
interval:
\begin{eqnarray}
\left[ \begin{array}{c}
\dot{R}_{orb}\\ m \ \dot{v}^{(o)}_{r}\\ m \ \dot{v}^{(o)}_{t}\\ M_{2} \ \dot{\omega_{(R)}}
\end{array} \right]
=
\left[ \begin{array}{cccc}
0  	& 1    	& 0      & 0 \\
-k_{r} 	& 0    	& - A    & - B \\
0 	& C    	& -k_{t} & 0 \\
0  	& D 	& 0      & -k_{\omega}		
\end{array} \right]
\cdot
\left[ \begin{array}{c}
R_{orb}\ \\ v^{(o)}_{r}\\ v^{(o)}_{t}\\ \omega_{(R)}
\end{array} \right]
+
\left[ \begin{array}{c}
0\\ E\\ 0 \\ 0
\end{array} \right],
\nonumber\\
\label{matricial_system_II}
\end{eqnarray}
defining the coefficients
\begin{eqnarray}
A &=&
\left[ + q B_1 + m_e\ \frac{ 2 (v^{(o)}_{t}) }{ R_{orb} } \right]_{ \mathbf{x} = \mathbf{x}_{st} },
\label{A_coef} \\ \nonumber\\
B &=&
\left[ + \frac{3 \hat{\mu} Q_{2}}{{R_{orb}^4}} \right]_{ \mathbf{x} = \mathbf{x}_{st} }, 		
\label{B_coef} \\ \nonumber\\
C &=&
\left[ + q B_1 \right]_{ \mathbf{x} = \mathbf{x}_{st} }, 	
\label{C_coef} \\ \nonumber\\
D &=&
\left[ + \frac{ 3 \hat{\mu} \ Q_{2}^{(*)} }{R_{orb}^4} \right]_{ \mathbf{x} = \mathbf{x}_{st} },
\label{D_coef} \\ \nonumber\\
E &=&
\left[ + k R_{orb} - q E_1 - m_e\ \frac{ (v^{(o)}_{t})^2 }{ R_{orb} } \right]_{ \mathbf{x} = \mathbf{x}_{st} }, 
\label{E_coef}
\end{eqnarray}
where, as indicated, each coefficient is evaluated in the point of stability (the point around
which we are linearizing), and therefore, $E_1=E_1(R_{orb}^{st})$, etc.
We must also note that all quantities are positive defined, i.e., $k_{r}, k_{t}, k_{\omega}, A, B, C, D > 0$,
except for $E$.
The eigenvalue equation for the coefficients matrix is, therefore:
\begin{eqnarray}
\left| \begin{array}{cccc}
\lambda 	& -1        & 0    		      & 0 \\
k_{r}       & \lambda 	& - A    	      & - B \\
0       	& - C	    & \lambda + k_{t} & 0 \\
0        	& - D 		& 0		          & \lambda + k_{\omega}					
\end{array} \right|
\ =\ 0.
\label{det_eqn}
\end{eqnarray}
With $k_{r}, k_{t}, k_{\omega}, A, B, C, D > 0$, all roots have a negative real part.
This is a necessary and sufficient condition for the point $\mathbf{x}_{st}$ to be a (local)
attractor in the configuration space of the (linearized) system.

\subsection{
Absence from the quantum spectrum: 2nd-Osc}

Again we stress that this lowest circular orbit is unrealistic, due to the fact that we have ignored
the secondary feedback loop (2nd-FL) that would introduce necessarily oscillations in $v^{(o)}_{r}$.
Later we establish a correspondence between a set of stable orbits and the actual quantum orbitals,
defining a labeling using certain integers $n_s,n_p=0,1,\ldots$ that would correspond to the principal
quantum number $n$. The case that we have just analyzed here would correspond to $n_p=0$.


\section{\label{s_stability}
Stability with vanishing (average) AM: towards the s-orbitals}


\subsection{Overview}


So far we have dealt with circular orbits. Clearly, these bear net (average) AM,
in contrast to the quantum mechanical lowest energy orbitals or s-orbitals.
There can be found, indeed, stationary (and stable) orbits (closed trajectories) with vanishing
AM, but it seemed much clearer to us to invert the presentation (addressing first
the more intuitive circular orbits) in the way we have done.
As an aside, the more or less clear relation between our classical orbits (or set of orbits)
and the quantum mechanical orbitals will be treated in detail in Sec.~\ref{Orbitals}.

Basically, we could formulate a qualitative argument in the way that we did before, for the
circular orbits. In this case the situation, however, seems a bit more complicated: this time
$\omega$ \cite{defs_omega} does not just oscillate slightly in response to external perturbations,
but does itself describe an stationary oscillatory curve with a certain amplitude (again,
nevertheless, the value $\omega=0$).
The amplitude of this oscillation should be determined working on the dynamical equations
of the system. For now we will be content to say that, assuming rather natural modifications
of the ``base'' trajectory, we can presume that this amplitude may result in a reasonable value.
For example, we can assume that the tangential velocity also oscillates leading to a trajectory
that resembles (in the simplest configuration that we can consider) an ``eight'', a trajectory
whose average AM still vanishes.
It is also important to say that to understand this kind of trajectory we need to take into
account the Lorentz interaction on a point particle (our stabilizing mechanism can only produce
forces in the radial direction).

A key question is that the 2nd-Osc is again ignored, its role being crucial in
our further justification of the full spectrum.
To conclude, again we leave any energy estimation for future sections.

\subsection{A ``model'' trajectory}

Harmonic oscillators are not (neither classically) eigenfunctions of an $\tfrac{1}{r}$-type
potential.
Nevertheless, considering the trajectories we propose are not strictly radial (straight lines)
at all, but they combine radial and tangential components, we will adopt the harmonic formulation,
with the intention of showing that
(i) they are a ``feasible'' approximation to the real trajectories,
(ii) they are stationary,
(iii) a balance of rad/abs can be attained, hence, they can be stable.

That last property (iii) will single out, amongst a continuum of possible stationary trajectories,
a particular one, the stable one or ``attractor''.
We will now name $\omega_{0}$ the basic frequency of oscillation.
This is no longer an angular velocity, as it was for p-type trajectories, and, therefore, it has
nothing to do with $\omega_{(o)}$.
Nevertheless, $\omega_{(o)}$ is still perfectly defined as one of the dynamical variables of the
system, and indeed it will still be useful to show that stability is possible.
In similarity with our previous treatment of stability for a circular trajectory, we will again
ignore the secondary feedback loop (2nd-FL).
We propose a set of stationary trajectories, parameterized by an amplitude $v^{(s)}_{r,0}$:
\begin{eqnarray}
v^{(o)}_{r} = v^{(s)}_{r,0} \cos( \omega_{0} t ), 	\label{v_r_s_gs}  
\end{eqnarray}
and, in terms of the radial coordinate $r=|\mathbf{r}_{1}|$,
\begin{equation}
r = R_{max} \cos( \omega_{0} t ), \\
\end{equation}
with $R_{max} = \frac{ v^{(s)}_{r} }{ \omega_{0} }$.



\subsection{Stationarity}

A key point is to see why we cannot have $\dot{v}^{(o)}_{t} = 0$. This is due to the term involving
$\omega$ in the equation for $\dot{v}^{(o)}_{t}$, at equation (\ref{vt_dynamic}) or (\ref{vt_dynamic_c}).
We will not extend much on this but to say that, in this kind of quasi-radial trajectory, the nucleus is
angularly accelerated back and forth due to the coupling between $\dot{v}^{(o)}_{r}$ and $\omega$ in the
corresponding equations.

Besides, to understand such a trajectory, that certainly goes beyond the purely radial one that we should
expect from a simple model, one has to think in the Lorentz interaction.
This Lorentz forces go normal to the instantaneous velocity, and can therefore explain (although we still
have not calculated with what amplitude) an oscillation such as the one we are suggesting for
$\dot{v}^{(o)}_{t}$.
Also, as said before, harmonic behavior is of course just a very rough approximation, but for the moment
it will be enough for our purposes.

\subsection{Stability: attraction dynamics} 

Let us leave aside a realistic estimation of the amplitude $v^{(s)}_{r,0}$. Instead of that,
let us also suppose that, for a certain values of that amplitude, a power balance can be attained.
Moreover, what we will do here is prove that the existence of that balance is itself enough for
stability, given the equations that we have for the system. We will do this by a very general
mathematical analysis.
The main point is that, although we have $\dot{v}^{(o)}_{t} \neq 0$, we do have
$\langle v^{(o)}_{t} \rangle_{(orbit)} = 0$.
This oscillation causes a dissipation/absorption game but is not any more dependent on a
particular value of $v^{(o)}_{t}$. Therefore, we make $k_{t}=0$.

Besides, no ``attraction'' dynamics can be supposed for $R_{orb}$, $v^{(o)}_{r}$
or $\omega_{(R)}$.
Indeed, what we seek for is an oscillatory one. Once the system is linearized,
this kind of behavior corresponds to roots on the imaginary axis.
Nevertheless, if those oscillations of $R_{orb}$, $v^{(o)}_{r}$ and $\omega_{(R)}$
are not very big (or reasonably small), we can linearize around the following point
this time:
\begin{equation}
\mathbf{x}_{st}^{\prime} =
\left[ \begin{array}{c} R_{orb} = 0\\ v_{r}^{(o)} = 0\\ v_{t}^{(o)} = 0\\ \omega_{(R)} = 0 \end{array} \right],
\label{pto_lineariz_s}
\end{equation}

First, stationarity is established. For that, we identify imaginary poles in
absence of rad/abs.
Therefore we do $k_{r}=0$, $k_{t}=0$ and $k_{\omega}=0$.
Our linearized system is similar to that of (\ref{matricial_system_II}), and
the matrix of linearized coefficients $M_{l}$ such that
$\dot{\mathbf{x}} = M_{l} \cdot ( \mathbf{x} - \mathbf{x}_{st}^{\prime} )$ is
written as
\begin{eqnarray}
M_{l} = \left[ \begin{array}{cccc}
0  	& 1    		& 0      	& 0 \\
0 	& 0    		& - A^{\prime}  & - B^{\prime} \\
0 	& C^{\prime}    & 0      	& 0 \\
0  	& D^{\prime} 	& 0      	& 0		
\end{array} \right],
\end{eqnarray}
and $A^{\prime}, B^{\prime}, C^{\prime}, D^{\prime}$ obey to the same expressions as
(\ref{A_coef})-- (\ref{D_coef}) but with $R_{orb} = 0$ this time.
Obviously, this assumes that the $1/R_{orb}$ dependence disappears for $R_{orb}$
sufficiently small, something that does occur when we include in the overall model
the inner structure of the nucleus, as we are doing here.

Again with $A^{\prime}, B^{\prime}, C^{\prime}, D^{\prime} > 0$, all roots $\lambda_i$
of the eigenvalue problem $|M_{l} - \lambda I| = 0$ are purely imaginary, a proof of
what is, once more, trivial and therefore not needed here.
That corresponds to the oscillatory movement we were seeking for.
In the same way that we had for our $p$-orbits, in absence of dissipation/absorption,
a perturbation sends the system from one stationary orbit to another, both belonging
to a continuum of stationary trajectories.
Once we have established the fact that there is such a continuum of stationary
trajectories, the element of dissipation and absorption from the background introduces
the attraction dynamics.


\section{\label{Ex_states}
Towards excited states}


\textit{We have to open space for an infinite, discrete spectrum, and this is done by the
introduction of the secondary oscillation (2nd-Osc), a natural consequence of the equations
of the system, in our analysis.}

Showing that that our framework here leaves reasonable space to account for the existence
of excited states is one of our main goals here.
In principle, the solution reduces to find new trajectories, first stationary, then necessarily
stable, too, whose (mean) energy is higher than the already seen.
So far, we have only proven the existence of one stationary circular orbit (a candidate to
generate, in 3D, a $p$-orbital) and another one with vanishing AM (a candidate
for the ground state of the system, an $s$-orbital).
This has been done disregarding the oscillation in $v^{(o)}_{r}$ coming from the 2nd-FL
that we have so far ignored.

Precisely that 2nd-Osc is the additional element that will allow for an extra freedom in the
power balance equation, yielding now an infinite (discrete) set of closed trajectories that
can attain that balance.
Now, we intend to prove that excited states, with different energies, can be built following
the same basic ideas:
first we would find a stationary trajectory consistent with the equations of the system, and
then we would have to prove that they are indeed stable against stochastic perturbations.
All we do here is propose some particular form for those stationary orbits, and make some
preliminary calculations on their energy difference with respect to the former ones.
Again we follow, because we think is more intuitive that way, the same structure: first
circular orbits, then the $s$-ones.

The trajectories that we present in the following are not stable, nor are they stationary,
if we stick to the simplified problem of an scalar central potential.
For the trajectories we propose here to be possible, we must add both the action of the
Lorentz force on a point particle, as well as our higher order terms arising from the multipole
orders that we wanted to include in the calculation (our ``stabilizing mechanism'').
We provide here two sets of stable trajectories, parameterized by two integer numbers
$n_s,n_p$, ranging in principle (in principle) from $0$ to infinity. The notation is
chosen so as to suggest that each group actually corresponds (all the necessary generalizations
done: extension to 3D, stochasticity) to the quantum spectrum of $s$ and $p$-orbitals.
Indeed, as we will see, those integers are directly related to a certain ``matching
condition'' between the principal orbital frequency of the trajectory and the frequency
of 2nd-Osc.

Later, in Sec.~\ref{quantum} we will establish the correspondence with the quantum number
$n$ or ``energy level''. Indeed, the correspondence is simply $n_s,n_p = n$, though it
also true that not all proposed values of $n_s$ and $n_p$ can give rise to real orbitals.
For example, we must first justify the exclusion of the $n=0$ solutions (in principle possible
in our scheme), as well as the one with $n_p=1$, as we know that for non-vanishing angular
momentum (in quantum terms $l>0$) we cannot have an orbital with $n=1$.
This questions will not be addressed yet in this section, though we will do in former ones.

In any case, the path to prove stability of each of our proposed trajectories is this:
First, prove stationarity with respect to the dynamical equations of the system, in absence of
rad/ABS (as we have seen, for our model, this stationarity is not satisfied by
just one trajectory but by a continuum set of them).
As a second step, we must evaluate the balance between looses and absorption from the background:
this additional condition will single out one trajectory from each former set of trajectories.
As we said, the correspondence with a real quantum orbital is left for Sec.~\ref{quantum}.


\subsection{\label{A parametrization}Parametrization of stationary orbits}

From now on we will do, for convenience, the following simplification, describing both the sets
of possible $s$ and $p$-orbits with the same parameters:

(i)
From now on, $R = |\mathbf{r}_{0}|$, representing the position vector for the electron respect
to the center of mass of the nucleus.
The definition of our two references frames, $RF_{0}$ and $RF_{1}$, after our last relocation of
mass, can be recalled in Sec.~\ref{Relocation}: both $RF_{0}$ and $RF_{1}$ anchored to
the center of mass of the nucleus, but only $RF_{0}$ retains inertiality, with $\mathbf{x}_{1}$
always in direction that joins that center of mass with the position of the electron
($\mathbf{R}_{orb} = R_{orb} \mathbf{x}_{1}$).

(ii)
$R_0^{(p)}$ and $\omega_0^{(p)}$ will be the orbital radius and orbital frequency for the
$p$-set of trajectories, therefore the amplitude and frequency for the primary $p$-oscillation.
Two other parameters, $R_1^{(p)}$ and $\omega_1^{(p)}$ will describe the 2nd-Osc.
Therefore, for $p$-orbits, omitting superscripts for clarity,
\begin{eqnarray}
\mathbf{r}_{0} &\approx&
\left[ \ R_0 \cos( \omega_0 t ) + R_1 \cos( \omega_1 t ) \ \right] \mathbf{x}_{1}, \\
|\mathbf{r}_{0}| &\approx&
R_0 + R_1 \cos( \omega_1 t ),
\end{eqnarray}
which corresponds to our expected radial oscillation.

(iii)
$R_0^{(s)}$ and $\omega_0^{(s)}$ will be the amplitude and frequency of the
orbital movement (primary oscillation or 1st-Osc) for the pendulum-like or
$s$-orbits.
The pair $R_1^{(s)}$ and $\omega_1^{(s)}$ will describe the 2nd-Osc; for
$s$-orbits, omitting superscripts for clarity,
\begin{eqnarray}
\mathbf{r}_{0} &\approx&
R_0 \cos( \omega_0 t ) \ \mathbf{x}_{1} + R_1 \cos( \omega_1 t ) \ \mathbf{y}_{1}, \\
|\mathbf{r}_{0}| &\approx&
\left[ \ R_0^2  + R_1^2 \cos^2( \omega_1 t ) \ \right]^{\frac{1}{2}},
\end{eqnarray}
which corresponds to an expected secondary oscillation in the tangential
direction, given always by $\mathbf{y}_{1}$.

(iv)
We complete the parametrization by renaming
$\omega_{(R)}|_{max} \in \{ \omega_{R,0}^{(s)},\ \omega_{R,0}^{(p)}\}$
depending on the type of trajectory.

\textit{
If no ambiguity is present, we will always drop superscripts (s),(p) for
clarity. We will only refer, henceforth, to $R_0$, $\omega_0$, $R_1$,
$\omega_1$ and $\omega_{R,0}$.}

\subsection{
The harmonic oscillator approximation: decoupling of
1st and 2nd-Osc}

\textit{Talk about Coulomb well, approximate linearization, etc.}
Harmonic oscillators are classical eigenstates of the system, once this is linearized.
Therefore, it is only an approximation. Indeed, that decoupling only takes place
under that linearization, due to the imaginary poles of the linearized matrix.

\subsection{\label{Model_t}A new continuum of stationary orbits}

Therefore, in principle a stationary trajectory or ``orbit'' is given by a quartet of parameters
$\{ \ R_0,\omega_0, R_1, \omega_1 \ \}$, and a fifth $\omega_{R,0}$ that is (as we will see)
determined by the former. Though we are, as announced, omitting superscripts (s) and (p), it
should be understood that these parameters do have different meaning depending if it is an $s$
or $p$-orbit, but for our purpose here we are not interested in making any distinction.

Now, we will impose stationarity on the equations of the system. In the first place, 
if we ignore the secondary oscillation (2nd-Osc), what we obtain is a relation on orbital
pair $(R_0,\omega_0)$: we have $\omega_0$ from $R_0$, for instance.
On the second place, if we do regard the secondary feedback loop (2nd-FL), we have is an
additional relation that determines $\omega_1$ from $R_0$. This accounts for a continuum
of orbits parameterized by $R_0$.
However, $R_1$ is still a free parameter. We bear in mind that.

\subsection{
Some necessary context: rad/abs in absence of 2nd-Osc}

We will not repeat Puthoff's calculations here, but just remark some questions of interest.
For circular orbits with a higher $R_0$, the dissipation decreases, as we can see from this
calculation...
Indeed, applying (\ref{p_stationarity_cond}) to the radiated power,
\begin{eqnarray}
R_0^3 &=& \frac{ q^2 }{ 4 \pi \epsilon_0 } \cdot \frac{1}{ m_e \omega_0^2 },  \\
\omega_0^2 &=& \frac{ q^2 }{ 4 \pi \epsilon_0 } \cdot \frac{1}{ m_e R_0^3},  \\
\langle P_{rad} \rangle &=&  \frac{ q^2 R_0^2 \omega_0^4 }{ 6 \pi \epsilon_0 c^3 } \propto \frac{1}{ R_0^4 },
\end{eqnarray}
which means that for feasible excited states corresponding to circular orbits, the
dissipation must be compensated by a significant contribution from the secondary
oscillation, that in turn will not cause much absorption.
A similar conclusion can be obtained for pendulum-type orbits.

\subsection{\label{Second_osc_int}2nd-Osc in the stationary orbits}

We know, because of the coupling between equations (\ref{vr_dynamic_c}) and (\ref{w_dynamic_c})
(the coupling is also present when we do not include the radiative corrections), any circular
stationary trajectory is not compatible with the condition $v^{(o)}_{r}=0$.
Nevertheless, it is interesting to do some preliminary study when we do impose $v^{(o)}_{r}=0$,
which itself will help understand how excited (higher energy) states can be identified with the
corresponding stationary trajectories.
Besides, for a pendulum-like trajectory, the coupling between (\ref{vr_dynamic_c}) and
(\ref{vt_dynamic_c}) forbids a stationary trajectory with $v^{(o)}_{t}=0$.
Both conditions will add a certain natural frequency $\omega_1^{(p)}$ and $\omega_1^{(s)}$
to the orbital movement.
As we have said, it is this 2nd-Osc that will make possible to deviate from Puthoff's unique
balance condition, allowing for an infinite set of stable orbits, discretized by a certain
matching condition.

Again, later, we will establish a correspondence between a labeling number $n$ (similar
to the principal quantum number) and each excited trajectory, and it is no less important
to remark that the presence of this 2nd-Osc will explain why no $p$ or $s$ orbits exists
for $n=0$.
Moreover, basing our argument in geometric considerations, we will exclude the
$p$-orbit for $n=1$, too, therefore in coherence with the actual atomic spectrum.


Puthoff's condition as given in \cite{Puthoff87} only allows for one unique stable circular
trajectory (and one unique pendulum one).
The only possibility to enhance the set of stable trajectories demands the introduction
of new degrees of freedom in the orbit:
that way, some degrees of freedom can radiate more or less than they absorb, counteracting
the excessive or defective absorption of the others.
This said, we have to realize that it is the structure of the system itself (the dynamical
equations) that provides us with those extra degrees of freedom. As we have seen, there is
a coupling between equations (\ref{vr_dynamic_c}) and (\ref{w_dynamic_c}), so trajectories
of the kind $v^{(o)}_{r} = 0$ can never be stationary at all if $\omega_{(R)} \neq 0$.

Besides, the coupling between (\ref{vr_dynamic_c}) and (\ref{vt_dynamic_c}), that stays
irrelevant in stationary circular trajectories because $v^{(o)}_{t} = 0$, acquires key
importance in trajectories of the pendulum type, making impossible that absence of radial
velocity any more.
To the orbital movement (either if it is circular or pendulum-like) we have to add an
oscillation on $v^{(o)}_{r}$.
An obvious candidate is a harmonic oscillation. Why? Because the (linearized) system
has imaginary poles in its frequency representation (extend on this), and harmonic
oscillation therefore is indeed an ``eigenstate'' (we mean a classical one, here) of
the system.
In the following, let $\omega_{1}$ be either $\omega_{osc}^{(p)}$ or $\omega_{osc}^{(s)}$,
i.e., the 2nd-Osc frequency respectively for the $s$ and $p$-types of stationary orbits.

\subsection{\label{Matching}Necessity of a matching condition}

In principle, stationarity requires a ``matching'' condition: $\omega_{0} = \omega_{1}/N$
for some integer $N=1,2\ldots$. Later, we will try to establish a correspondence between
the integer $N$ and the well known principal quantum number.

\subsection{\label{feasible_p}Feasible $p$-orbits}

We provide now an expression for the whole discrete $p$-spectra, introducing now, as
already advanced, a parametrization by a certain integer $n_p$, that later we will
make correspond to the well known principal quantum number. Initially,
\begin{eqnarray}
v^{(o)}_{t} &\approx& \omega_{(o)} R_{orb}, \\
v^{(o)}_{r} &\approx& v^{(p)}_{r,0} cos( \omega_1 t ). 	\label{v_t_p_ex}
\end{eqnarray}

\subsubsection{\label{p_matching}$p$-matching condition} 

Now, our matching condition will be 
\begin{eqnarray}
\omega_1 = n_p \omega_{(o)}, \quad n_p =2,3,\ldots, \label{p_matching_eq}
\end{eqnarray}
where the absence of $n_p=1$ corresponds to the exclusion of the $p$-orbitals for the
first atomic level (later we will give some more justification on this),
yielding
\begin{eqnarray}
v^{(o)}_{t} &\approx& \omega_{(o)} R_{orb}, \\
v^{(o)}_{r} &\approx& v^{(p)}_{r,0} \cos( n_p \omega_{(o)} t ),  \label{v_t_p_ex_II}
\end{eqnarray}
where, clearly, the higher the $n_p$, clearly the higher the energy.
Besides, the amplitude of secondary oscillation will possibly depend on $n_p$, determined
by the dynamical equations of the system:
\begin{eqnarray}
v^{(p)}_{r,0} = v^{(p)}_{r,0}(n_p), \quad n_p = 2,3,\ldots
\end{eqnarray}

\begin{figure}[ht!] \includegraphics[width=0.60 \columnwidth,clip]{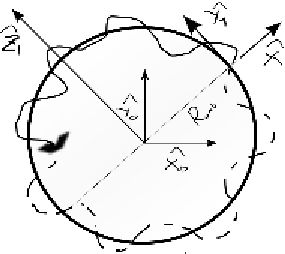} \caption{
In our proposed $p$-orbits, the orbital distance undergoes a ``secondary'' oscillation
(2nd-Osc) in the way $R = R_0 + R_1 \cos(\omega_1 t)$ (which is simultaneous with an
oscillation in the AM (or angular velocity) $\omega$ of the nucleus, vanishing on average).
\textbf{It may look a counterintuitive trajectory for a central potential, but mind the
intervention of the ``inner'' degree of freedom!}
The frequency matching condition for a $p$-orbital is $\omega_1 = n_p \omega_0$.
\textit{This basic $p$-orbit gives rise to a rotationally invariant electronic density
around the axis $\mathbf{z}_{1}$.
When we analyze the situation in 3D, we have to consider a modulation of our stabilizing
interaction from the misalignment of that axis $\mathbf{z}_{1}$ and the magnetic moment
of the electron $\mu_e$ (this misalignment is nothing but the classical counterpart of
the quantum LS term).}
} \label{Fig_p_orbital_I} \end{figure}

\subsubsection{Stationarity and stability} 

To proof stationarity, it is enough to show that the system, in absence of rad/abs
terms that will further on privilege a discrete set of orbits and explain
their attractive behavior, has a frequency description where all poles are purely
imaginary.
To analyze it, we first had to linearize around a particular point of equilibrium, but
once the secondary oscillation dynamics is included, neither of the dynamical variables
used in our description of the system (see...) adopt a constant value around the stationary
orbit.
Anyway, we can keep on considering that the oscillations in $v^{(o)}_{t}$ are small, and
the description already used in Sec.\ref{p_stability} still remains adequate.
The attraction behavior is again provided once we include the rad/abs terms.
We would need to prove that this new balance is attainable.
We do not consider this a major difficulty, however, and we leave any calculations for the
Appendix.

\subsection{\label{feasible_s}Feasible $s$-orbits}

In the way that we have done for the $p$-spectrum, we introduce a parameter $n_s$,
which will later allow us to establish a correspondence with the well known principal
quantum number.
Initially,
\begin{eqnarray}
R &\approx& R_0 \cos( \omega_{0} t ), \\
v^{(o)}_{t} &\approx& v^{(s)}_{t,0} \cos( \omega_{1} t ). \label{v_t_s_ex}
\end{eqnarray}

\subsubsection{\label{s_matching}$s$-matching condition} 

Now we have
\begin{eqnarray}
\omega_1 = 2 n_s \omega_{0}, \quad n_s = 1,2,\ldots, \label{s_matching_eq}
\end{eqnarray}
yielding
\begin{eqnarray}
R &\approx& R_0 \cos( \omega_{0} t ), \\
v^{(o)}_{t} &\approx& v^{(s)}_{t,0} \cos( 2 n_s \omega_{0}), \label{v_t_s_ex_II}
\end{eqnarray}
where, clearly, the higher the $n_s$ the higher the energy, and where the amplitude of the
2nd-Osc, this time affecting what we defined as the tangential velocity component, will
possibly depend on $n_s$, determined by the dynamical equations of the system:
\begin{eqnarray}
v^{(s)}_{t,0} = v^{(s)}_{t,0}(n_s), \quad n_s = 1,2,\ldots,
\end{eqnarray}

\textit{Special attention should be taken to the factor $2$:
we need a trajectory that ``crosses '' with itself, so the average AM over
a whole cycle yields exactly zero.}
Again, the reason for the absence of solution $n_s=0$ is that it disregards
the 2nd-Osc so typical of our system.
Following our program, as we did for the p-case, we now have to check if they
are indeed stationary, and then if they can be stable (strict stability happens
whenever a power balance can be attained, singling out a particular trajectory
from the set of all stationary ones).

\begin{figure}[ht!] \includegraphics[width=0.60 \columnwidth,clip]{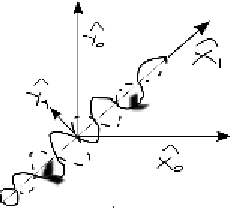} \caption{
In the proposed $s$-orbits, the secondary oscillation (2nd-Osc) takes place in $\mathbf{y}_{1}$,
as well in small variations of the angular momentum (AM) of the nucleus (that vanish, in average).
The average overall AM of this orbit is zero as expected for the quantum mechanical
counterpart.
The frequency matching condition for an $s$-orbital is $\omega_1 = 2 n_s \omega_0$,
with $n_s = 4$ in this example.
\textbf{It may look a counterintuitive trajectory for a central potential, but mind
the intervention of the ``inner'' degree of freedom!}
Of course this trajectory is not stationary in the purely electrostatic picture, but
it is feasible when we include both the Lorentz and higher order terms.
\textit{On the other hand,
this basic $s$-orbit gives rise to a rotationally invariant electronic density in
3D, when the axis $\mathbf{x}_{1}$ is allowed to precess freely in response to
stochastic perturbations.}
} \label{Fig_s_orbital_I} \end{figure}

\subsubsection{Stationarity and stability} 

Our reasoning is the same as for the $p$-spectrum: if we consider that the secondary
oscillation (2nd-Osc) is small enough, then the linearized description of the system
used in Sec.~\ref{s_stability} remains approximately correct, and the purely imaginary
character of the poles in the frequency representation, hence the purely oscillatory
behavior, guaranteed.
Once the existence of a continuum of stationary orbits is proved for the equations
of the system in absence of dissipation/absorption terms, their inclusion privileging
a discrete set of orbits (and explaining the attraction behavior).

Our next task will be to do a first approach to the energy spectrum.
We have already given the basic equations for the stable orbits (for the linearized
system, in 2D and etc.), parameterized them by a natural number $n_s$ or $n_p$, and
now we want to estimate the quantity
$\Delta E (n_{\alpha}, n_{\alpha} + 1) = E(n_{\alpha} + 1) - E(n_{\alpha})$.
I.e., we want to estimate the energy gap between adjacent states.


\section{\label{Spectrum_I}
A first approach to the E-spectrum}


In principle, we would have to calculate the following energies, for $p$ and $s$-orbits,
respectively,
\begin{eqnarray}
E^{(p)}(n_p) = E_1^{(p)}(n_p) + E_2^{(p)}(n_p), \\
E^{(s)}(n_s) = E_1^{(s)}(n_s) + E_2^{(s)}(n_s),
\end{eqnarray}
where the subscripts $1,2$  make reference to the principal (orbital) or secondary
oscillatory movements that characterize each orbit.
We will hereon omit superscripts for simplicity (the dependence on $n_s$ or $n_p$ eliminates
any ambiguity).

\subsection{\label{Spectrum_Ia}
The $p$-spectrum (in absence of 2nd-Osc)}

For $p$-orbits, disregarding the 2nd-Osc, we know from (\ref{En_p_1}),
\begin{eqnarray}
E_1(n_p) = - E_{kin}(n_p) =
- \frac{1}{2} m_e (v^{(o)}_{r})^2 = - \frac{1}{2} m_e \omega_0^2 R_0^2, \nonumber\\
\end{eqnarray}
and now, applying the frequency matching condition, $\omega_1 = n_p \omega_0$,
\begin{eqnarray}
E_1(n_p) = - \frac{1}{2} m_e \frac{\omega_1^2}{n_p^2} R_0^2,
\end{eqnarray}
but, from (\ref{omega_1_p}),
\begin{eqnarray}
\omega_1 \propto \frac{1}{R_0^4},
\end{eqnarray}
so, therefore
\begin{eqnarray}
E_1(n_p) \propto - \frac{1}{n_p^2} \frac{1}{R_0^2}.
\end{eqnarray}
Considering that
\begin{eqnarray}
\frac{1}{R_0^2} \approx \frac{1}{R_0^2} - 2 \frac{1}{R_0^3} ( R - R_0 ) + \ldots,
\end{eqnarray}
with $R(n_p) - R(n_p=2) << R(n_p=2)^3, \ \forall n_p>2$, we can make the approximation
\begin{eqnarray}
\frac{1}{R_0(n_p)^2} \approx \frac{1}{R_0(n_p=2)^2}, \quad n_p \geq 2,
\end{eqnarray}
and finally arrive to
\begin{eqnarray}
\Delta E_1 (n_p, n_p + 1)
&=& E_1(n_p + 1) - E_1(n_p) \nonumber\\
&\propto& \left[\ \frac{1}{n_p^2} -  \frac{1}{(n_p+1)^2} \ \right],
\end{eqnarray}
and therefore $\Delta E_1 (n_p, n_p + 1) > 0$ as expected.

\emph{
Doing a finer estimate, however, from Appendix \ref{Ap_B3} we can see that actually}
\begin{eqnarray}
\frac{1}{R_0^2} &\propto&
n_p^\frac{4}{5}, \quad \Rightarrow E_1(n_p) \propto - \frac{n_p^\frac{4}{5}}{n_p^2}.
\end{eqnarray}
\emph{but one has to bear in mind that all these estimates are done on a two-dimensional
model of the problem, so their significance is only relative.}

\subsection{\label{Spectrum_Ib}The $s$-spectrum (in absence of 2nd-Osc)}

For $s$-orbits, disregarding the 2nd-Osc, we know from (\ref{En_s_1}),
\begin{eqnarray}
E_1(n_s) = - E_{kin}(R = 0) = - \frac{1}{2} m_e \omega_0^2 R_0^2,
\end{eqnarray}
but applying the frequency matching condition, $\omega_1 = n_s \omega_0$,
\begin{eqnarray}
E_1(n_p) = \frac{1}{2} m_e \frac{\omega_1^2}{n_s^2} R_0^2 + E_{pot}(R = 0),
\end{eqnarray}
but, from (\ref{omega_1_p}),
\begin{eqnarray}
\omega_1 \propto \frac{1}{R_0^3},
\end{eqnarray}
so, therefore
\begin{eqnarray}
E_1(n_s) \propto  - \frac{1}{n_s^2} \frac{1}{R_0}.
\end{eqnarray}

Considering that
\begin{eqnarray}
\frac{1}{R_0} \approx \frac{1}{R_0} - \frac{1}{R_0^2} ( R - R_0 ) + \ldots,
\end{eqnarray}
with $R(n_s) - R(n_s = 1) << R(n_s = 1)^2, \ \forall n_s>1$, we can make the
approximation
\begin{eqnarray}
\frac{1}{R_0(n_s)^2} \approx \frac{1}{R_0(n_s=1)^2}, \quad n_p \geq 1,
\end{eqnarray}
and finally arrive to
\begin{eqnarray}
\Delta E_1 (n_s, n_s + 1)
&=& E_1(n_s + 1) - E_1(n_s) \nonumber\\
&\propto& \left[\ \frac{1}{n_s^2} -  \frac{1}{(n_s + 1)^2} \ \right].
\end{eqnarray}

\emph{
Actually, if we do a finer estimate using the results of Appendix \ref{Ap_B3}
we can see that actually}
\begin{eqnarray}
\frac{1}{R_0} &\propto&
n_s^\frac{2}{3}, \quad \Rightarrow E_1(n_s) \propto - \frac{n_s^\frac{2}{3}}{n_s^2},
\end{eqnarray}
\emph{where again we remind the reader that all these estimates are done on a
two-dimensional model of the problem, and under many simplifications, so their
significance is only relative.}

\subsection{\label{Spectrum_Ic}2nd-Osc in the E-spectrum?}

For the moment, we will not include $E_2^{(p)}(n_p), E_2^{(s)}(n_s)$ in the calculation.
There is a reason for this. Our secondary oscillation (2nd-Osc) expresses a resonance
behavior: the system has the capacity of absorbing/dissipating whatever power is necessary
to keep its orbital movement at a particular frequency ($\omega_1$).
Therefore, the energy corresponding to that 2nd-Osc is, always, ``immediately'' absorbed
or given to the background, and with this we mean that in a much smaller time scale than
the one corresponding to the principal oscillation (that, from the point of view of this
2nd-Osc, exhibits a practically constant behavior).
This is the main reason why we believe this energy should not be included in the estimate
of the energy spectrum.



\section{\label{Non_circular}
Non circular orbits in 2D: elliptical. Orbits with higher AM}


We have so far only addressed the (trivial) case of circular orbits in the plane
(2-dimensional problem). Bohr used elliptical orbits to account for states with
quantum number $l > 1$. This would be the starting point in our approach, as well.
More complicated stable orbits (``limit cycles'' in the literature of systems theory),
given by the nonlinear nature of the dynamical system, require a more sophisticated
analysis,
but some simple ``mind experiment'' (giving some initial condition $v_{r}^{(o)} \neq 0$
in the planar problem, for instance) may surely convince the reader that these do
actually exist.

\emph{Of course, once we have a nonlinear behavior, stability is restricted to
certain ranges for initial conditions. This is another task.
Comment on the possibility of elliptical attractors. Use of the Poincaré method.}


\section{\label{3D_model}A 3D model} 


\textit{Only some general, preliminary ideas here. Convenient to include it anyway.}

\subsection{\label{3D_approach_p}
A first approach to $p$-orbits in 3D}

When we consider general initial conditions, the new forces and/or torques (to the
leading order) are strictly normal to the ones in the planar case, so the stability
mechanism remains valid.
To make our picture consistent, the classical counterpart of the quantum LS interaction
must appear here as one of the new terms. This interaction would modulate the strength
of our stabilizing mechanism: it decreases its intensity if the magnetic moment
$\boldsymbol{\mu}_e$ is not aligned with the axis $\mathbf{z}_{1}$ for the circular
orbit.
On the other hand, we know that a dipoling interaction between magnetic moments of
electron and proton must be also present (although it does not appear as a primary
term in the Hamiltonian expansion), and it tends to anchor the relative direction
of those two moments.
The combination of this last dipoling interaction with the classical counterpart
of the quantum LS must be the key to determine the probability distribution of
some ``preliminary'' $p$-orbitals in 3D.
We say preliminary because we are also considering all the way here that the inner
structure of the nucleus is perfectly isotropic: nothing further from truth.
Perhaps those differences may well explain the lack of rotational symmetry (not
even around the axis of the orbit) of the ``real'', quantum, $p$-orbitals.

\subsection{\label{3D_approach_s}
A first approach to $s$-orbits in 3D}

For this kind of trajectories we directly include here some ideas on the extension
to 3D. On physical terms, the situation we have is one where the electron oscillates
back and forward, moving in a direction that, due to stochastic interaction with the
background, can freely precess giving rise to an spherically symmetric set of trajectories.
Of course, reorientation of the spin of the electron is needed, but this is the
consequence of the terms in Sec.~\ref{General_ic}. This is the kind of behavior
we would expect for a quantum s-orbital.
\textit{Hay que elaborar más la interpretación de cada término en Sec.~\ref{General_ic}.}


\section{\label{Simulations}Simulations} 


Numerical simulations. No simulations for the moment.


\section{\label{quantum}The quantum hydrogen: atomic orbitals}


We now relate our results to strictly quantum features of the real hydrogen atom.
With this section we complete the bridge from our SED context to QM.

\subsection{\label{Rel_AM}Classical and quantum AM}

Classical and quantum angular momentum (AM) are often treated as different
concepts with only a limited relation between them. If we let aside rotations that
are not (globally) isomorphic to the group $SO(3)$ of rotations in ordinary space, the
main difference between the concepts of classical and quantum AM is that the former can
not in general be added, because of their dependence on the chosen point in respect
to which they are evaluated.
\emph{However, the average value of a classical AM on a closed, periodic trajectory
is independent of the origin, as we will see from the following calculation.}

Consider an origin of coordinates $\mathbf{O}_1$ and let $\mathbf{r}_{1}$ be the
position vector defined in respect to it, as well as a second origin of coordinates
$\mathbf{O}_2$ such that the position vector is now $\mathbf{r}_{2}$.
Obviously, the following difference
\begin{eqnarray}
\mathbf{r}_{1,2} = \mathbf{r}_{2} - \mathbf{r}_{1} = \mathbf{O}_1 - \mathbf{O}_2,
\end{eqnarray}
is clearly a constant vector.
Now, let $\Lambda$ be a closed trajectory, with a period $T$, so
$\mathbf{r}_{i}(T)=\mathbf{r}_{i}(0)$, for $i=1,2$.
For the movement of a system of mass $m$ whose position is (simultaneously) given by
$\mathbf{r}_{1}(t)$ and $\mathbf{r}_{2}(t)$, in the respective system of reference,
it is easy to prove:
\begin{eqnarray}
\langle\mathbf{L}_{2}\rangle_{\Lambda}
&=& \frac{1}{T}\int_{0}^{T} \mathbf{L}_{2} \ \mathrm{d}t = \frac{1}{T}\int_{0}^{T} \mathbf{r}_{2} \wedge
\mathbf{p} \ \mathrm{d}t \nonumber\\
&=& \frac{1}{T}\int_{0}^{T} (\ \mathbf{r}_{1} + \mathbf{r}_{1,2} \ ) \wedge \mathbf{p} \ \mathrm{d}t \nonumber\\
&=& \frac{1}{T}\int_{0}^{T}  \mathbf{r}_{1} \wedge \mathbf{p} \ \mathrm{d}t + \frac{1}{T} \mathbf{r}_{1,2}
\wedge \int_{0}^{T} m \frac{ \mathrm{d} \mathbf{r}_{1} }{ \mathrm{d} t } \mathrm{d} t \nonumber\\
&=& \langle\mathbf{L}_{1}\rangle_{\Lambda}  + \frac{1}{T} \mathbf{r}_{1,2} \wedge
m \left[\ \mathbf{r}_{1} (T) - \mathbf{r}_{1}(0) \ \right] \nonumber\\
&=& \langle\mathbf{L}_{1}\rangle_{\Lambda},
\end{eqnarray}
Thus, the classical AM averaged over a closed orbit is independent of the
origin and can be added up, just like its quantum counterpart, which suggests
a possibility of relating both concepts.
For instance, for a $p$-orbit $\Lambda_p$,
$\langle \mathbf{L}_{1} \rangle_{\Lambda_p} = \langle \mathbf{L}_{2} \rangle_{\Lambda_p} = \hbar/2$,
and for an $s$-orbit $\Lambda_s$,
$\langle \mathbf{L}_{1} \rangle_{\Lambda_s} = \langle \mathbf{L}_{2} \rangle_{\Lambda_s} = 0$.

\subsection{\label{Orbitals}On the concept of classical ``orbital''}


Most of the time, we have been talking about trajectories, rather than orbitals.
We feel in need to establish a clearer and more convincing bridge from one concept
to the other.
We have worked with orbits, or closed trajectories, in the plane.
As a matter of fact and keeping an eye on the quantum orbitals, the extension to the
3D problem arises the need to generalize that concept to a new one, that of an
``orbital''.
We probably said before that with the term ``orbital'' we mean a collection of
points in the configuration space of the system, meeting the condition that, if
the system is inside, it would probably stay inside with overwhelming probability.
With this concept we are going two steps forward. On one side, where needed we
generalize from one particular trajectory to a set of them (for example, a set
of s-orbits containing all possible directions, so the resulting orbital exhibits
spherical symmetry). On the other, we introduce the stochastic character;
those stationary orbits, when subjected to the action of a perturbating
background, gives rise to a probability distribution.

Besides, so far we have only been always talking about ``orbits'' as ``closed''
or ``periodic'' trajectories.
In regard to this, we must say that ``stability'' can also be seen as a consequence
of a more ``chaotic'' dynamics, for example through trajectories that adopt the
so-called ``fractal'' kind of behavior:
it never goes through the same point (initial condition or state) twice, but the
evolution stays forever into some particular region of the configuration space
(space of values for the dynamical or state variables) of the system.
At this point we have already shown that, at least in an idealized model, this
purely classical stable orbitals exist, and they have a discrete character, in
the sense that each of them correspond to a particular stationary trajectory
(or a set of them).
This is indeed the case of our s-orbitals. By contrary, p-orbitals in 2D arise
from one particular circular trajectory.
Their extension to 3D requires to consider a modulation through the classical
counterpart of the quantum LS effect.

\subsection{\label{quantum_eq}
Equivalence to quantum orbitals}

\subsubsection{
The ``node'' problem}

\begin{center}
\textit{(to be updated: the LS argument does not work in the $l=0$ case, but
we can invoke Fritsche's reasoning: pure states cannot be prepared)}
\end{center}

Ultimately, our classical orbitals are characterized by a probability distribution
of the electron around the nucleus.
The density of probability may extend to infinite, due to the fact that there
is a non-vanishing probability that high values of the background field introduce
such strong oscillations in the particle movement (anyway, the bulk of the
distribution must be confined to a finite region of space).
Therefore, our picture can find quite a consistency with the purely quantum
one.
\textit{Besides, it must be said here that, while non relativistic orbitals
in the hydrogen problem show disconnected regions and nodes where the probability
density vanishes, this is not the case of the fully relativistic ones (for
instance see \cite{Szabo69}).}

\subsubsection{Phase averaging and symmetry}

\begin{center} \textit{(to be completed)} \end{center}

\begin{figure}[ht!]
\includegraphics[width=0.60 \columnwidth,clip]{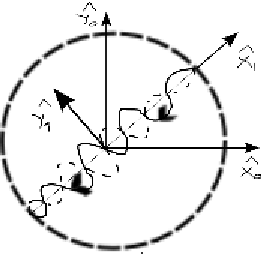} \caption{
The basic $s$-orbit that gives rise to a rotationally invariant electronic
density in 2D, when the axis $\mathbf{x}_{1}$ is allowed to precess freely
in response to stochastic perturbations.
\textbf{It may look a counterintuitive trajectory for a central potential,
but mind the intervention of the ``inner'' degree of freedom!}
} \label{Fig_s_orbital_II} \end{figure}

\begin{figure}[ht!]
\includegraphics[width=0.50 \columnwidth,clip]{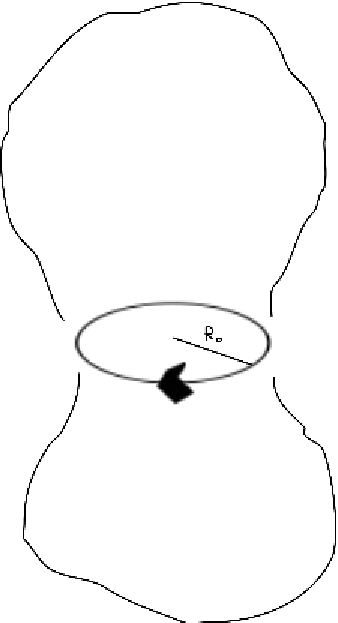} \caption{
The basic $p$-orbit that gives rise to a rotationally invariant electronic
density around the axis $\mathbf{z}_{1}$.
When we analyze the situation in 3D, we have to consider a modulation of our
stabilizing interaction from the misalignment of that axis $\mathbf{z}_{1}$
and the magnetic moment of the electron $\mu_e$ (this misalignment is nothing
but the classical counterpart of the quantum LS term).
} \label{Fig_p_orbital_II} \end{figure}

For both types of stationary orbits, the whole spectrum of stable trajectories
we just proposed exhibits several free parameters: the phase (for a given principal
axis) in the case of the circular ones, the principal direction in the second
case (s-orbits).
Now, in both cases, when averaged over all possible angles and phases, they recover
the symmetry of the lowest one.
Their associated distribution of probability is centered, nevertheless, at a
different orbital radius, as the point of balance of radiated and absorbed
power is this time different.
This is an important step towards the quantum orbitals.
Now we recall Figs. \ref{Fig_s_orbital_I} and \ref{Fig_p_orbital_I}.
With those in mind we present the following two figures: \ref{Fig_s_orbital_II} and
\ref{Fig_p_orbital_II}.

The extension of the $s$-orbits to 3D had already been discussed before,
in Sec.~\ref{3D_approach_s}. 
About the $p$-orbitals in 3D, we can only say, for the moment, that clearly
we have a trajectory that is associated with a particular direction is space.
Whether there may be only three orthogonal directions giving rise to simultaneous
stable trajectories is a question that we will left aside for the moment.

\subsubsection{$n_s,n_p$ against the quantum number $n$}

With $n$ the principal quantum number, the correspondence is completely
straightforward,
\begin{eqnarray}
n_p &\rightarrow& n, \\
n_s &\rightarrow& n,
\end{eqnarray}
with $n_p$ and $n_s$ as defined in Secs. \ref{feasible_p} and \ref{feasible_s}.
It is remarkable that in our framework, no $n_s=0$ or $n_p=0$ stationary (hence not
stable either) orbits exists, due to the presence in the equations of the 2nd-FL
that we have already commented in Sec. \ref{Feedback_2}, for instance, and therefore
we do not have to exclude that possibility as an ``ad hoc'' hypothesis.
Another question is why should we exclude the $n_p=1$ case (the first quantum
$p$-orbital corresponds to $n=2$). A closer inspection of the equation for the
orbit leads to think that, due to obvious the lack of inversion symmetry when
$n_p=1$, the implicit decoupling of the movement in an orbital (primary) and
secondary oscillations (2nd-Osc) cannot be assumed any more.

\subsection{Orbital double occupancy: ideas}

A flip of the electron spin corresponds to a sign inversion for the classical
magnetic moment of the electron, from $\mu_e \mathbf{z}$ to $- \mu_e \mathbf{z}$.
As a consequence, terms corresponding to the Lorentz force and our Inverse
Magnitude Spin Orbit (IMSO) coupling reverse sign.
This accounts for a shift in the energies (the electrostatic terms obviously
remains constant), but in a way that stability remains equally valid, both
for the $s$ and $p$-case.
In our conclusions we will comment more on this issue, even suggesting an
interpretation in terms of a classical Pauli principle, that now arises as
a consequence of our model rather than being an ``ad-hoc'' hypothesis.


\section{\label{Disc}Discussion}


Here, we try to add clarity on some points. We may also comment on alternative
approaches to some of the questions treated so far.

\subsection{\label{Disc_I}
Rad/abs in the dynamical equations}

Perhaps we would need to dedicate some more attention to the way in which we
include rad/abs in the dynamical equations of the system.
This is done in equations (\ref{vt_dynamic_c}), (\ref{vr_dynamic_c_II}) and
(\ref{w_dynamic_c}), by means of the new terms (\ref{linear_chi_t}), (\ref{linear_chi_r})
and (\ref{linear_chi_w}).
Those terms represent an unbalance of radiated and absorbed power in a quasi-instantaneous,
but not instantaneous, basis.
Therefore, they are averages, but over an interval small enough so that the dynamic
included in (\ref{vt_dynamic_c}), (\ref{vr_dynamic_c_II}) and (\ref{w_dynamic_c})
stays, effectively, ``frozen''.

Looking at (\ref{P_rad})--(\ref{P_ab}), we see the first of them is a cubic
polynomial on $\omega_{(o)}$, while the second is a fourth order one (both
positively defined for $\omega_{(o)}$ positive).
Clearly, this two curves do cross each other, the point where they do so being
modified by the extra factor $R_{orb}^{2}$. This obviously gives rise to some
``attraction'' around a certain value $\omega_{(o)}^{st}$.
Above this value, the probability of loss prevails. Below that value, the probability
of absorption prevails.
We must recall here that, in our formulation, $\omega_{(o)}$ is not one of the state
variables of the model, but the tangential velocity component
$v^{(o)}_{t}=\omega_{(o)} R_{orb}$, and also $\omega = \omega_{(R)} + \omega_{(o)}$
are so themselves.

We are interested, at least primarily, on circular orbits. Stationarity
of these imposes an extra relation on the pair $R_{orb}$, $\omega_{(o)}$,
aside from angular momentum (AM) quantization, already obtained from the
balance of average power around a closed orbit.
This condition results in the corresponding (discrete) set of pairs of values,
and therefore, it is justified to employ a prevailing dependence on $R_{orb}$ for
$\chi_{r}$, as it is done in (\ref{linear_chi_r}).
This dependence can be linearized around the point of interest.
Strictly speaking, we know that Puthoff's equations (\ref{P_rad})--(\ref{P_ab})
only apply to those circular orbits, but we have extrapolated that behavior to
a situation where the system is varying dynamically the parameters of its orbit,
and therefore this does not have to remain circular or stable at all (its stability
or instability being precisely the ultimate object of our study here).
The extrapolation seems natural in the case of (\ref{linear_chi_r}), anyway.

Less evident is our choice for (\ref{linear_chi_t}) and (\ref{linear_chi_w}).
For the first of them, we can simply state that because we have, as defined,
$v^{(o)}_{t} = \omega_{(o)} R_{orb}$, an attractor in $\omega_{(o)}^{st}$
induces one in $v^{(o)}_{t}|_{st} = \omega_{(o)}^{st} R_{orb}^{st}$.
In the case of (\ref{linear_chi_w}), we have that $\omega = \omega_{(R)} + \omega_{(o)}$
is none other than the angular velocity of rotation of a charge distribution around its
axis, always in respect to an inertial frame of reference. 
Therefore, the rad/abs game applies again, and it is natural to assume
that an attractor must exist at a certain value $\omega = \omega|^{st}$.
Indeed, we have seen, for the simplest of stationary states,
$\omega|^{st}=\omega_{(o)}|^{st}$.

The reader will agree that this is still hardly a rigorous treatment,
but we are content to show that stability is not only possible, but arises as
a natural behavior purely from the dynamical equations of the system.
Though we have made some assumptions, they are rather natural.





\subsection{\label{Disc_IV}
Relativistic considerations, ``rigidity'' and self-reaction forces}

Within our context here, all the discussion about the implications of a point particle
in electromagnetism is unnecessary. On one hand, and sufficiently away, the electron
is seen as a point-like magnetic dipole (which is conventional, again, sufficiently far).
On the other, far but not infinitely far, the proton is seen as a distribution of charge
(and magnetic momentum) that can ``rotate'' as a whole.
Besides, any reference to the rigidity of the distribution can be regarded as a
device to make calculations simpler, but, qualitatively speaking, makes no difference
in relation to our main results.
As matter of fact, throughout our development in Sec. \ref{2D}, in no moment we have
considered the forces that some elements of charge in the distribution exert over
other elements in that same distribution;
indeed, the mechanism we expose is dominant up to a certain order, whenever the two
objects remain sufficiently apart. Higher orders may deform the shape of the
distribution, or have other effects that, whenever the range of distance is the
appropriate, do not affect our main argument.

We have nevertheless already briefly commented on the (relativistic) difficulties
that arise, not only
(i) when we deal with ``rigidity'', but also
(ii) with the picture of spin as a spinning sphere, rotating around its axis,
if we try to adjust the angular velocity so as to reproduce the (phenomenological)
magnetic moment of the proton.
The difference there is made in what kind of distribution we assume.
Simple models such as those of a rotating solid sphere or a spherical shell of
charge are ruled out; on the contrary the quark model, where some discrete entities
carry charge, and more important to our purposes, all the magnetic momentum, seems
to solve those difficulties.


\section{\label{Conc}Conclusions}


\subsection{\label{Conc_1}General conclusions}

In this paper, we have tried to reasonably establish a framework to support the idea that
SED (i.e., classical Maxwellian electromagnetism plus a background of radiation), in
combination with a model of the nucleus where this presents some very elementary inner
structure (specifically, a model of the proton as a spatial distribution of charges),
is perfectly capable of explaining atomic stability.
Perhaps we should remark, before any other feature of our model, the following point:
an atomic transition implies an emission/absorption of a wave-packet with an energy
$\hbar(\omega_i - \omega_j)$, where $\omega_i,\omega_j$ are simply the orbital angular
frequencies of the basic stable orbits corresponding to the orbitals involved.
This comes as a consequence of our very basic hypothesis, rather than being introduced as
a principle as it does in QM.

Along with the ideas, we have done many specific calculations that, for the moment, reinforce
our previous qualitative approach.
They must, nevertheless, be regarded only as preliminary and we hope they will help, on future
developments, to check whether we can go forward and converge to more accurate estimates of
quantities such as the hydrogen atomic spectra.
The ultimate goal of this paper is, therefore, to settle a plausible framework, not necessarily
completely quantitative.

\subsection{\label{Conc_2}
The IMSO coupling}

The first main result presented in this paper is the identification of a classical mechanism
that appears to be capable of stabilizing the electronic orbitals against (small) perturbations.
This mechanism becomes exposed once second order moments are considered for a distribution of
charges, under the action of a point-like magnetic source.
This interaction couples the magnitude of the angular momenta of the orbital movement, with the
magnitude of the inner AM carried by the charge distribution rotating around its
axis of symmetry.

In Sec.~\ref{At_Hamiltonian} we did a quick overview of the main terms in the classical (also
quantum mechanical) Hamiltonian for the hydrogen atom. These range, as a result of a multipole
expansion for the proton, from (i) the purely Coulomb term, (ii) the Lorentz interaction,
(iii) the dipoling (or in QM, spin-spin) interaction, to (iv) the spin-orbit interaction
(or LS in QM).
That said, if we now include the second order (spatial extension) for the nucleus, we get the
terms corresponding to what we have called IMSO coupling.
The identification of this higher order interaction is done in two steps.
Having worked with a previous electromechanical model, with, in principle, no resemblance with the
actual hydrogen atom, we later depart from it and reinterpret our equations in
Sec.~\ref{Mass_reloc}.
Its presence in a realistic system (a hydrogen atom) is now established.

Another question is why we name it by ``spin-orbit''. This is done because, certainly,
it bears quite a relation to the conventional spin-orbit coupling (LS term in QM): it is,
in order, the following correction to orbital movement of the electron around the nucleus.
Nevertheless, we must stress several facts here:

(a)
\emph{First, we are talking about a coupling between the AM of the electronic
orbit and a classical inner AM of the proton (nucleus), not that of the electron.}
Through more or less complicated arguments, we link this classical AM to a residual
oscillation of the proton as a composed object (the quarks inside being point-like entities
carrying charge and spin - inner AM), in the form of a rotation around its axis
of symmetry.
The integral of this oscillation would vanish, so in average the value of the classical (inner)
AM would correspond to the quantum mechanical value: from here, our choice of
the terms ``residual spin'' (RS) and ``Inverse Magnitude Spin Orbit coupling'' (IMSO).

(b)
We have to insist in that hey are terms of a different nature.
In QM, the LS term couples an effective magnetic field, resulting from the orbital movement
of the proton from a reference frame attached to the electron, to the orientation of a
magnetic dipole (that of the electron), not changing its magnitude.
\emph{Our IMSO interaction couples the magnitude of the AM of the orbit (given
by the pair of variables $\omega_{(o)}$ and $R_{orb}$) to the magnitude (magnitude of the
oscillations) of that  ``residual spin'' of the proton (coming from a non-zero value of
$\boldsymbol{\omega}$), not to its orientation, in contrast to what the well-known LS
term in quantum atomic physics does (there, the coupled spin is that of the electron).}

(c)
\emph{Probably, the contribution on an average over a closed orbit of this new term is always
vanishing} (again in contrast to the quantum - or classical - LS shift).
A convincing general proof of this is not within our reach yet.
On a quasi-instantaneous scale (transparent to QM) this kind of effect would indeed be observable,
but at least not by conventional atomic spectrography.

Aside, because our high order interaction takes place exclusively via the Lorentz law, it is
reasonable to expect that such correction in the energy spectrum vanishes identically, although
this is a matter still to be proven more rigorously.
Indeed, we tried to shed some more light on this question with some calculations in
Sec.~\ref{En_correction_p} (Appendix).

\subsection{\label{Conc_3}Why is it stabilizing?}

Stability of the atomic orbitals is, at least at present, an exclusively quantum mechanical
prediction.
On the other hand, within the SED framework and up to now, the ZPF background provides a mean
to counteract the loss of energy that a charge undergoes when accelerated, through radiation.
However, given the stochastic nature of the ZPF, the cancellation is just on average.
Nothing, however, had been said yet on how could this balance be attained on a quasi instantaneous
basis,  or how the orbit could accommodate the instantaneous excess or defect of energy without
modifying the pair $R_{orb}$, $\omega_{(o)}$ (therefore keeping the stationarity of the trajectory,
losing it otherwise).
In other words, it does not explain why the system should remain confined to a reduced portion
of its possible configuration space, that portion corresponding to what we know by an orbital.

It does not explain, neither, why there should exist a discrete set of energies for the orbitals
with a given AM.
However, if we provide the system with the capacity of storing energy on the fluctuations in that
classical (residual) ``spin'' of one of the particles, the former picture can be now made to work.
\emph{This classical residual spin of the particle must nevertheless be able to participate in an
interplay of energy exchange with the orbital degrees of freedom, and in this paper we have shown
that the forces and torques for this are present, once that one simply allows for a slightly more
complex model of the situation, where the nucleus is permitted to have a non vanishing second
order moment.}

Now, the instantaneous excess or defect of energy is stored in the proton ``residual spin'', so
to say, the (residual) AM due to orbital movement of the quarks within the nucleon
structure.
Adopting the quark model is necessary because the quarks are carriers of net AM
and magnetic moment, avoiding that way difficulties with special relativity (those difficulties
arise when the more simple model of an spherical distribution of charge is assumed for the
electron, and also for the proton).

\subsection{\label{Conc_4}Why a discrete spectra?}

Puthoff's work accounted for the necessity of the quantization of AM,
but now, the spectrum of parameters for the stable orbits is also discrete, when we impose a
condition of ``stationarity'' over the dynamical equations of the system (Puthoff's argument did
not take them into account).
For instance, for a stable circular orbit, the possible values of both orbital radius $R_{orb}$
and the angular velocity $\omega_{(o)}$ are also discrete, aside from the overall AM.
This point is of key importance, as the previous picture only accounted for the discretization of
the spectrum of AM.
\emph{But yet, we could go the opposite way.
If, in the first place, we impose stationarity what we have is a continuum set of stationary orbits,
all of them plausible candidates to be stable, then, as a second step we can demand Puthoff's power
balance, and end up with only one orbit that would have, therefore, perfectly defined values for
its main parameters ($R_{orb}$ and the angular velocity $\omega_{(o)}$), as well as for its overall
average AM.}

Of course, this does not mean that real orbitals are reducible to neat classical orbits in that way:
under the bath of radiation, they oscillate around that privileged trajectory or set of trajectories
(this is the natural situation when we formulate the three dimensional problem).

\subsection{\label{Conc_4b}And infinite?}

We had explained the discreteness of the spectrum, but this accounts for only one (feasible)
stable trajectory for each particular value of AM.
The matter of how can we account for an infinite set of stable orbits for each of those values
of AM was the following question to be addressed.

So far, we had only addressed the orbital movement of the electron. This movement, that we had
called ``primary oscillations'', could be stabilized by a feedback loop (primary) relating radial
velocity and relative angular velocity in respect to the nucleus.
But moreover, the dynamical equations of the system also include another feedback loop, that
we had called secondary, relating the radial and tangential components of velocity.
This secondary loop introduces an additional oscillation that takes place entwined with the
principal, orbital oscillation.
\emph{Puthoff's initial calculation only took into account the primary oscillation, therefore
yielding a unique point of equilibrium of rad/abs processes.}
When the presence of the secondary oscillation (2nd-Osc) is taken into account, an infinite
set of points of equilibrium becomes possible: the excess or defect of power in the primary
oscillation can be exactly canceled by the absorption or dissipation through the secondary
movement.

In principle, that infinite set would be continuum, as well, but the inclusion of an additional,
and rather obvious from the point of view of the stationarity of the orbit, restriction relating
the frequencies of both oscillations.
This restriction recovers the discreteness of the spectrum, but now that spectrum includes an
infinite number of stable (classical) states, with increasing energies.

\subsection{\label{Conc_5}
Stable orbits with and without net AM}

The second of the main results of this paper consists of a necessary and sufficient proof
of the existence of stable orbits, though only applicable, for the moment, to the lowest
energy one for two different sets of closed trajectories.
By stable we mean an orbit that, under the effect of a (small) perturbation, tends naturally
to go back to its initial configuration. Those perturbations would correspond, in this framework,
to the instantaneous differences between emitted and absorbed radiation, once a stochastic
background is included..

As a first step, in Sec.~\ref{p_stability}, a rigorous treatment is provided only for the trivial
case of circular orbits in the plane, for which we provide a necessary and sufficient condition.
These orbits have, nevertheless, non-vanishing AM, and because of this we can presume
they do not correspond to the ground state, but to the so-called $p$-orbitals.
Another question is that those circular orbits are isotropic (around their axis of symmetry),
while real $p$-orbitals are clearly not.
This could be due to the fact that our simplistic model of the atomic nucleus assumes its isotropy,
whereas the real hydrogen nucleus posses a clear break of isotropy inherent to the quark model.
In Sec.~\ref{s_stability}, an additional (infinite) set of stable solutions is provided to account
for the whole (discrete) $s$-orbitals spectrum. These trajectories show vanishing (average) angular
momentum.
The main axis of each of these $s$-trajectories may randomly precess around the nucleus giving
rise to the expected spherically symmetric quantum $s$-orbitals.

Particularly in the case of our $s$-orbits, it is convenient to bear in mind that those trajectories
are not stable, nor they are stationary, in the simplified problem of a scalar central potential.
For them trajectories to be possible, we must add both the action of the Lorentz force on a point
particle, as well as our higher order terms arising from the multipole orders we include in the
calculation (our ``stabilizing mechanism'').
The identification of that last set of stable trajectories with vanishing AM, a set of
``candidates'' for the s-orbitals, is maybe one of the most important contributions in our present work.
Indeed, to our knowledge, it is something not present in any previous treatment on the subject (and of
course not in the Bohr model!).


\subsection{\label{Conc_5b}
Equivalence with QM orbitals}

Beyond the fact that this SED picture may give rise to stable orbits for the hydrogen
atom, \emph{the equivalence with the well-known quantum orbitals has been established
in the following ways}:

(i)
\emph{Average angular momentum (AM) over the orbit}: we have shown the existence of stable orbits
whose average AM either vanishes (for the $s$-orbits) or not (for our $p$-orbits).
See Secs. \ref{s_stability}, \ref{p_stability} and \ref{Ex_states}.

(ii)
\emph{Symmetry properties}: as seen in Sec.~\ref{quantum_eq}, once phase averaging is done,
rotational symmetry is recovered, three dimensional for the $s$-case and around an axis for the
$p$-cases..

(iii)
\emph{Parametrization of the discrete spectrum by a counterpart of the principal quantum
number}, obtained from an additional frequency matching condition.
See Secs. \ref{Matching}, \ref{s_matching} and \ref{p_matching}.


\subsection{\label{Conc_5c}Energy spectrum}

A remarkable approximation to the typical $1/n^2$, at least when we include only the principal
oscillation, is also obtained. See Secs. \ref{Spectrum_I} and the Appendix, Sec.~\ref{Ap_Spectrum_II}.
It is, anyway, a very preliminary result, given that the quantitative analysis is reduced to two
dimensions, an isotropic model for the structure of the proton (very far from a realistic one) and
the lack of a rigorous approach to the inclusion of the stochastic background.
Some of these convenient extensions are discussed briefly in Sec.~\ref{Future}.

\subsection{\label{Conc_6}A possible relation with Pauli statistics?}

This is an incidental question, but not devoid of interest at all. Basically, in our framework,
stable orbits correspond to modes of oscillation. That correspondence is one to one, which suggest
an interpretation in terms of a sort of ``classical'' Pauli principle.

\emph{
There is a very interesting interpretation of our results here. Stability of an orbit (or the orbital
generated from it) requires vibration of the nucleus in a certain mode. The presence of one of these
modes can exclude (or not) the existence of more instances of the same mode, giving rise to a sort of
Pauli principle that nevertheless we will only suggest here.}
For example, for the $s$-case everything seems to suggest that the only possible configuration for
two electrons is the one in which their spins are anti-aligned, and that no third one can be included
in the stable picture, for a given value of our principal number $n_s$. The question would require,
as said before, further attention.

\subsection{\label{Conc_6b}Extension to 3D}

A great deal of our conclusions are immediately generalizable to 3D.
For instance, we have justified three dimensional rotational symmetry
from our proposed stable $s$-orbits.
However, in general, extension to the 3D problem requires some extra
calculations that for the moment we have not faced here. In particular, the
effects of the mechanism we expose here are clearly modulated by the classical
counterpart of the LS interaction in QM, between the electron and the nucleus, when
the system is allowed to evolve in 3D.

\subsection{\label{Conc_7}Additional conclusions}

With the aim of shedding some more light on our work, in Sec.~\ref{Disc} we
have discussed the approach we adopted to include, in the dynamic equations
of the system, the effect of rad/abs as a result of the interaction with the
stochastic background.
Certainly, our approach is not the most rigorous one, although it is enough
so as to show how the equations can lead naturally to a set of privileged
trajectories or attractors.

Of course, all this only constitutes a preliminary effort, and more detailed
calculations about the specific predicted parameters of those stable orbits
(or orbitals, in a wider sense) are awaiting to be done.
Moreover, even for the case of elliptical orbits, a strict proof of stability
is still lacking. The necessary elements are there, though, and at least in
this particularly point, further work is ongoing and will be presented 
elsewhere.
\emph{
A point that is convenient to stress is that our mechanism belongs to a
``simplest one'' class. Therefore, calculations based exclusively on our
picture here do not guarantee for reasonable estimates, for instance
of the $s$ or $p$ energy spectra.
Nothing forbids, nevertheless, that further refinements of the model could
adjust more to the experimental results.
We have seen, already, that special relativity arises the need to assume a
``discrete'' charge (and magnetic moment) distribution for the proton, and
space for much more sophistication is clearly still open in this issue.}


As a last remark, David Regalado, student, has performed some calculations
on the mean speed of the electron in our proposed $s$-orbitals that agree
quite well with the accepted values for the quantum case.
This calculations will be presented elsewhere.

\section{\label{Future}On future developments}


The main problem that will occupy our time in the next future will consist of trying to extract,
in a more rigorous way than what we have done, the well known $\tfrac{1}{n^2}$ dependent-spectrum
from our framework.
In this regard, the calculations presented here can only be considered as very preliminary estimates.

For the moment, what we have proven is that the existence of an infinite, discrete, spectrum is the
natural consequence of our model, and we have even done some estimates that bear some resemblance
with the real quantities.
In any case, from here several main problems are to be addressed, each of them will presumably
further correct that first estimate:

(I) The extension of the model to 3D, that we have already treated tangentially
in some sections.

(II) The inclusion of anisotropies in the model: mainly, the evident anisotropy in
the structure of the proton, and its lack of inversion symmetry.
This correction should make our classical orbits (or sets of them, in the three dimensions,
and their associated probability distribution once background noise is included) look more
like the actual quantum orbitals.

(III) On the other hand, we have not so far predicted any specifical probability distributions,
which would in the last term define an ``orbital''.
Indeed, a very interesting extension of that work would be to predict the density of probability
that we would find considering a ``broad band'' (high frequency-cutoff) model for the fluctuations.
Up to now, we have used a ``low band'' model where the value of the ZPF was averaged over an
interval (we did this through the inclusion of the $\chi$-functions, see Sec.~\ref{rad_incl}),
small enough not to mask the dynamics of the system as an electromechanical mechanism, so we could
calculate the ``privileged'' trajectories or attractors.






\section*{Acknowledgements}


D. Rodríguez thanks R. Risco-Delgado, A. Casado, F. Barranco, J. Mart{\'i}nez,
A. Gonz{\'a}lez for useful comments on my idea, thought at a much earlier stage
of it.
This could not have been (at least) initiated without the financial support
from the Department of Applied Physics II, University of Sevilla.









\appendix


\section{\label{Ap_A}
Additional calc. on the IMSO} 

We include here several additional calculations, regarding our idealized situation, with no necessary mention
to the actual hydrogen atom. Therefore, notation here is in coherence with that of the first part of the paper:
we always use $R_{orb}$ and $\omega_{(o)}$ because they are more general and perfectly defined variables (in
contrast to $R_0, \omega_0, \ldots$, which have different meaning for $s$ or $p$-orbits).

Though not relevant enough to be included in the main body of the paper, they may be useful,
however, to interpret further results.

\subsection{\label{En_correction_p}
E-correction for $p$-orbits}

\textit{The aim of this section was to do a preliminary exploration: it seems, up to now,
that the high order terms under our focus do not produce any correction... on average over
a whole circular (closed) orbit. This seems reasonable as all the new terms stem from the
Lorentz force, for which ``work'' always vanishes (it is normal to velocity).
This question should be treated more carefully in future versions of the paper.}

We want to evaluate what shifts in energy are introduced by our IMSO coupling.
It is important to state again here that this new correction is of a completely different
nature from the well known quantum LS.
We have a coupling between (oscillating) magnitudes of classical angular momenta, not
just the one between their relative orientations in space.

This said, we will confront the fact that (in some ways ``as expected''), our correction
vanishes over a whole cyclic orbit. This is a natural consequence of our calculations.
Nevertheless, it would be interesting to think how this could be reformulated in terms of
work, taking into account the special nature of the Lorentz forces.

Specifically, we insist on the fact that a moving charge in an electrostatic field will
``see'' a magnetic field, due to Lorentz covariance.
This magnetic field produces a force that couples to the charge. In a purely classical
picture, we have seen we need to go up to a second order moment to be able to couple this
field to the AM with which  that charge is rotating around its axis of symmetry.

We have, therefore, a classical coupling between the magnitude of the orbital angular
momentum and of what we have called, with the aim of accommodating our reasoning to a
physical situation that could resemble a realistic hydrogen atom, ``residual spin''.
Hence, after all what we have is a new correction term to the classical Hamiltonian (in
the same way that one adds new terms to the quantum one).
We focus now only on this new energy correction.


On the other hand, and based on the formal equivalence of the equations exposed in
Sec.~\ref{Equivalence}, we can consider that the following calculations are valid for
either of the treated situations, the first one comprising two systems of charges, a
heavy pointlike one and a lighter extended one, the second being a first model of the
hydrogen atom.

As a first approach to characterize quantitatively this new energy correction, we
begin by equating, classically, the (radial) force with the centrifugal term. 
We will first focus on the order of the correction (then more detailed conclusions can
be obtained by using the electron, proton mass, charge, magnetic moment, etc).

For a perfectly circular orbit, we have, from (\ref{vr_dynamic}) with $\dot{v}^{(o)}_{r}=0$
and the perfect compensation (as an expectation value) of radiative loss and background pressure,
\begin{eqnarray}
F_{e} + F_{m}^{(o)} + F_{m}^{(R)} = m\ \frac{ (v^{(o)}_{t})^{2} } {R_{orb}},
\end{eqnarray}
The terms $F_{e}$ and $F_{m}^{(o)}$ are already present at any classical treatment of
the hydrogen atom, and represent, respectively, the electrostatic and the Lorentz (reciprocal,
comment on this!) forces. The terms that give rise to $F_{m}^{(R)}$ are the interesting ones.

\subsubsection{Kinetic E-correction}

The kinetic energy $T = \frac{1}{2}\ m\ (v^{(o)}_{t})^{2}$, and we can compute each
correction by:
\begin{eqnarray}
\Delta (T)_{i} \approx \frac{R_{orb}}{2}\ F_{i},
\end{eqnarray}

Specifically:
\begin{eqnarray}
F_{e} \propto R_{orb}^{-2} \ &\Rightarrow &  \Delta T_{e} \propto R_{orb}^{-1},
\label{Delta_T_e}\\
F_{m}^{(o)} \propto q B_1 \ v^{(o)}_{t} \propto R_{orb}^{-2} \ &\Rightarrow & \Delta T_{m}^{(o)}
\propto R_{orb}^{-1},
\label{Delta_T_m}\\
F_{m}^{(R)} \propto \frac{3 Q_{2} \omega }{R_{orb}^4} \propto R_{orb}^{-4} \ &\Rightarrow & \Delta T_{m}^{(R)}
\propto R_{orb}^{-3},
\label{Delta_T_m_R}
\label{kinetic_terms_order}
\end{eqnarray}
where we use the sign $\Delta$ all the time because a complete evaluation of the
kinetic energy would have to include the term in $\omega^2$.
From the perturbative point of view, we are only interested in the
former ones.

\subsubsection{Potential E-correction}

We need to compute $\Delta R_{orb}$ as a result of $F_{m}^{(R)}$, and then do,
to first order,
\begin{eqnarray}
\Delta V \ \propto\ \frac{1}{R_{orb}} - \frac{1}{R_{orb} + \Delta R_{orb}}.
\end{eqnarray}
This accounts for solving a polynomial equation.
We multiply by $R_{orb}\ \ +\ \Delta R_{orb}$ on the right,
\begin{eqnarray}
(R_{orb}\ \ +\ \Delta R_{orb})\ \Delta V \ &\propto& \ \frac{\Delta R_{orb}}{R_{orb}}, \\
(R_{orb}^2\ +\ R_{orb}\ \cdot\Delta R_{orb})\ \Delta V \ &\propto& \ \Delta R_{orb},
\end{eqnarray}
and therefore
\begin{eqnarray}
\Delta V \ \propto \ \Delta R_{orb}\ \cdot R_{orb}^{-2},
\end{eqnarray}
but we also had, from (\ref{Delta_T_m_R}), $\Delta T_{m}^{(R)} \propto R_{orb}^{-3}$.
Now, in first order to perturbation (we consider the other terms of the kinetic energy
$T_{e}$ and $T_{m}^{(o)}$ remain fixed), again selecting the leading terms,
\begin{eqnarray}
\Delta T_{m}^{(R)}
&\propto& \frac{1}{2} m \ \omega_{orb}^2 \ \left[\ ( R_{orb} + \Delta R_{orb} )^2 - R_{orb}^2 \ \right] \nonumber\\
&\propto& \frac{1}{2} m \ \omega_{orb}^2 \ \left[\ 2 R_{orb} \Delta R_{orb} \ \right] \nonumber\\
&\propto& m \ \omega_{(o)}^2 \ R_{orb} \ \Delta R_{orb},
\end{eqnarray}
and therefore,
\begin{eqnarray}
\Delta R_{orb} \propto R_{orb}^{-4},
\end{eqnarray}
so we can neglect the correction to the potential energy $\Delta V \propto R_{orb}^{-6}$.

\subsubsection{Order of the full correction}


Our first idea was to compare our estimate with the value of $\langle\psi_{GS}| \ L \cdot S\  |\psi_{GS}\rangle$,
the first correction in the perturbation expansion of the energy predicted by non-relativistic QM for the ground
state $|\psi_{GS}\rangle$ of the system.
Now, from \cite{Sakurai_ad} (page 88), we have, for a central potential,
the leading term with $L\cdot S$ has the magnitude
\begin{eqnarray}
\frac{1}{mc^2} \cdot \frac{1}{r} \cdot \frac{dV}{dr} \propto R_{orb}^{-3}, \ \ \ r=R_{orb}, \ \ \ V \propto R_{orb}^{-1},
\nonumber\\
\end{eqnarray}
for a Coulombian potential. It does agree, therefore, with the order of our
leading correction $\Delta T_{m}^{(R)}$ (!) in (\ref{Delta_T_m_R}).
Nevertheless, we must stress our correction is of a completely different nature,
coupling oscillations on the magnitude of a classical AM vector,
not on its orientation as the quantum LS does (see Sec.~\ref{At_Hamiltonian}).

\textit{Then, what is the interest of this comparison?} In nuclear physics,
phenomenological (ad hoc) terms with the LS form are introduced in order to
explain the negative binding energies (therefore, the stability) of most nuclei.
This is a very interesting question that we should treat elsewhere. 


There was, anyway, a triviality waiting to be acknowledged here: for a perfectly
circular orbit, we simply have $\Delta T_{m}^{(R)}=0$.
But still we were interested in that calculation, due to, for an elliptical orbit,
it still remains valid in the points where the radial velocity vanishes.
In the rest of the points, our correction to the kinetic energy needs to include
the velocity component in the radial direction, $v^{(o)}_{r}$ From (\ref{vr_dynamic}) we have
\begin{eqnarray}
|\dot{v}^{(o)}_{r}| \propto R_{orb}^{-4} \ &\Rightarrow&  \  |v^{(o)}_{r}| = \int{ |\dot{v}^{(o)}_{r}| \ dt } \propto R_{orb}^{-4} \nonumber\\
&\Rightarrow&\ \Delta T \propto |v^{(o)}_{r}|^2 \propto R_{orb}^{-8}, \nonumber\\
\end{eqnarray}
so therefore we can neglect that contribution and simply extend our former analysis
to any possible orbit (we have only strictly proved the existence of circular ones).

All this analysis justifies us to stick just to the leading term, and give a complete
estimate, using (\ref{F_m_R_truncated}):
\begin{eqnarray}
\Delta T_{m}^{(R)} &=&  \frac{R_{orb}}{2}\ F_{m}^{(R)} \nonumber\\
&=& \frac{3 \hat{\mu} Q_{2}} {2 \ R_{orb}^3} \ \omega_{(R)} \nonumber\\
&=& \frac{3 \hat{\mu} Q_{2}} {2 \ R_{orb}^3} \ \left[\ \omega - \omega_{(o)} \ \right],
\label{LS_estimate_prev}
\end{eqnarray}
where in the last line we have used (\ref{omega_R}). To compute the correction, we
would average the quantity $\Delta T_{m}^{(R)}$ on all the points of the orbit.
We do not tackle this calculation because we have not yet found those stable elliptical
orbits.


\section{\label{Ap_B}
Additional calculations on the H-model}


Here we provide additional calculations on the model, once this is completely particularized
to account for the hydrogen properties.
Whenever the results here are used in the body of the paper, it is always properly referenced.
Notation here corresponds to that of the last part of the paper: for convenience, and because
there is no possible ambiguity, we use $R_0, \omega_0, R_1, \omega_1$ instead of $R_0^{(s)},
R_0^{(p)}, \ldots$.

\subsection{\label{Ex_states_prev_energies}
Energies without 2nd-Osc}

\subsubsection{Energy of a stationary $p$-orbit}

A circular stationary trajectory $\Lambda_p$ is given by the pair of parameters
$\omega_0 \equiv \omega_{(o)}$ and $R_0 \equiv R_{orb}$.
Assuming $v^{(o)}_{r}=0$, we can accept the following expression, from
(\ref{p_stationarity_cond}),
\begin{equation}
R_0^3 = \frac{ q^2 }{ 4 \pi \epsilon_0 } \cdot \frac{1}{ m_e \omega_0^2 },  
\label{R_p_dependence}
\end{equation}
and therefore
\begin{equation}
R_0 \propto (\omega_0)^{ -\frac{2}{3} },  \label{R_p_dependence_II}
\end{equation}
which means that, in general, if $\omega_0$ decreases, $R_0$ will increase, which
is intuitive for a higher energy orbit.
After the inclusion of the secondary oscillation the situation will be, however,
slightly more subtle than that, with $R_0$ keeping an almost constant (and perhaps
even decreasing) value, that nevertheless will not be associated with a decrease
in energy.
Now, for a circular stationary orbit, $v^{(o)}_{t} = \omega_{(o)} R_0$, and we have
\begin{equation}
m_e \frac{ (v^{(o)}_{t})^2 }{ R_0 } = m_e \omega_0^2 R_0 = \frac{ q^2 }{ 4 \pi \epsilon_0 R_0^2 }, 
\label{p_stationarity_cond}
\end{equation}
and, therefore, in terms of energy,
\begin{eqnarray}
E_{kin}
&=&  \frac{1}{2} m_e (v^{(o)}_{t})^2  \nonumber\\
&=&  \frac{1}{2} m_e \omega_{(o)}^2 R_0^2  \nonumber\\
&=&  \frac{1}{2} \frac{ q^2 }{ 4 \pi \epsilon_0 R_0 } \nonumber\\
&=&  \frac{1}{2} q V( R_0 ) = - \frac{1}{2} E_{pot},
\label{energy_stat_prop_I}
\end{eqnarray}
with $V(R_{orb})$ the potential created by a charge with value $q>0$ at $\mathbf{r}_{0} = 0$
($\mathbf{r}_{1} = - R_{orb} \mathbf{x}_{1}$), and therefore, with,
\begin{eqnarray}
E &=& E_{kin} +  E_{pot} \nonumber\\
&=& - \frac{1}{2} E_{pot} + E_{pot} = \frac{1}{2} E_{pot},
\end{eqnarray}
where both $E_{kin}$ and $E_{pot}$ stay constant along the orbit.
On the other hand, we can also reverse (\ref{energy_stat_prop_I}), obtaining
\begin{eqnarray}
E_{pot} = - 2 E_{kin},
\end{eqnarray}
and finally,
\begin{eqnarray}
E &=& - E_{kin}, \label{En_p_1}\\
\Delta E &=&  - \Delta E_{kin} = \frac{1}{2} E_{pot}.
\end{eqnarray}

\subsubsection{Energy of a stationary $s$-orbit}

On the other hand, an $s$-type trajectory $\Lambda_s$ is for instance determined by the
pair of parameters $v_{max} = \max_{\Lambda_s} v^{(o)}_{r}$ and
$R_{max} = \max_{\Lambda_s} R_{orb} \equiv R_0$
(this last equivalence only under the approximation of harmonic dynamics both in the radial
and tangential directions).
Now, we can equate kinetic and potential energies (for this second one we adopt a convenient
absolute reference so it vanishes when it is minimum) at the extremal points (when the velocity
is maximum and when it is zero):
\begin{equation}
E_{kin}(R = 0) + E_{pot}(R = 0) = E_{pot}(R = R_0),
\end{equation}
and (with $I_z$ the $z$-moment of inertia of the nucleus),
\begin{eqnarray}
E_{kin}(R = 0) 	     &=& \frac{1}{2} m_e (\omega_0 R_0)^2 + \frac{1}{2} I_z \omega^2,
\end{eqnarray}
and
\begin{eqnarray}
E_{pot}(R = R_0) &=& - q V(R = R_0) - E_{pot}(R = 0) \nonumber\\
		 &=& - \frac{ q^2 }{ 4 \pi \epsilon_0 R_0 } - E_{pot}(R = 0), \nonumber \\
\end{eqnarray}
and finally,
\begin{eqnarray}
&E_{pot}(R = R_0) = E_{kin}(R = 0) + E_{pot}(R = 0),  \nonumber \\ \\
&E_{kin}(R = 0)   = E_{pot}(R = R_0) - E_{pot}(R = 0),\nonumber \\
\end{eqnarray}
but now, disregarding the contribution to the kinetic energy coming from $\omega$,
\begin{equation}
\frac{1}{2} m_e (\omega_0 R_0)^2 = \frac{ q^2 }{ 4 \pi \epsilon_0 R_0 }, \label{s_stationarity_cond} \\
\end{equation}
and indeed, if no other contribution to the energy is considered (no oscillation in
$v^{(o)}_{t}$ or $\omega$ are present), the average energy is
\begin{eqnarray}
E_{kin}(R = 0) &=& - E_{pot}(R = R_0), \\
E &=& - E_{kin}(R = 0),  \label{En_s_1}\\
\Delta E &=&  - \Delta E_{kin} = \frac{1}{2} E_{pot}.
\end{eqnarray}

Of course, the former calculation must assume that the potential energy is bound
in the origin $\mathbf{r}_{0} = 0$, somehow. This is however compatible with one
of the main ideas behind this paper: the introduction of the inner structure of
the nucleus in the atomic model.
If we adopt now the harmonic oscillator approximation, with $R = R_0 \cos( \omega_0 t )$
(where from $\omega_0$ does not stand for a rotation in respect to $RF_0$ but can still
be defined as an angular frequency), then $v_{r}|_{max} = R_0 \omega_0$.
Also working with (\ref{s_stationarity_cond}),
\begin{eqnarray}
\frac{1}{2} m_e (\omega_0 R_0 )^2 &=& - \frac{ q^2 }{ 4 \pi \epsilon_0 R_0 }, \\
R_0^3  &=& \frac{ q^2 }{ 2 \pi \epsilon_0 m_e \omega_0^2 },
\label{R_s_dependence}
\end{eqnarray}
and again to arrive at a similar condition as in the circular case, this time
for the parameter $R_0$,
\begin{equation}
R_0^2 \propto (\omega_0)^{ -\frac{1}{3} }.  \label{R_s_dependence_II}
\end{equation}

\subsubsection{Potential E-gap between adjacent orbits}

We recall once more that, either for the $p$ or $s$ case, and with no rad/abs,
our model admits an infinite set of stable trajectories, therefore an infinite
set of pairs $\omega_{(o)},R_{orb}$ and $v_{max}, R_{max}$.
That continuous set will be reduced to a discrete one by means of the imposition
of that balance.
Let us now consider (feasible) excited states. The main idea here is that,
either for the $s$ or $p$ cases,
\begin{equation}
\Delta E \propto \Delta E_{pot},
\end{equation}
and moreover, if $R_{orb}$ ($R_{max}$) increases, $\Delta E > 0$ because
$\Delta E_{pot} > 0$ as $E_{pot} < 0$.

See where we are going?
The same spectrum for $s$ and $p$-type trajectories demands $R_{orb} \neq R_{max}$
for a given value of the principal quantum number $n$, but this is no problem.
Let us, for the moment, just work with a generic parameter $R_n$ (corresponding
to an excited state with quantum number $n$).
It is interesting to quantify this increment between the trajectories $n$ and $n+1$ as
\begin{equation}
\Delta E \propto \Delta E_{pot} 
= \frac{ q^2 }{ 4 \pi \epsilon_0 } \left[ \frac{1}{ R_{n+1} } -  \frac{1}{ R_{n} } \right].
\end{equation}

\subsection{\label{Ap_B2}
Determining the parameters of 2nd-Osc}

\begin{center}
\textit{(correction: we write here $\hat{\mu}$ instead of $\hat{\mu}_e$... there are
some missing factors here)}
\end{center}

Aside from the harmonic oscillator approximation, we neglect the change in the
electrostatic and Lorentz parts of the potential as a result of the change in
the radial coordinate $R = |R_{orb}|$.
Those electrostatic and Lorentz terms contribute to the mean centripetal
acceleration, but nothing else, at least under our approximation.
Therefore, only the term corresponding to our high order model enters in the
following calculations.

\subsubsection{2nd-Osc for $p$-orbits}

For circular orbits, the equations of interest are (\ref{vr_dynamic}) and (\ref{w_dynamic}).
Considering that, as already stated, the secondary oscillation (2nd-Osc) can be
expressed assuming the radial direction as $\Delta R = R - R_0 = R_1 \cos{ \omega_1 t}$,
this oscillation univocally determines an associated one for $\omega_R \equiv \omega_{(R)}$
(see footnote \cite{defs_omega}):
\begin{eqnarray}
\omega_R = - \omega_0 + \omega_{R,0} \cos( \omega_1 t ), \label{sec_p_w_R}
\end{eqnarray}
where $\omega_{R,0}$ is determined from $\omega_0, R_0 \omega_1, R_1$.
Now, we can apply the following two conditions, evaluating the point of maximum acceleration,
measured as a variation of the radial velocity (we are already eliminating the centripetal
contribution, associated to $\omega_0$ in the former expression):
\begin{eqnarray}
\left[\ m_e \dot{v}^{(o)}_{r} \ \right]_{max} &=&  \left[\ \frac{ 3 \hat{\mu} Q_{2} }{ R_0^4 }\ \omega_{(R)} \right]_{max}
\Rightarrow \nonumber\\
m_e R_1 \omega_1^2 &=& \frac{ 3 \hat{\mu} Q_{2} }{ R_0^4 }\ \omega_{R,0},
\end{eqnarray}
and
\begin{eqnarray}
\left[\ M_{2} \ \dot{\omega_{(R)}} \ \right]_{max} &=&  \left[\ \frac{ 3 \hat{\mu} \ Q_{2}^{(*)} }{ R_0^4 } \ v^{(o)}_{r} \ \right]_{max}
\Rightarrow \nonumber\\
M_{2} \omega_{R,0} \omega_1  &=& \frac{ 3 \hat{\mu} \ Q_{2}^{(*)} }{ R_0^4 } \ R_1 \omega_1, \\
M_{2} \omega_{R,0} &=& \frac{ 3 \hat{\mu} \ Q_{2}^{(*)} }{ R_0^4 } \ R_1,
\label{sec_p_w_R_0}
\end{eqnarray}
naturally with $v^{(o)}_{r}|_{max} = R_1 \omega_1$.
The two former equations allow us to eliminate $ \omega_{R,0}$ and produce a relation
between $R_1$ and $\omega_1$, which is what we were looking for:
\begin{eqnarray}
\omega_1^2 
&=& \frac{3 \hat{\mu} Q_{2}}{ m_e R_0^4 }\ \omega_{R,0}
= \frac{9 \hat{\mu}^2 Q_2 Q_{2}^{(*)} }{ M_2 R_0^8 }, \\
\omega_1   &=& \frac{3 \hat{\mu} \sqrt{ Q_2 Q_{2}^{(*)}/ m_e M_2 } }{ R_0^4 }. \label{omega_1_p}
\end{eqnarray}

\subsubsection{2nd-Osc for $s$-orbits}


For circular orbits, the equations of interest are (\ref{vr_dynamic}) and (\ref{vt_dynamic}).
Considering that, as already stated, the secondary oscillation can be expressed for the
tangential direction as $\mathbf{R}_{orb} \cdot \mathbf{y}_{1} = R_1 \cos{ \omega_1 t}$.
As it already happened for circular orbit, this oscillation univocally determines another
one, in this case associated with $v^{(o)}_{r}$:
\begin{eqnarray}
v^{(o)}_{r} 		&=& \omega_0 R_o   \cos( \omega_0 t ) + \Delta v^{(o)}_{r}, 	\\
\Delta v^{(o)}_{r} 	&=& \Delta v_{r,0} \cos( \omega_1 t ),
\end{eqnarray}
where $\Delta v_{r,0}$ is determined from $\omega_1,R_1$.

Now, we can apply the following two conditions, evaluating the point of maximum
acceleration, measured as a variation of the radial velocity (again we are already
eliminating the main, centripetal contribution):
\begin{eqnarray}
\left[\ m_e \dot{v}^{(o)}_{t} \ \right]_{max} &=& \left[\ q B_1 \ \Delta v^{(o)}_{r} \right]_{max}
\Rightarrow \nonumber\\
m_e R_1 \omega_1^2 &=& q B_1 \ \Delta v_{r,0} ,
\end{eqnarray}
and
\begin{eqnarray}
\left[\ \Delta( m \dot{v}^{(o)}_{r} )\ \right]_{max} &=&  \left[\ q B_1 \ v^{(o)}_{t} \ \right]_{max}
\Rightarrow \nonumber\\
m_e \omega_1 \Delta v_{r,0}  &=& q B_1 \ R_1 \omega_1, \\
m_e \Delta v_{r,0}  &=& q B_1 \ R_1.
\end{eqnarray}

Now, the two equations allow us to eliminate $\Delta v_{r,0}$ and produce a
relation between $R_1$ and $\omega_1$, which is what we were looking for,
\begin{eqnarray}
\omega_1^2 
&=& \frac{ q B_1 }{m }\ \omega_{R,0} = \frac{ q^2 B_1^2 }{m^2 } = \frac{ q^2 \hat{\mu}^2 }{ R_0^6 }, \\
\omega_1   
&=& \frac{ q \hat{\mu} }{ R_0^3 }. \label{omega_1_s}
\end{eqnarray}


\subsection{\label{Ap_B3}
Approach to the orbital radius spectrum}

\subsubsection{$p$-orbits}

We know
\begin{eqnarray}
m_e \omega_0^2 R_0 &=& \frac{ q^2 }{ 4 \pi \epsilon_0 R_0^2 },          \nonumber\\
\omega_1 &=& \frac{3 \hat{\mu} \sqrt{ Q_{2}^{(*)}/m_e } }{ {R_0}^4 }, \nonumber\\
\omega_0 &=& \frac{1}{n_p} \omega_1 ,  					                \nonumber\\
\end{eqnarray}
and using (\ref{omega_1_p}) as well as our external (it is a supplementary
assumption) ``matching condition'' $\omega_0 = \frac{1}{n_p} \omega_1$,
\begin{eqnarray}
&\left[\ \frac{1}{n_p} \frac{3 \hat{\mu} \sqrt{ Q_{2}^{(*)}/m_e } }{ R_0^4 } \ \right]^2 R_0  = \frac{ q^2 }{ 4 \pi \epsilon_0 R_0^2 }, \\
&\frac{ q^2 }{ 4 \pi \epsilon_0 } R_0^5 = \left[\ \frac{1}{n_p} 3 \hat{\mu} \sqrt{ Q_{2}^{(*)} } \ \right]^2, \\
&R_0^5 = \frac{ 4 \pi \epsilon_0 }{q^2} \left[\ \frac{1}{n_p} 3 \hat{\mu} \sqrt{ Q_{2}^{(*)} } \ \right]^2,
\end{eqnarray}
so therefore
\begin{eqnarray}
R_0^5 \propto \frac{1}{n_p^2}, \quad 
R_0  \propto \frac{1}{ n_p^{\frac{2}{5}} }, 	\quad 
\frac{1}{R_0}   \propto n_p^{\frac{2}{5}}, 	\quad 
\frac{1}{R_0^2} \propto n_p^{\frac{4}{5}}, 	\nonumber\\
\end{eqnarray}
and so
\begin{eqnarray}
\omega_0 \propto \frac{1}{R_0^3} \propto n_p^{\frac{6}{5}}.
\end{eqnarray}

\subsubsection{$s$-orbits}


Under the harmonic oscillator approximation, the spectrum of the pendulum orbit
maximum radius follows the same rule as for circular orbits.
Indeed, from the stationarity equation, where we disregard higher order terms:
\begin{eqnarray}
m_e \omega_0^2 R_0 = \frac{ q^2 }{ 4 \pi \epsilon_0 R_0^2 },    \quad
\omega_1 = \frac{ q \hat{\mu} }{ {R_0}^3 }, 			\quad
\omega_0 = \frac{1}{2 n_p} \omega_1, 				\nonumber\\
\end{eqnarray}
and using (\ref{omega_1_p}) and the matching condition $\omega_0 = \frac{1}{n_p} \omega_1$,
\begin{eqnarray}
\left[\ \frac{1}{2 n_p} \frac{ q \hat{\mu} }{ {R_0}^3 } \ \right]^2 R_0
&=& \frac{ q^2 }{ 4 \pi \epsilon_0 R_0^2 },  \\
R_0^3 &=& \ldots,
\end{eqnarray}
and so
\begin{eqnarray}
R_0^3 \propto \frac{1}{n_p^2}, \quad 
R_0^2 \propto \frac{1}{ n_p^{\frac{4}{3}} }, \quad 
R_0   \propto \frac{1}{ n_p^{\frac{2}{3}} }, \quad 
\frac{1}{R_0} \propto n_p^\frac{2}{3}, 	     \nonumber\\
\end{eqnarray}
and therefore
\begin{eqnarray}
\omega_0 \propto \frac{1}{R_0^3} \propto n_p^2.
\end{eqnarray}

\subsubsection{Comments}

\textit{
It is surprising that $R_0$ is actually a decreasing function of $n_p,n_s$.
Nevertheless, what we do here is simply provide a feasible stable orbit, from which
a realistic probability distribution (extended, in principle, to the whole space)
could be generated once the stochastic background is included in the picture.
Actually, the inner regions or the quantum $s$ and $p$-orbitals do tend to decrease!}
Besides, in a real atom repulsion between electrons can well explain that the actual
mean distances in the orbitals increase with the principal quantum number.

\subsection{\label{Ap_B4}
Modified power balance}

So far, we have seen that given $R_0$, $\omega_0$ and $\omega_1$ are determined (the
second through our ``matching'' condition). Therefore, the only free parameter is $R_1$.
Imposing the radiated/absrobed power balance will fix it.

\subsubsection{$p$-orbits: first approach}

Firs, we notice that (\ref{P_rad}) and (\ref{P_ab}) provided instant averages,
rather than averages over the whole orbit, which is important to bear in mind
for the following calculation (otherwise, due to $\omega_0 < \omega_1$, some
extra factors would be necessary).
Now, recovering (\ref{P_rad}) and (\ref{P_ab}), we have,
\begin{eqnarray}
\langle P_{rad} \rangle
&=& \frac{ q^2 \omega_0^4 R_0^2 }{ 6 \pi \epsilon_0 c^3 } +  \frac{ q^2 \omega_1^4 R_1^2 }{ 12 \pi \epsilon_0 c^3 }, \nonumber\\
&=& \frac{ q^2 }{ 12 \pi \epsilon_0 c^3 } \left[ 2 \omega_0^4 R_0^2 + R_1 \omega_1^4 R_1^2 \right], \nonumber\\
\end{eqnarray}
as well as
\begin{eqnarray}
\langle P_{ab} \rangle
&=& \frac{ q^2 \hbar \omega_0^3 }{ 6 \pi \epsilon_0 m_e c^3 } +  \frac{ q^2 \hbar \omega_1^3 }{ 12 \pi \epsilon_0 m_e c^3 }, \nonumber\\
&=& \frac{ q^2 \hbar }{ 12 \pi \epsilon_0 m_e c^3 } \left[ \omega_0^3 + \omega_1^3 \right]. \nonumber\\
\end{eqnarray}

Now, applying $P_{rad} = P_{ab}$ and $\omega_1 = n_p \omega_0$,
\begin{eqnarray}
2 \omega_0^4 R_0^2 + \omega_1^4 R_1^2 &=& \frac{ \hbar }{m_e} \cdot \left[ 2 \omega_0^3 + \omega_1^3 \right], 		 \label{c_pow_bal_p}\\
2 \omega_0^4 R_0^2 + n_p^4 \omega_0^4 R_1^2 &=& \frac{ \hbar }{m_e} \cdot \left[ 2 \omega_0^3 + n_p^3 \omega_0^3 \right],\\
2 \omega_0 R_0^2 + n_p^4 \omega_0 R_1^2 &=& \frac{ \hbar }{m_e} \cdot \left[ 2 + n_p^3 \right],
\end{eqnarray}
and
\begin{eqnarray}
R_1^2 = \frac{1}{ n_p^4 \omega_0 } \left[\ \frac{ \hbar }{m_e} \cdot (\ 2 + n_p^3 \ ) - 2 \omega_0 R_0^2 \ \right],
\end{eqnarray}
and now we can do a preliminary calculation regarding only the highest
order terms, and in the limit where $n_p \rightarrow \infty$, taking into
account that $R_0^2 \propto \frac{1}{n_p^{4/5}}$,
\begin{eqnarray}
R_1^2 &\rightarrow& \frac{1}{ n_p^4 \omega_0 } \left[ \frac{ \hbar }{m_e} n_p^3  +  2 \omega_0 R_0^2 \ \right], \\
R_1^2 &\rightarrow& \frac{1}{ n_p \omega_0 } \cdot \frac{ \hbar }{m_e} ,
\end{eqnarray}
and therefore $R_1 \rightarrow 0$ when $n_p \rightarrow \infty$.

\subsubsection{$p$-orbits: detailed calculation}

A rigorous calculation of absorbed power $P_{ab}$ for the $p$-case implies
taking into account the following expression for the trajectory:
\begin{eqnarray}
x = R_0 \left[\ 1 + R_1 \cos (\omega_1 t) \ \right] \cos( \omega_0 t ), \\
y = R_0 \left[\ 1 + R_1 \cos (\omega_1 t) \ \right] \cos( \omega_0 t ),
\end{eqnarray}
and now we would have to write the radiation damping equation for each of
this components.
An easy path can be developing the cosine products in an addition of cosines:
we would have the sum of three harmonic oscillators for each $x$ or $y$ dimension.

\begin{center} (\textit{complete}) \end{center}

On the other hand, the former expression for radiated power can be justified
starting from the following general formula for instantaneous loss due to
radiation (Abraham-Lorentz, non-relativistic):
\begin{equation}
P^{rad} = \frac{ \mu_0 q^2 a^2} {6 \pi c},
\end{equation}
where $a$ is the acceleration modulus.
Now, for our $p$-orbits we would have
\begin{equation}
a = \frac{ (v^{(o)}_{t})^2 }{ R_{orb} } - \dot{v}^{(o)}_{r},
\end{equation}
from where we can see that, when doing the square and integrating along
a whole closed orbit, cross terms clearly vanish, and we have just the
addition of two independent contributions.

\begin{center} (\textit{complete}) \end{center}

\subsubsection{$s$-orbits}

For $s$-orbits, no departure from Puthoff's initial calculation, except
for the inclusion of a factor, is necessary, with
\begin{eqnarray}
x \ = \  R_0 \cos( \omega_0 t ), \\
y \ = \  R_1 \cos( \omega_1 t ),
\end{eqnarray}
and this time we have the typical case of two harmonic oscillators in
quadrature. 
The radiation damping equation for on oscillator in one dimension
is well known, and the calculation can follow Puthoff's steps making the
necessary distinctions. We have, at the end, two contributions, one coming from the
radial velocity and the other coming from the tangential one.
The calculation is similar to the $p$-case, but some factors change, as now the
frequency matching condition is $\omega_1 = 2 n_s \omega_0$, to keep zero average
AM.
Applying $P_{rad} = P_{ab}$ and $\omega_1 = n_p \omega_0$,
\begin{eqnarray}
\omega_0^4 R_0^2 + \omega_1^4 R_1^2 &=& \frac{ \hbar }{m_e} \cdot \left[ \omega_0^3 + \omega_1^3 \right], 			 
\label{c_pow_bal_s}\\
\omega_0^4 R_0^2 + 16 n_s^4 \omega_0^4 R_1^2 &=& \frac{ \hbar }{m_e} \cdot \left[ \omega_0^3 + 8 n_s^3 \omega_0^3 \right], \\
\omega_0 R_0^2 + 16 n_s^4 \omega_0 R_1^2 &=& \frac{ \hbar }{m_e} \cdot \left[ 1 + 8 n_s^3 \right],
\end{eqnarray}
and
\begin{eqnarray}
R_1^2 =
\frac{1}{ 16 n_s^4 \omega_0 } \left[\ \frac{ \hbar }{m_e} \cdot (\ 1 + 8 n_s^3 \ ) - 2 \omega_0 R_0^2 \ \right].
\end{eqnarray}



\subsection{\label{Ap_Spectrum_II}
2nd-Osc in the E-spectrum?}


In \ref{Spectrum_Ic}, we have commented on why we would not include the 2nd-Osc
into the energy spectrum.
Here we nevertheless do some extra calculations to have better elements of judgement.
Before proceeding, we recall that once the secondary oscillation is included in the
picture, there are (at least) two main contributions to consider: 
(i) a kinetic energy coming from the movement of the electron as a point particle
(contributions to the potential energy vanish on average for this movement), and
(ii) a kinetic contribution arising from oscillations in $\omega$, i.e., the nucleus
(proton) rotating as a whole.
As usual, the sum of the two terms is constant, and at the extreme points of this
oscillation, always one of them vanishes, simplifying the calculation.


In particular, following our calculations in Appendix \ref{Ap_B4}, eq. (\ref{omega_1_p}),
we could work on the power balance equations of the $p$-orbits, estimating the (kinetic)
energy of that secondary oscillation.
From (\ref{c_pow_bal_p}),
\begin{eqnarray}
2 \omega_0^4 R_0^2 + \omega_1^4 R_1^2 = \frac{ \hbar }{m_e} \cdot \left[ 2 \omega_0^3 + \omega_1^3 \right],
\end{eqnarray}
and now we can isolate
\begin{eqnarray}
E_2(n_p) &=&
\frac{1}{2} m_e  \omega_1^2 R_1^2 = \frac{1}{2} \frac{1}{\omega_1^2} \left[ \hbar (2 \omega_0^3 + \omega_1^3) - m_e \omega_0^4 R_0^2 \right],
\nonumber\\ \\
E_2(n_p) &=&
\frac{1}{2} m_e  \omega_1^2 R_1^2 = \frac{1}{2} \frac{1}{n_p^2} \left[ \hbar (2 + n_p^3) \omega_0 - m_e \omega_0^2 R_0^2 \right],
\nonumber\\
\end{eqnarray}
where the only interesting term is the one that grows like $n_p \omega_0$.
This is not a cause of concern, however, because $\omega_0 << m_e \omega_0^2 R_0^2$
in general, and we should also bear in mind that all this is no more than a very
gross approximation.
Besides, at higher frequencies higher order terms in the multipolar development
must take action, surely with the consequence that a high-frequency cut-off can be
effectively introduced in the model.
That cut-off would eliminate the problem that the term in $\omega_1^3$ poses for
our prediction of the spectrum.

Also, with $m_e \omega_0 R_0^2 \approx \hbar$ (now Puthoff's condition is not
exactly fulfilled for the orbital movement, but we may still be able to approximate),
\begin{eqnarray}
E_2(n_p) \approx \frac{1}{2} m_e  \omega_1^2 R_1^2 \approx \frac{1}{2} m_e \omega_0^2 R_0^2, \\
\nonumber
\end{eqnarray}
where the variables on the right hand side, $R_0$, $\omega_0$, etc would also
depend on $n_p$.
Also, from (\ref{c_pow_bal_s}), we do the same for the $s$-orbits, arriving to
a similar result,
\begin{eqnarray}
E_2(n_s)
\ \approx\ \frac{1}{2} m_e  \omega_1^2 R_1^2
\ \approx\ \frac{1}{2} m_e  \omega_0^2 R_0^2,
\end{eqnarray}
but, as seen in Appendix \ref{Ap_B}, either for circular or pendulum orbits,
this means precisely
\begin{eqnarray}
E_2 \ \approx\ E_{kin,1} \ =\  - E_1,
\end{eqnarray}
and therefore here, surprisingly only in principle (read below), with $n\equiv n_p,n_s$,
the gap between adjacent states obtained from the primary oscillation would simply get
cancelled out: $\Delta E_2(n) = E_2(n+1) - E_2(n) = - \Delta E_1(n, n+1)$.

\textit{\textbf{
The interpretation of this last result is still an open question for us. Our guess
is that the secondary oscillation must be regarded as as a phenomenon that is
``transparent'' to QM:}
there needs to be, certainly, some secondary wave-packet emission/absorption
associated to an atomic transition, though at a higher range of frequency (such
high frequency perhaps explains why they are not detected, remaining completely
transparent to ordinary QM).}





\begin{center}
(bibliography is not complete)
\end{center}




\begin{thebibliography}{99}
\bibitem{rel_spin}
The picture of spin from a classically spinning - structured - particle collides
with special relativity, this has been known for eighty years or so.
Anyway, taking into account that special relativity is nothing more than a
reinterpretation of Maxwell's laws... are those QED-renormalizacion procedures
(cut-offs on high momentum transfers) something different than renouncing (or
tuning down) the laws of electromagnetism at the smallest range of distances?

Anyway, the issue does not affect our developements, as we define the electron
and quarks as point particles carrying charge and magnetic moment, the classical
 ``resiual spin'' that we deal with here having a negligible contribution to
phenomenological values of that magnetic moment.
\end{thebibliography}
\end{document}